\documentclass{jfm}
\usepackage{graphicx}
\usepackage{epstopdf, epsfig}
\usepackage[utf8]{inputenc}
\usepackage{amssymb}
\usepackage{amsmath}
\usepackage{setspace}
\usepackage{float}
\usepackage{fancyhdr}
\usepackage{subfigure}
\usepackage{mathrsfs}
\usepackage{natbib}

\usepackage{color}

\newcommand{\ii}[0]{\mathrm{i}}

\newcommand{\ee}[0]{\mathrm{e}}

\shorttitle{Forcing statistics in resolvent analysis}
\shortauthor{P. A. S. Nogueira et al.}

\title{Forcing statistics in resolvent analysis: application in minimal turbulent Couette flow}

\author{Petrônio A. S. Nogueira \aff{1}
  \corresp{\email{petronio@ita.br}},
  Pierluigi Morra\aff{2},
  Eduardo Martini\aff{1},
  André V. G. Cavalieri \aff{1},
 \and Dan S. Henningson\aff{2}}

\affiliation{\aff{1}Divisão de Engenharia Aeronáutica, Instituto Tecnológico de Aeronáutica, São José dos Campos, SP, Brazil
\aff{2}Department of Mechanics, Linn\'{e} FLOW Centre, SeRC, KTH Royal Institute of Technology, SE-100 44 Stockholm, Sweden}

\graphicspath{{FiguresNew/}}

\begin{document}

\maketitle

\begin{abstract}
    An analysis of the statistics of the non-linear terms in resolvent analysis is performed in this work for turbulent Couette flow at low Reynolds number. Data from a direct numerical simulation of a minimal flow unit, at Reynolds number 400, is post-processed using Fourier analysis in both time and space, leading to the covariance matrix of the velocity. From the same data, we computed the non-linear terms of the Navier-Stokes equations (treated as forcing in the present formulation), which allowed us to compute the covariance matrix of the forcing for this case. The two covariances are related exactly by the resolvent operator; based on this, we explore the recovery of the velocity statistics from the statistics of the forcing as a function of the components of the forcing term. This is carried out for the dominant structures in this flow, which participate in the self-sustaining cycle of turbulence: (i) streamwise vortices and streaks, and (ii) spanwise coherent fluctuations of spanwise velocity. The present results show a dominance by four of the non-linear terms for the prediction of the full statistics of streamwise vortices and streaks; a single term is seen to be dominant for spanwise motions. A relevant feature observed in these cases is that forcing terms have significant coherence in space; moreover, different forcing components are also coherent between them. This leads to constructive and destructive interferences that greatly modify the flow response, and should thus be accounted for in modelling work.
\end{abstract}


\section{Introduction}
Coherent structures have been studied in turbulent flows for some time now. Findings in that area led to a change in view: instead of considering turbulence as completely stochastic, this is now seen as having a clear coherent motion among the apparent unsteady, chaotic field. In turbulent jets, for instance, structures governed by the Kelvin-Helmholtz instability were found to be important not only for transition \citep{michalke1964inviscid}, but also for the sound radiation at shallow angles \citep{cavalieri2012axisymmetric}. In wall-bounded flows, streamwise elongated, spanwise organised structures (called streaks), found firstly by \cite{kline1967structure}, are also ubiquitous in shear flows. These structures are found for all kinds of turbulent shear flows, including channels \citep{gustavsson1991energy}, pipes \citep{hellstrom2011visualizing}, and even round jets \citep{nogueira2019large}.

The mechanism behind streak formation was firstly studied by \cite{ellingsenpalm1975}, later complemented by \cite{landahl1980note}. In their study, they concluded that the presence of shear in the mean flow and a non-zero wall-normal velocity induce momentum transfer between different layers of the fluid. If streamwise vortices are present, for instance, these generate streaks via a non-modal, linear mechanism such that, when fluid is lifted from high- to low-speed regions of the flow, a high-speed streak is formed (the opposite happening for a low-speed streak). This is called the lift-up effect and, considering that streaks are present in several shear flows, this effect should be an important part of the turbulent dynamics. This is explored, for example, by \cite{hamilton1995regeneration, waleffe1997} where a self-sustaining process for wall bounded turbulence was proposed. This can be summarised as following: (i) streamwise vortices in the turbulent medium generate streamwise streaks via the lift-up effect; (ii) streaks grow until the instability is triggered, leading to their breakdown; (iii) the resulting flow interacts non-linearly in order to  regenerate the streamwise vortices, thus restarting the process. The first step is well understood, considering the work of \cite{ellingsenpalm1975}, but the other stages have also received a good deal of attention. The streak breakdown was studied by \cite{schoppa1999formation} using numerical simulation and linear stability theory, revealing that the breakdown of a low-speed streak directly generates new streamwise vortices in the end of the process. This was further explored numerically \citep{jimenez1999autonomous,andersson2001breakdown}, experimentally \citep{asai2002instability} and even theoretically \citep{kawahara2003linear}, using a simplified vortex sheet model. As synthesised by \cite{brandt2007numerical}, this process can occur via either a varicose or a sinuous mode, the latter being the dominant mechanism; in both cases, the final structures resulting from the instability of a low-speed streak are elongated in the streamwise direction (quasi-streamwise vortices).

The cited works focused on the steps leading to the breakdown of streaks, looking at the self-sustaining cycle in the time domain (in order to evaluate the sequence of events); in most cases results confirm qualitatively trends observed in turbulent flows, but quantitative comparisons are difficult due to the simplifications introduced in the modelling process. Another approach is based on the analysis of the linearised Navier-Stokes operator, considering the mean field as the base flow. In this framework, linear models such as resolvent analysis \citep{jovanovic2005componentwise,mckeon2010critical} and transient growth \citep{butler1992three,schmid2001stability} are useful to obtain optimal responses from the forced linearised Navier-Stokes system (resolvent) or the structure resulting from a non-modal growth of initial disturbances (transient growth). Both analyses, albeit linear, reproduce some experimental trends \citep{del2006linear,pujals2009note, cossu2009optimal, sharma_mckeon_2013,morra2019relevance, abreu2019tcfp}. Nevertheless, some limitations are intrinsic of the models. Transient growth does not consider non-linearities in any way and focuses only on obtaining the initial disturbance that will lead to the maximum growth. Resolvent analysis, instead, considers the non-linear terms as forcing; therefore, the optimal forcing from resolvent analysis suggests the shape that the non-linear terms should have to maximise the gain between forcing and response. Usually, this analysis assumes that the statistics of the non-linear terms are uncorrelated in space (spatial white noise), which is an assumption never verified in practice, considering , for instance, that non-linear terms are zero on a wall, and should be negligible in regions of uniform flow. Moreover, non-linear terms are expected to have a specific structure, as they result from the turbulent flow field which itself has a level of organisation; hence, attempts have been made to identify the colour of such "forcing" terms using some flow statistics \citep{zare_jovanovic_georgiou_2017}. Another way to consider the structure of non-linear terms is by introducing an eddy viscosity model on the linearised Navier-Stokes operators \citep{del2006linear,pujals2009note,illingworth_monty_marusic_2018,morra2019relevance}, which models at least part of non-linear effects as turbulent diffusion on the flow structures. 

Finally, a method avoiding the simplifications above is to consider the exact covariance of the non-linear terms in the analysis. By taking the exact covariance of the non-linear terms into account (or the cross-spectral density if the analysis is carried out in the frequency domain), the exact response of the system can be obtained. To the best of the authors' knowledge, this has not been attempted for turbulent flows, but only for model problems such as the forced Ginzburg-Landau equation \citep{towne2018spectral,cavalieri2019amr}. A notable exception is the work of \cite{chevalier_hoepffner_bewley_henningson_2006}, where time covariances of the non-linear terms are obtained from a simulation of turbulent channel flow, and subsequently used, assuming white noise in time, to design estimators of flow fluctuations using Kalman filters. However, the use of the exact covariance of the non-linear forcing term can be prohibitive for complex flow cases. Therefore, an approximation or modelling of the forcing term is usually needed to reduce the number of variables to be computed/modelled, and may also lead to insight on the relevant physics, as the dominant mechanisms of excitation of flow structures may be thus isolated. An analysis of forcing covariances should reveal the degree of organisation of the excitation, showing how significant are the departures from the simplified white-noise assumption.


This work focuses on the analysis of the non-linear forcing term in the resolvent analysis of a turbulent Couette flow, here studied in the minimal computational box of \cite{hamilton1995regeneration}, which allows a full determination of the cross-spectral densities of forcing and response without any simplifying assumption. Reduction of the complexity of forcing statistics can be performed subsequently in modelling steps, whose accuracy may be evaluated by computing flow responses to simplified forcings using the resolvent operator. The paper is divided as follows: In \S~\ref{sec:numericalapproach} the relevant parameters of the simulation are defined, followed by the methods showing how the covariance of the velocity fluctuations is obtained from the covariance of the non-linear terms of the Navier-Stokes equations. After that, in \S~\ref{sec:reducedforcing} we focus on obtaining the relevant parts of the non-linear terms (the ones that will generate the bulk of the covariance of the response) for two different cases (streaks, and streamwise oscillations of spanwise velocity). The paper closes in \S~\ref{sec:selfsust}  with a connection between this analysis and the self-sustaining cycle characteristic of turbulent Couette flow.

\section{Numerical approach}
\label{sec:numericalapproach}
As a first attempt in such a study, we consider a turbulent flow, with a small number of relevant scales, which implies a small number of dominant frequencies and wavenumbers. The case chosen for the study is the Couette flow in the minimal flow unit. Such a flow case retains salient features of turbulent flows, like the dominance of streaks by the lift-up effect and the self-sustaining process defined by \cite{hamilton1995regeneration,waleffe1997}, and minimises computational power requirements. The minimal flow unit for Couette flow is defined below.

\subsection{The Minimal Flow Unit}
The flow case chosen is the one explored by \cite{hamilton1995regeneration}. It is defined by the smallest box and the smallest Reynolds number in which turbulence can be sustained without any external forcing. For Couette flow, the minimal box has dimensions $(L_x,L_y,L_z)=(1.75\pi h,2h,1.2\pi h)$, denoting lengths in the streamwise, wall-normal and spanwise directions, respectively; $h$ is the channel half-height. The discretisation was chosen as $(n_x,n_y,n_z)=(32,65,32)$ before dealiasing in the wall-parallel directions, which gives a slightly higher resolution than the one used in \cite{hamilton1995regeneration}. We consider Couette flow with walls moving at velocity $\pm U_w$; the Reynolds number for this case is $Re=400$, based on wall velocity $U_w$ and half-height $h$. For this flow this leads to a friction Reynolds number $Re_\tau \approx 34$, based on the friction velocity. From now on, all quantities are non-dimensional values based on outer units, with $h$ and $U_w$ as reference length and velocity. The simulation is performed in SIMSON, a pseudo-spectral solver for incompressible flows \citep{chevalier2007simson}, with discretisation in Fourier modes in streamwise and spanwise directions, and in Chebyshev polynomials in the wall-normal direction. 

\begin{figure} 
\centering
\subfigure[Mean flow]{\includegraphics[clip=true, trim= 0 0 0 0, width=0.5\textwidth]{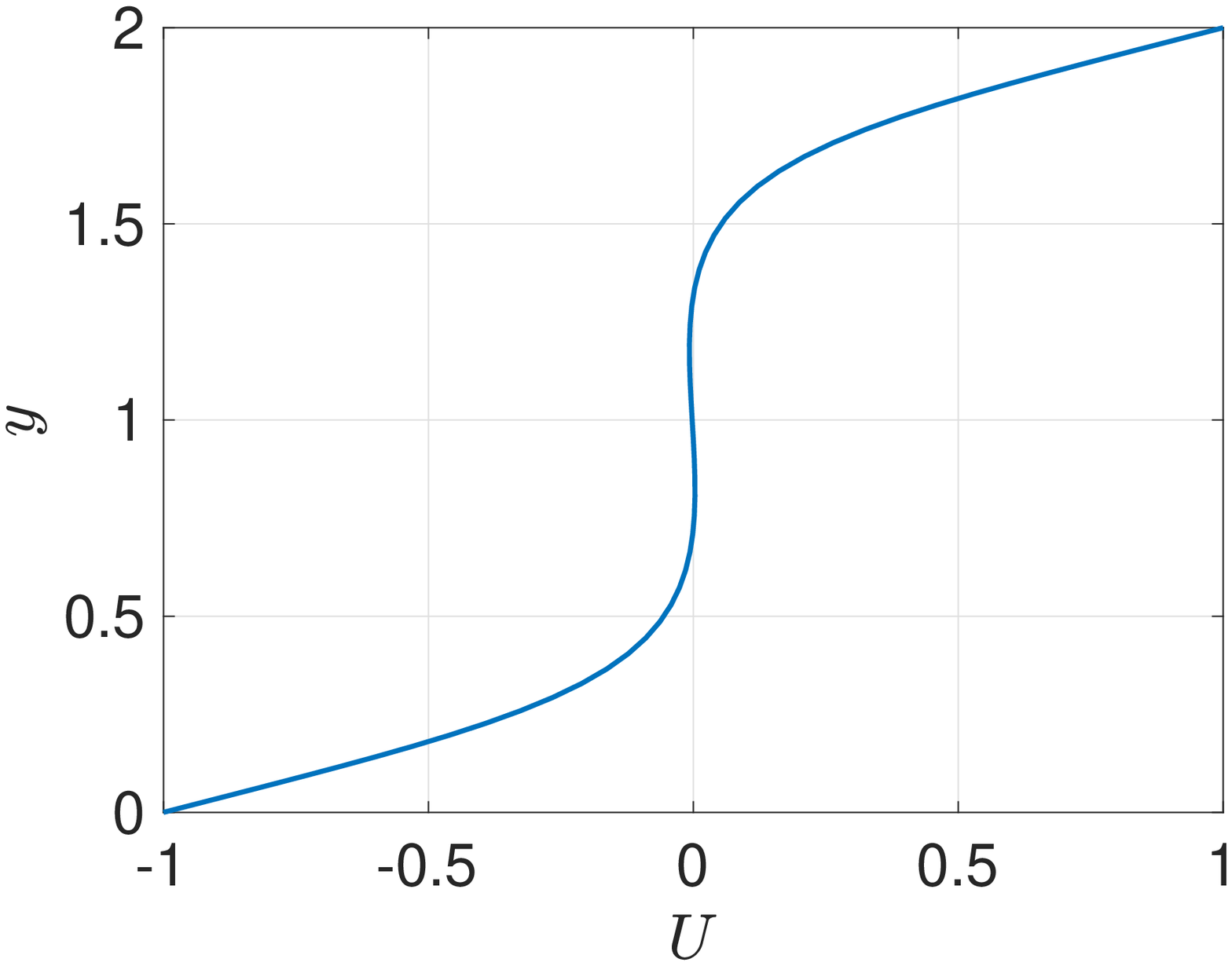}}\subfigure[Snapshot of velocity field]{\includegraphics[clip=true, trim= 0 0 0 0, width=0.5\textwidth]{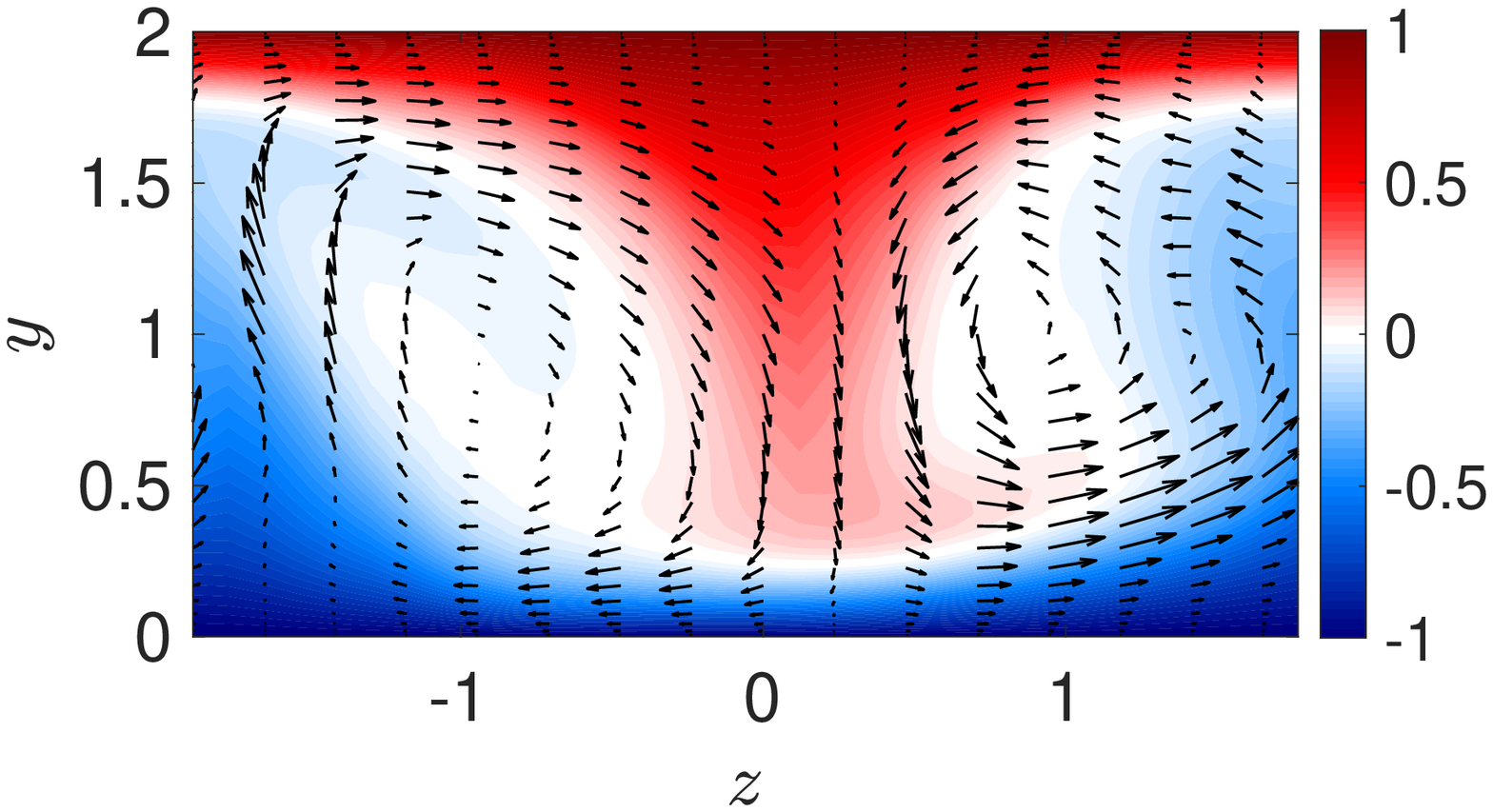}}
\caption{Mean velocity for Couette flow and snapshot of the velocity field in the $(y,z)$ plane (colours: streamwise velocity; arrows: spanwise and wall-normal velocities).}
\label{fig:Umean}
\end{figure}

The simulation is initialized with white noise in space for $Re=625$, which becomes turbulent after a few timesteps; the Reynolds number is, then, slowly decreased until the desired value of $400$. After reaching the desired value of $Re$ and discarding initial transients, flow fields are stored every $\Delta t=0.5$ in the interval $10000<t<25000$. During this period, several streak regeneration cycles are observed, and the flow is expected to be statistically stationary. The mean turbulent velocity profile is shown in figure \ref{fig:Umean}(a), which has the usual ``S'' profile typical of turbulent Couette flow. The mean profile and fluctuation levels match previous results for the same computational domain and Reynolds number~\citep{gibson_halcrow_cvitanovic_2008}. A snapshot of the streamwise velocity fluctuations for a $(y,z)$ plane is shown in figure \ref{fig:Umean}(b). As expected, streaks arise clearly in the velocity field, since these structures are the most relevant in the turbulent dynamics of this sheared flow; they are flanked by streamwise vortices with the expected lift-up behaviour: positions with downward (respectively upward) motion display lifted high (low) momentum, leading to a high-speed (low-speed) streak.

Using the results from the DNS, the velocity fluctuations around the mean flow profile (treated as response of a linearised system) were computed; these, in turn, were used to compute the non-linear terms of the Navier-Stokes equations, which will be treated as a forcing term in the N-S equations linearised around the mean flow. From these terms, we computed the cross-spectral densities of the response ($\mathrm{S}$) and of the forcing ($\mathrm{P}$). This is detailed in the next section.

\subsection{Recovering response statistics from the forcing cross-spectral density}
\label{sec:PqqfromPff}

\begin{figure} 
\centering
\includegraphics[clip=true, trim= 0 0 0 0, width=0.5\textwidth]{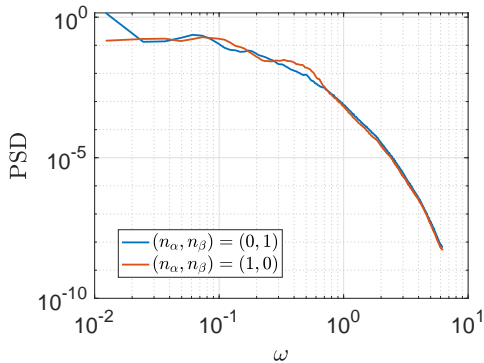}
\caption{Power spectral density of the wall-normal integrated kinetic energy for the different modes studied herein.}
\label{fig:SpectrumBase}
\end{figure}

Since Couette flow is homogeneous in $x$ and $z$ directions, the first step for the analysis of this flow is to perform a spatial Fourier transform of both response and forcing, so the analysis can be performed separately for each wavenumber. Modes are defined by the integers $(n_\alpha,n_\beta)$, with $\alpha=2 \pi n_\alpha/L_x$ and $\beta=2 \pi n_\beta/L_z$ being the wavenumbers, following the same notation of \cite{hamilton1995regeneration}. The integrated kinetic energy of the first two modes, some of the most relevant ones in the analysis of \cite{hamilton1995regeneration}, is shown in figure \ref{fig:SpectrumBase}, showing the fluctuations peak at vanishing frequency $\omega \to 0$ {for mode $(0,1)$, and a plateau is observed at low frequencies for mode $(1,0)$}. Since we consider a Couette flow with walls moving at opposite velocities, $\pm U_w$, this reflects that the two modes peak at {approximately} zero phase speed, without preferred motion following some streamwise or spanwise direction. The time power spectral density (PSD) for each mode was estimated using the Welch method, with the signal divided in segments of $n_{fft}=1024$ with $75\%$ overlap, which led to 114 blocks for the analysis. A Hanning window was used to reduce spectral leakage, allowing accurate relation between forcing and response of the system.

We will study the two most energetic modes at low frequencies, as showed in \cite{hamilton1995regeneration}: mode $(0,1)$, which is related to the appearance of streaks; and mode $(1,0)$, which emerge once the amplitude of mode $(0,1)$ decreases, characterising streak breakdown \citep{hamilton1995regeneration}. Applying the Reynolds decomposition, we can write the incompressible Navier-Stokes equations as

\begin{equation}
    \frac{\partial \tilde{u}_i}{\partial t} + U_j \frac{\partial \tilde{u}_i}{\partial x_j} + \tilde{u}_j \frac{\partial U_i}{\partial x_j} = -\frac{\partial \tilde{p}}{\partial x_i} + \frac{1}{Re}\frac{\partial^2 \tilde{u}_i}{\partial x_j \partial x_j} + \tilde{f}_i,
    \label{eqn:LNS1}
\end{equation}

\noindent and the continuity equation 

\begin{equation}
    \frac{\partial \tilde{u}_j}{\partial x_j}=0,
    \label{eqn:ContLNS1}
\end{equation}

\noindent where Einstein summation is implied,  $\tilde{u}_i$ are the three velocity components, $\tilde{p}$ is the pressure and the forcing term $\tilde{f}_i$ is considered to gather the non-linear terms of the Navier-Stokes equations,

\begin{equation}
\tilde{f}_i = -\tilde{u}_j\partial \tilde{u}_i/\partial x_j.
\label{eqn:nonlinearfull}
\end{equation}

We consider an expansion around the mean flow, averaged over streamwise and spanwise directions, so that each component of the velocity can be written as the sum of mean and fluctuation fields ($U_i+\tilde{u}_i$), with the tilde indicating variables in time domain. Taking the Fourier transform in $x$, $z$ and $t$, and considering that the mean flow has only the streamwise component $U(y)$, leads to

\begin{subequations}
\begin{eqnarray}
    -\ii \omega u + \ii \alpha U u + v \frac{\partial U}{\partial y} = -\ii \alpha p + \frac{1}{Re}\left(-\alpha^2-\beta^2+\frac{\partial^2}{\partial y^2}\right) u + f_x \label{eqn:momentum1} \\
    -\ii \omega v + \ii \alpha U v = -\frac{\partial p}{\partial y} + \frac{1}{Re}\left(-\alpha^2-\beta^2+\frac{\partial^2}{\partial y^2}\right) v + f_y \label{eqn:momentum2} \\
    -\ii \omega w + \ii \alpha U w = -\ii \beta p + \frac{1}{Re}\left(-\alpha^2-\beta^2+\frac{\partial^2}{\partial y^2}\right) w + f_z \label{eqn:momentum3} \\
    \ii \alpha u + \frac{\partial v}{\partial y} + \ii \beta w =0,  \label{eqn:continuity}
\end{eqnarray}
\end{subequations}

\noindent where $\omega$ is the frequency, $(u,v,w,p)$ are the Fourier transformed perturbation quantities, and $(f_x,f_y,f_z)$ the Fourier transforms of $\tilde{f}_i$. From the equations above, we can write the incompressible Navier-Stokes equations for the perturbations in the wavenumber and frequency domain in an input-output form as

\begin{equation}
-\ii \omega \mathrm{H} \mathbf{q} =\mathrm{A} \mathbf{q} + \mathbf{f},
\end{equation}

\noindent where $ \mathbf{q} =[u \ v \ w \ p]^T$ is the output and $\mathbf{f}=[f_x \ f_y \ f_z \ 0]^T$ is the forcing term (input). The matrix $\mathrm{A}$ is the linear operator defined by Navier-Stokes and continuity equations (which is a function of the streamwise and spanwise wavenumbers $\alpha$ and $\beta$) and the matrix $\mathrm{H}$ is defined to zero the time derivative of the pressure. This can be rewritten in the usual resolvent form

\begin{eqnarray}
(-\ii \omega \mathrm{H} - \mathrm{A}) \mathbf{q} = \mathbf{f} \\
\Rightarrow \mathrm{L} \mathbf{q} = \mathbf{f} \\
\Rightarrow \mathbf{q} = \mathrm{L}^{-1} \mathbf{f} = \mathrm{R} \mathbf{f}.
\end{eqnarray}

With the equations written in this shape, optimal forcings and responses can be obtained for a turbulent flow by performing a singular value decomposition of the resolvent operator, which leads to orthogonal bases for responses $\mathbf{q}_i$ and forcings $\mathbf{f}_i$, related by gains $s_i$. {In other words, resolvent analysis amounts to solving the linear optimisation problem given by}

\begin{equation}
     s_1^2 = \max_\mathbf{{f}}\frac{\langle \mathrm{R}\mathbf{{f}},\mathrm{R}\mathbf{{f}} \rangle}{\langle \mathbf{{f}},\mathbf{{f}} \rangle},
    \label{eqn:Resolventoptim}
\end{equation}

\noindent {where $\langle \ , \ \rangle$ is the Euclidean inner product associated to the energy norm, and $s_1$ is the highest resolvent gain. The solution for equation \ref{eqn:Resolventoptim} is obtained by solving the equivalent eigenproblem}

\begin{equation}
     \mathrm{R}^H\mathrm{R}\mathbf{{f}}_i = s_i^2 \mathbf{{f}}_i \Rightarrow \mathrm{R} \mathrm{R}^H\mathbf{{q}} = s^2 \mathbf{{q}}.
    \label{eqn:cap3ResolventoptimEig}
\end{equation}

{Optimal response and the respective gain from resolvent analysis are directly comparable to the most energetic structures in the flow if the forcing is considered to be statistically white in space \citep{towne2018spectral}. In order to consider the actual statistics of the flow in this framework, we define the covariance matrices of the response $\mathrm{S}=\mathcal{E}(q q^H)$ and of the forcing $\mathrm{P}=\mathcal{E}(f f^H)$ where the $H$ superscript denotes the Hermitian, and $\mathcal{E}$ representing the expected value of a signal, which is computed by averaging realisations in the Welch method (taking the block-average of $\mathbf{q} \mathbf{q}^H$ as a function of frequency). Following \cite{cavalieri2019amr}, these quantities are related via}

\begin{eqnarray}
\mathbf{q} \mathbf{q}^H = \mathrm{R} \mathbf{f} \mathbf{f}^H \mathrm{R}^H \nonumber \\ 
\mathcal{E}(\mathbf{q} \mathbf{q}^H) = \mathrm{R} \mathcal{E}(\mathbf{f} \mathbf{f}^H) \mathrm{R}^H \nonumber \\ 
\mathrm{S}=\mathrm{R} \mathrm{P} \mathrm{R}^H. \label{eqn:PqqRPffR}
\end{eqnarray}

{If it is assumed that the covariance of the forcing is uncorrelated in space, then $\mathrm{P} = I$ and the covariance of the response is given by $\mathrm{S}=\mathrm{R} \mathrm{R}^H$. Therefore, an eigendecomposition of both sides of equation \ref{eqn:PqqRPffR} considering this hypothesis leads to}

\begin{equation}
eig(\mathrm{S})=eig(\mathrm{R} \mathrm{R}^H), \label{eqn:eigPqqRPffR}
\end{equation}

\noindent {showing that, if the statistics of the forcing follows this hypothesis, the SPOD modes (eigenfunctions of $\mathrm{S}$) are identical to response modes (left singular vectors of $\mathrm{R}$, or eigenfunctions of $\mathrm{R} \mathrm{R}^H)$.}

However, non-white forcing covariance (which must be the case in turbulent flows) must be included in the formulation in order to correctly educe the response statistics. The inclusion of $\mathrm{P}$ can be done in several ways. One of them is to directly compute the non-linear terms of the Navier-Stokes equations from simulation data. Other approaches include modelling the forcing starting from specific assumptions on the flow case \citep{moarref_sharma_tropp_mckeon_2013}, its identification from limited flow information \citep{zare_jovanovic_georgiou_2017,towne_yang_lozanoduran2018}, and modelling by means of an eddy viscosity included in the linear operator \citep{tammisola_juniper_2016,morra2019relevance}.

The present work focuses on exploring different forcing covariances and their effect on the covariance of the response. The first approach is to simply consider the forcing to be uncorrelated in space; for this case, previous analyses for wall-bounded flows has shown that even though reasonable agreement may be obtained for near-wall fluctuations \citep{abreu2019tcfp}, a mismatch is observed for larger-scale structures \citep{morra2019relevance}. One can also compute the non-linear terms directly from a numerical simulation and then determine accurately $\mathrm{P}$. If the simulation is converged and the signal processing is done properly, the result of equation \ref{eqn:PqqRPffR} using the actual forcing covariance $\mathrm{P}$ (in the sense that this quantity is directly obtained from the simulation) should be equal to the response covariance $\mathrm{S}$ computed from the same simulation.

\section{Reconstruction of response statistics from the full forcing CSD}

\subsection{Connection between forcing and response statistics -- role of the correction due to windowing}

{Although equation \ref{eqn:PqqRPffR} is exact, if $\mathrm{P}$ and $\mathrm{S}$ are estimated from a finite set of data, large errors arise in the velocity statistics recovery process using the resolvent operator, leading to non-negligible values for $|\mathrm{S}-\mathrm{R}\mathrm{P}\mathrm{R}^H|$. From the analysis of \cite{martini2019accurate} (summarised in Appendix \ref{appA} for the present case), application of windowing in the data (which is necessary for applying Welch's method) generates extra force-like terms that must be accounted, so that equation \ref{eqn:PqqRPffR} holds for the estimated covariances. The impact of these extra terms can be measured as}

\begin{equation}
Error = \frac{||\mathrm{S}_{rec} - \mathrm{S}||}{||\mathrm{S}||},
\label{eqn:errormetric}
\end{equation}

\begin{figure} 
\centering
{\includegraphics[clip=true, trim= 0 0 0 0, width=0.5\textwidth]{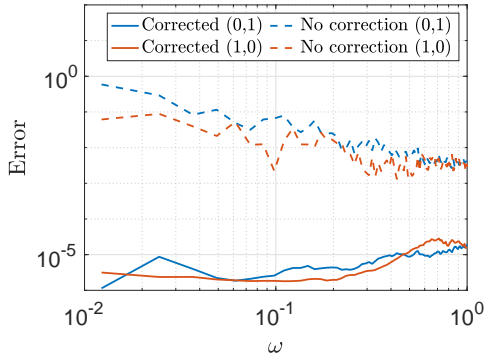}}
\caption{Effect of the correction term in the recovery of $\mathrm{S}$. Error spectrum normalised by $\Delta \omega$.}
\label{fig:errorPqq}
\end{figure}

\noindent {where $\mathrm{S}_{rec}$ is the covariance of the response recovered from the statistics of the forcing using eq. \ref{eqn:PqqRPffR}, applied for both uncorrected and corrected forcing terms; the norm was chosen as the standard $L^2$ for matrices. With this metric, we can evaluate how this correction term affects the process of recovering $\mathrm{S}$ from $\mathrm{P}$. The comparison of the errors with and without this correction term is shown in figure \ref{fig:errorPqq} for the two Fourier modes studied in this work.}

From figure \ref{fig:errorPqq}, we can see that the correction affects a wide range of frequencies, with a slightly lower effect in higher frequencies. Following \cite{martini2019accurate}, we can divide the aliasing error into two components: the first is due to spectral content at frequencies higher than the Nyquist frequency, and the other is due to the window spectral leakage, which generates extra content above the Nyquist frequency. Only the latter can be reduced  with a proper choice of windowing function. The aliasing behaviour of the error explain the larger errors obtained at higher frequencies in figure \ref{fig:errorPqq}, and indicates that the first type of aliasing is dominant in that region. As this work will focus on the study of low frequency structures, no further effort is made in order to reduce the error in higher frequencies; moreover, since the normalised errors are below $10^{-4}$ in all cases, an optimisation of the window was deemed unnecessary for our purposes.

The correction greatly improves the recovery process for all modes, leading to a reduction of the error of more than one order of about 5 orders of magnitude for the two considered modes, which is possibly related to the low-order dynamics of the flow for this combination of wavenumbers, a feature that will be further explored in \S \ref{sec:mode01} and \ref{sec:mode10}. The effect of the correction on the shapes of each case will be studied in the next section.

The role of the forcing terms is studied throughout this work. All the comparisons between forces and responses will implicitly consider the correction term, unless otherwise stated. In all other contexts ``external force'' will refer only to the term $\mathbf{f}$, without considering the correction term.

\subsection{Comparison with statistics from white-noise forcing and no correction}
\label{sec:statisticswihtorwithout}
Here, we analyse the effect of the different choices of forcing CSD $\mathrm{P}$ on the velocity statistics. As stated previously, we focus on very low frequencies in order to evaluate the effect of the statistics of the non-linear terms on the most energetic streaks, and on the equivalent structures for other combinations of wavenumbers. For that reason, we chose $\omega=0.0123$ for the analysis of modes $(0,1)$ and $(1,0)$; this is the first non-zero frequency from the Welch method for the present case (this frequency can be seen as  the limit $\omega \to 0$ in this analysis). 
This frequency was chosen so as to analyse the behaviour of the very-large, almost time-invariant structures observed in the minimal flow unit. Analogous structures have been identified by a number of authors \citeauthor{komminaho_lundbladh_johansson_1996,Tsukahara2006,pirozzoli_bernardini_orlandi_2011,pirozzoli_bernardini_orlandi_2014,lee_moser_2018} for turbulent Couette flow at higher Reynolds numbers. For the present low Reynolds number, these large structures have the same characteristic length of the near-wall streaks; since they have the same overall behaviour, it becomes harder to separate these in such a small box. The analysis of \citeauthor{rawat_cossu_hwang_rincon_2015} indicates nonetheless that the minimal flow unit streaks become very large scale structures in turbulent Couette flow if continuation methods are applied. Therefore, the analysis of mode $(n_\alpha,n_\beta)=(0,1)$ should be seen as a study of the overall dynamics of the largest streaks in Couette flow.

\begin{figure} 
\centering
\subfigure[Streamwise component]{\includegraphics[clip=true, trim= 0 0 0 0, width=0.5\textwidth]{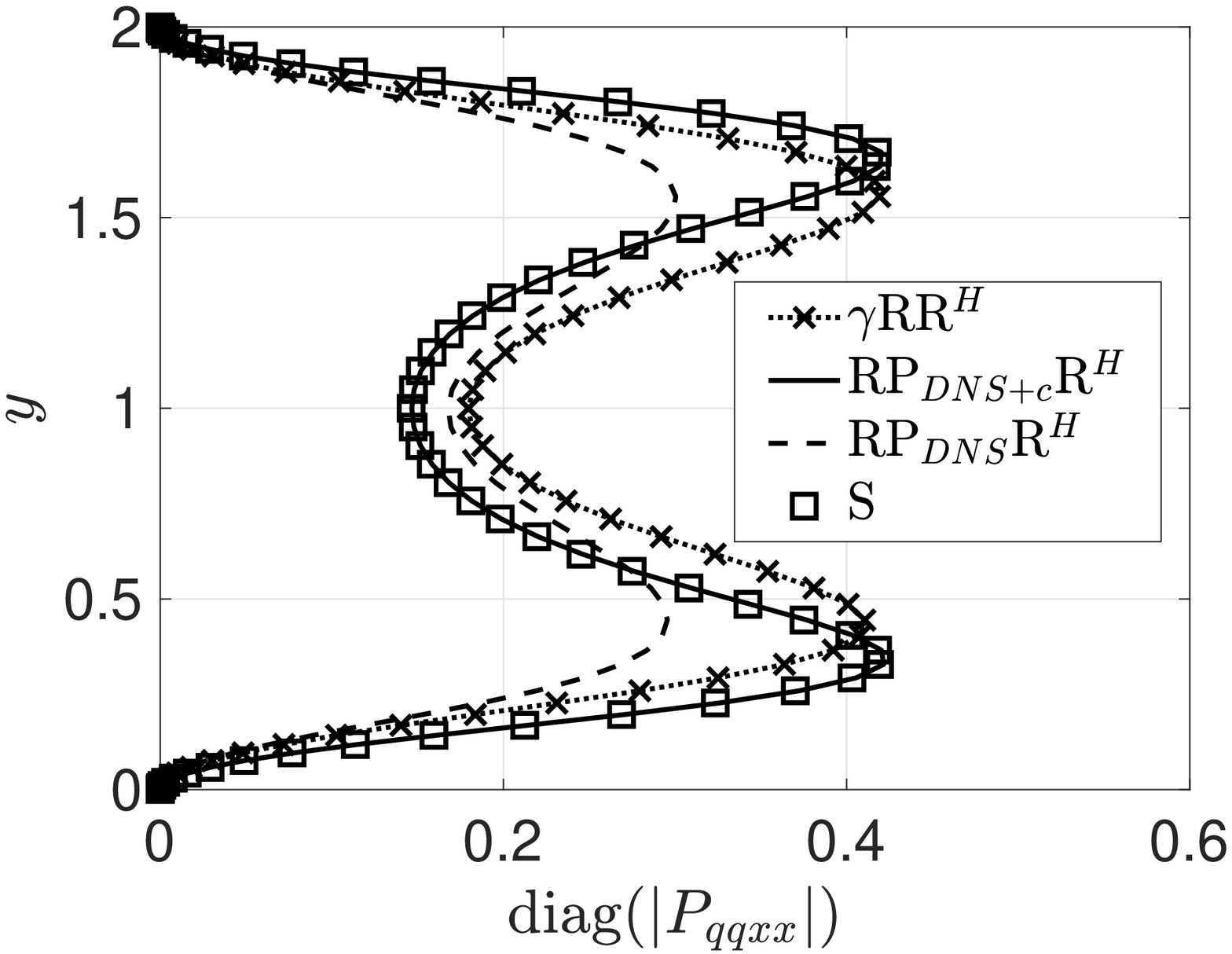}}\subfigure[Wall-normal component]{\includegraphics[clip=true, trim= 0 0 0 0, width=0.5\textwidth]{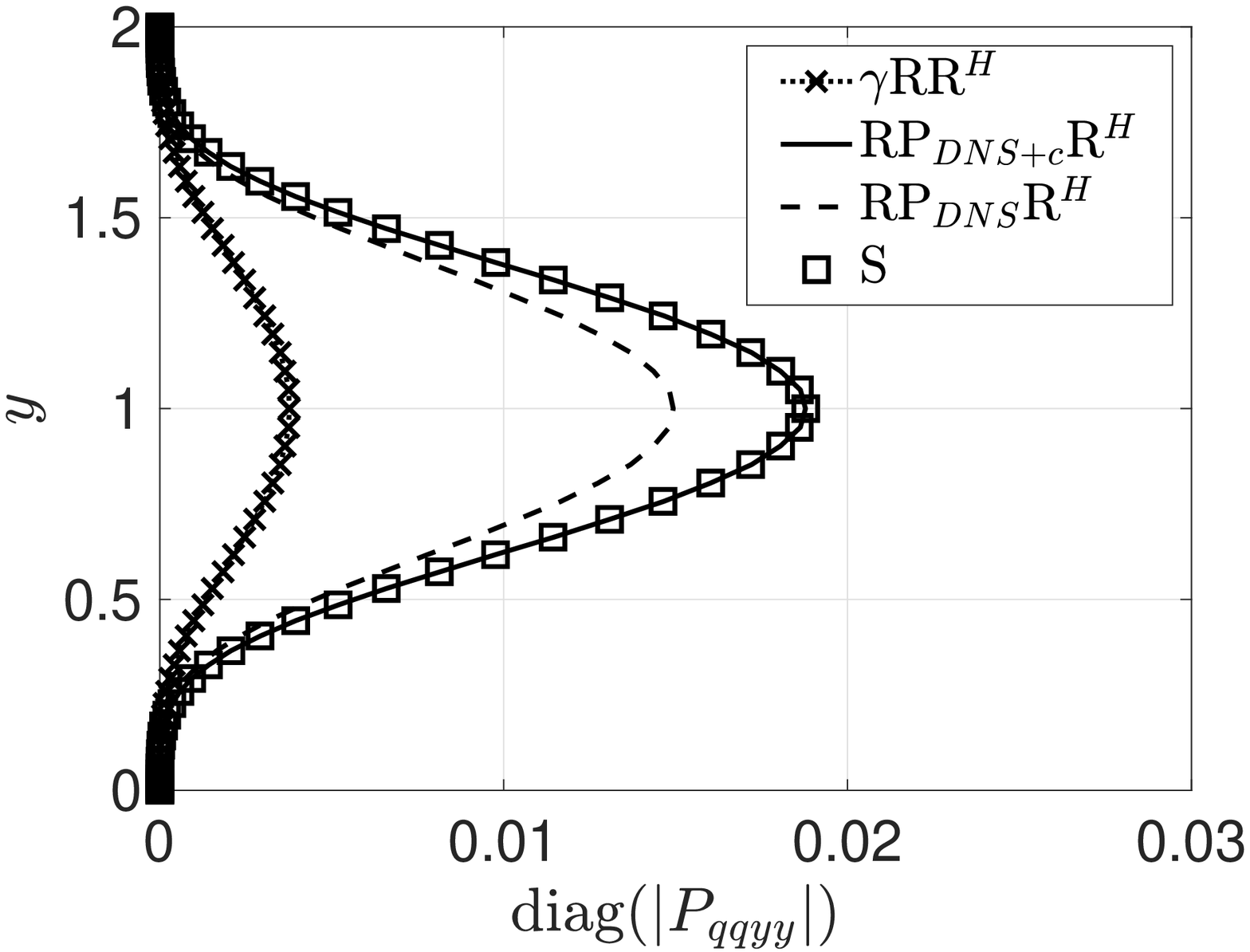}}
\subfigure[Spanwise component]{\includegraphics[clip=true, trim= 0 0 0 0, width=0.5\textwidth]{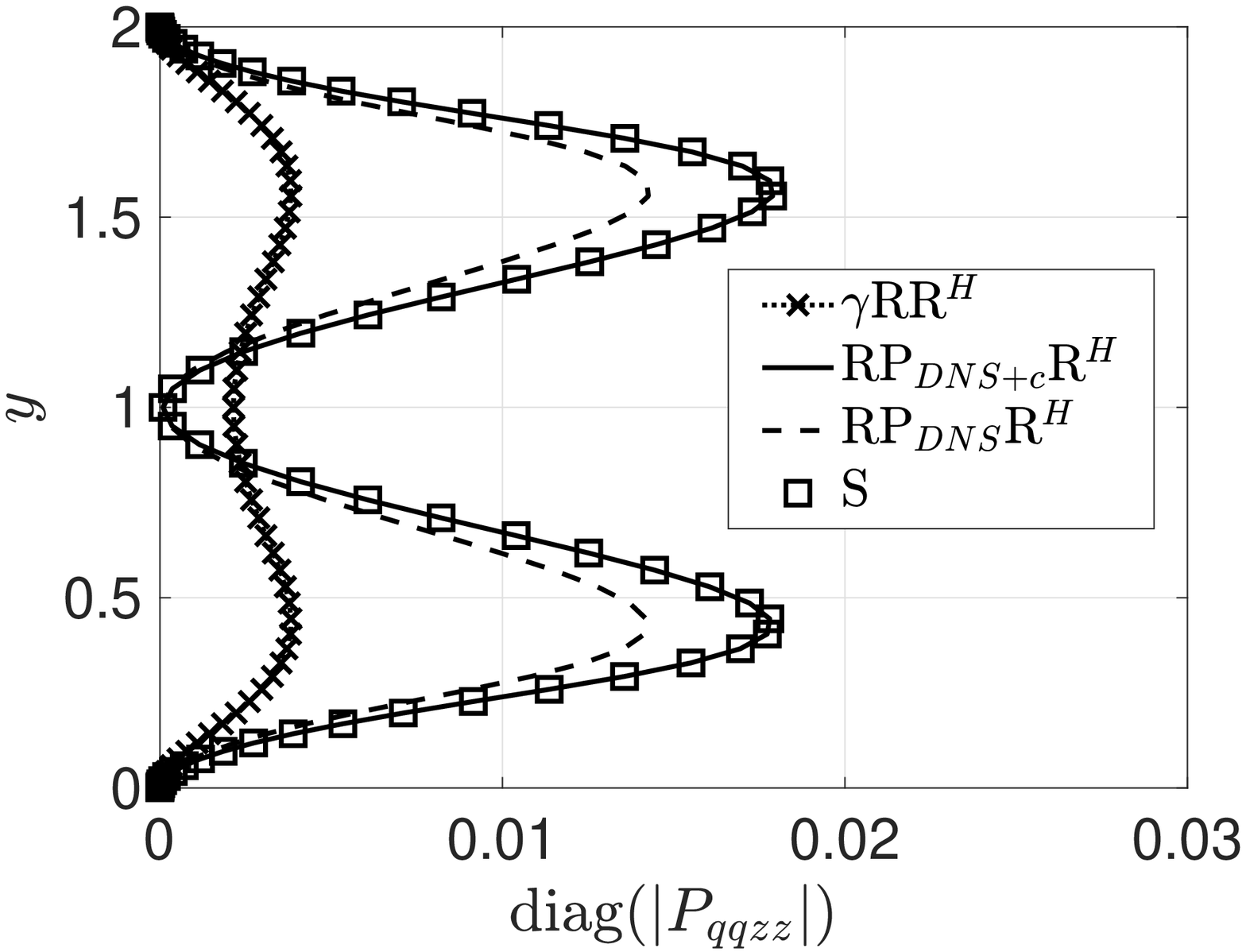}}
\caption{$\mathrm{S}$ from simulation and from the recovery process using different $\mathrm{P}$ (white-noise, uncorrected and corrected) for case $(n_\alpha,n_\beta)=(0,1)$. Prediction using white-noise $\mathrm{P}$ was scaled to match the maximum of $\mathrm{S}$ by the factor $\gamma=1.029\times 10^{-4}$.}
\label{fig:diagabsPqq01}
\end{figure}

Figure \ref{fig:diagabsPqq01} shows the comparison between the absolute value of the main diagonal of $\mathrm{S}$, which corresponds to the power spectral density (PSD) of the three velocity components. We consider the response CSD of case $(n_\alpha,n_\beta)=(0,1)$ using the statistics of the forcing obtained from the simulation without any correction  ($\mathrm{P}_{DNS}$), the one considering the correction term ($\mathrm{P}_{DNS+c}
$), and the one using the white-noise $\mathrm{P}=\gamma\mathrm{I}$ (with constant $\gamma$ chosen so as to match the maximum PSD for this frequency). It can be seen that the shape of the main component (streamwise velocity $u$) can be fairly well reproduced by simply using $\mathrm{P}=\gamma \mathrm{I}$. Streaks are represented, with peak amplitudes distant of about $0.4$ from the wall; this may be related to the higher shear near the walls, as shown in figure \ref{fig:Umean}, leading to a stronger lift-up mechanism in that region. Still, considering white-noise forcing, we can see some discrepancies in the amplitudes and in some details of the shapes of the streamwise vortices ($v$ and $w$ components). Standard resolvent analysis, considering white-noise forcing, predicts thus streamwise vortices and streaks that are in reasonable agreement with the DNS, but with a mismatch in the relative amplitudes of $u$, $v$ and $w$ components; similar results were obtained for turbulent pipe flow by \cite{abreu2019tcfp}. This will be explored in more detail in \S~\ref{sec:mode01}. For an accurate quantitative comparison, the statistics of the non-linear terms should be used; by doing that without the correction, the overall relative levels are closer to the one from the simulation, but the amplitudes are still off, especially at the peak of each component (which explains the errors seen in figure \ref{fig:errorPqq}). When the correction is taken into account, the exact shapes are recovered for all components, without any need of rescaling.

\begin{figure} 
\centering
\subfigure[Streamwise component]{\includegraphics[clip=true, trim= 0 0 0 0, width=0.5\textwidth]{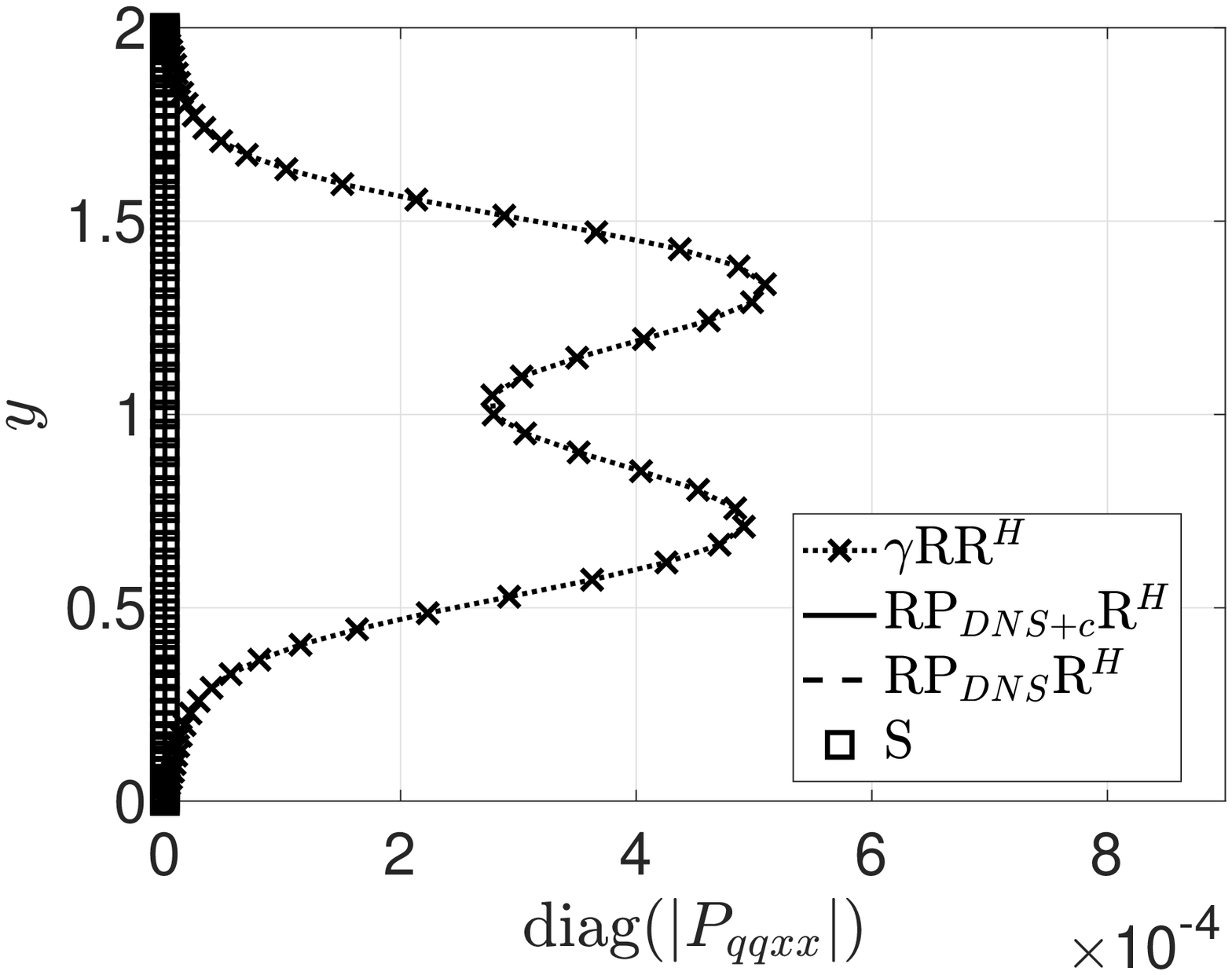}}\subfigure[Wall-normal component]{\includegraphics[clip=true, trim= 0 0 0 0, width=0.5\textwidth]{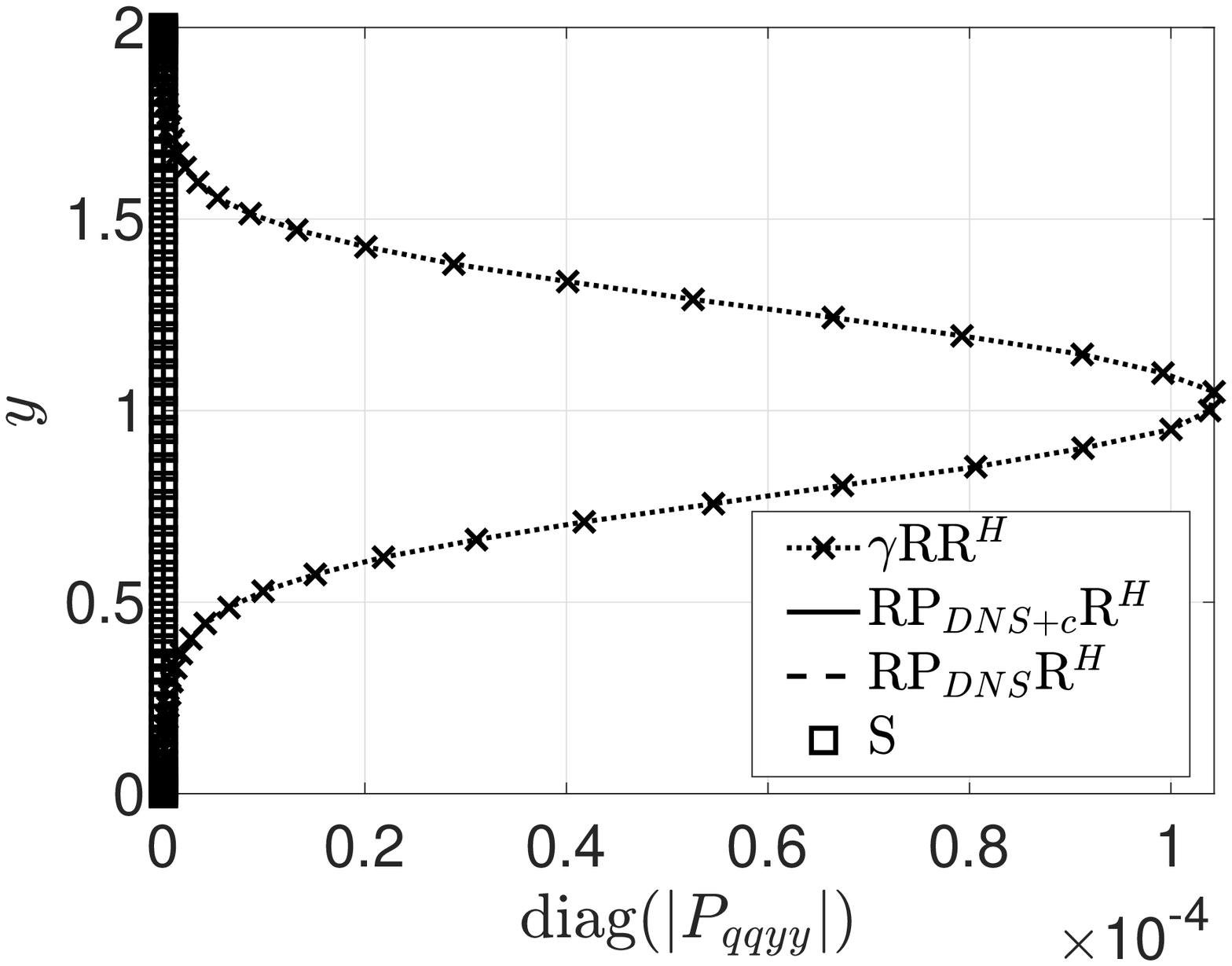}}
\subfigure[Spanwise component]{\includegraphics[clip=true, trim= 0 0 0 0, width=0.5\textwidth]{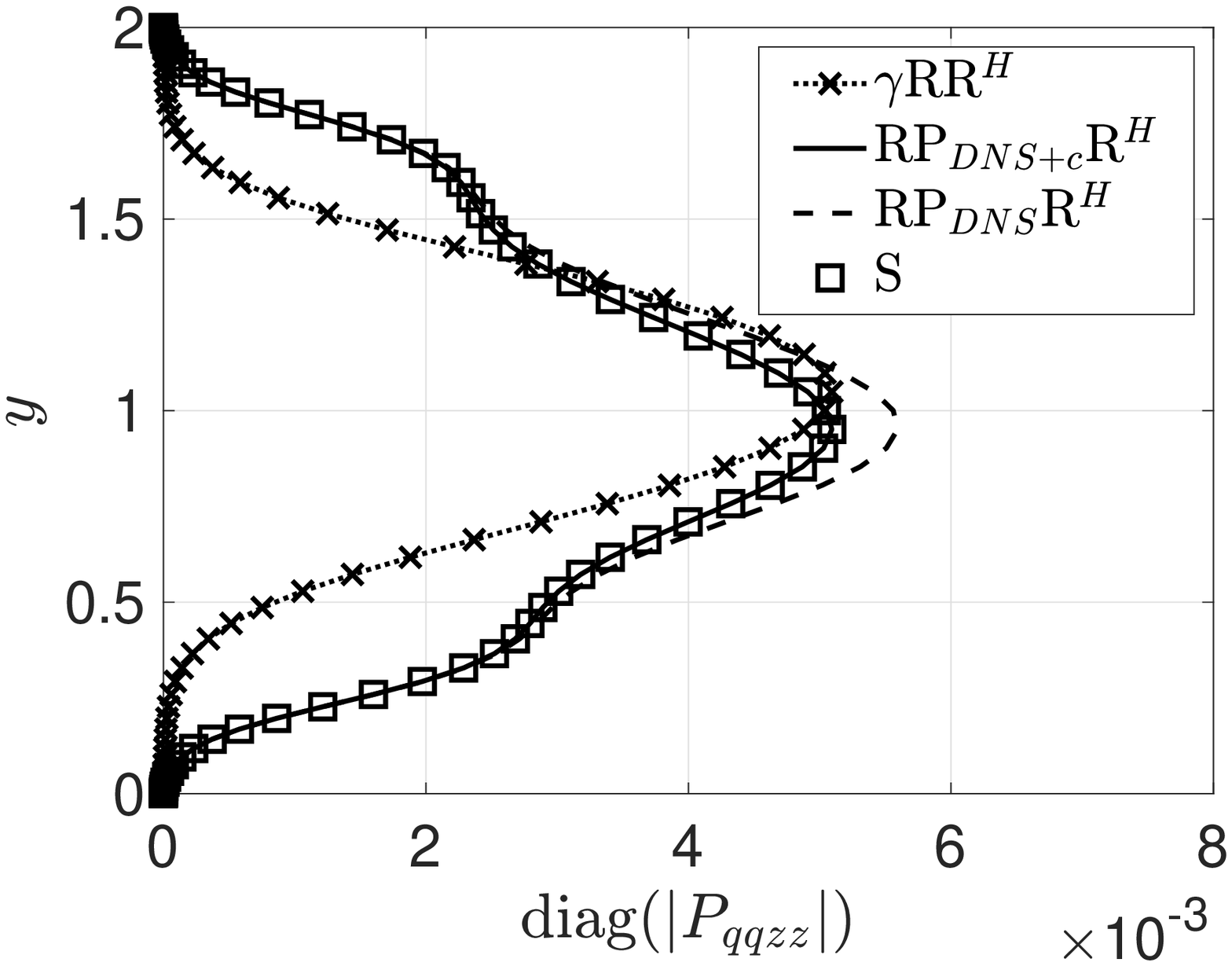}}
\caption{$\mathrm{S}$ from simulation and from the recovery process using different $\mathrm{P}$ (white-noise, uncorrected and corrected) for case $(n_\alpha,n_\beta)=(1,0)$. Prediction using white-noise $\mathrm{P}$ was scaled to match the maximum of $\mathrm{S}$ by the factor $\gamma=3.6295 \times 10^{-5}$.}
\label{fig:diagabsPqq10}
\end{figure}

The same process was carried out for case $(1,0)$, and the results can be seen in figure \ref{fig:diagabsPqq10}. For this combination of wavenumbers, the flow is dominated by the spanwise velocity component, and the energy of other ones are several orders of magnitude lower, as shown in figure \ref{fig:diagabsPqq10}. The dominance of $w$ for the (1,0) mode was also observed by \cite{smith2005low}; analysis of wavenumber spectra of turbulent channels also shows that structures with large spanwise extent are observed for $w$ \citep{delalamo_jimenez2003}. By considering white-noise statistics of the forcing, this dominance of spanwise velocity fluctuations could not be captured, and all components predicted by $\mathrm{R}\mathrm{R}^H$ have the same overall levels, leading to large errors for the $(u,v)$ components. For the spanwise component, however, the position of the peak and the shape of $\mathrm{S}$ at the centre of the domain is roughly captured by the model, even though larger errors are found for regions closer to the wall. These facts altogether points out to an action of the non-linear terms in forcing mainly in the spanwise direction, with a more effective action closer to the wall. By including the uncorrected statistics of the forcing, $\mathrm{P}_{DNS}$, the problem of the relative amplitudes of the different components is solved; now the spanwise component dominates the response, and its shape resembles the one obtained directly from the simulation, even though a slight mismatch is found in the centre of the channel. This comparison is further improved by including the correction term in the statistics of the forcing; by using it, a perfect match between $\mathrm{S}$ computed from the simulation and the one from the forcing statistics is obtained, and the normalised error for this case is lower than $10^{-5}$.

\section{Simplifications of the forcing CSD}
\label{sec:reducedforcing}

The previous analysis detailed the full recovery process, aiming at obtaining accurately the response statistics from the full statistics of the non-linear terms of the Navier-Stokes equations. However, the structure of the non-linear terms can be complex, and analysis of the full non-linear term does not lead directly to physical insight. For this reason, it would be interesting to simplify the forcing, with an evaluation of which components are mostly responsible for the energy of the response. This is performed in this section for wavenumbers $(n_\alpha,n_\beta)=(0,1)$ and $(1,0)$.

\subsection{Case $(n_\alpha,n_\beta)=(0,1)$}
\label{sec:mode01}

\subsubsection{Contribution of each component of the non-linear terms}

In this section we analyse more closely the structure of the non-linear term for the case $(0,1)$ for $\omega \to 0$ ($\omega=0.0123$). Our objective here is to look closely at the forcing statistics in order to isolate the important parts for this simple case. The number of components of the forcing makes an \emph{ad hoc} modelling approach prohibitive; still, if only certain parts of the forcing are necessary to reproduce the statistics of the response, modelling can be considered an option. Specifically for the case $(n_\alpha,n_\beta)=(0,1)$, the momentum equations \ref{eqn:momentum2} and \ref{eqn:momentum3} ($v$ and $w$) decouple from the streamwise velocity, and these equations become also independent of the mean flow $U$. Using this, we can rearrange the system in order to obtain separate equations for streamwise vortices (concentrated in the $v,w$ components) and streaks (concentrated in the $u$ component) as

\begin{eqnarray}
\label{eqn:LNSvort}
    \underbrace{\left[-\ii \omega +\frac{1}{Re}\left(\beta^2-\frac{\partial^2}{\partial y^2}\right)\right]}_{\mathrm{L}_{01}} \underbrace{\left(\beta^2-\frac{\partial^2}{\partial y^2}\right)}_{\mathrm{B}_{01}} v = \ii\beta\left(-\ii\beta f_y+\frac{\partial f_z}{\partial y}\right) \nonumber \\
    = \left(\beta^2-\frac{\partial^2}{\partial y^2}\right) f_y + \frac{\partial}{\partial y} \underbrace{\left( \frac{\partial f_y}{\partial y} + \ii\beta f_z \right)}_{\text{zero response}} \nonumber \\ 
    \Rightarrow \mathrm{L}_{01} \mathrm{B}_{01} v=\mathrm{B}_{01} f_y ,
\end{eqnarray}

\begin{equation}
    \left[-\ii \omega +\frac{1}{Re}\left(\beta^2-\frac{\partial^2}{\partial y^2}\right)\right] u = f_x - v \frac{\partial U}{\partial y} \Rightarrow \mathrm{L}_{01} u = f_x - v \frac{\partial U}{\partial y},
    \label{eqn:LNSstreaks}
\end{equation}

\noindent where the influence of the spanwise component of the forcing is already considered in equation \ref{eqn:LNSvort}, by considering that only the rotational part of the forcing leads to a response in velocity; inspection of the linear operators in the Orr-Sommerfeld-Squire formulation \citep{jovanovic2005componentwise} confirms that hypothesis, which was also verified by \cite{Rosenberg_2019}. The spanwise component $w$ can be obtained from $v$ using the continuity equation. The optimal response of equation \ref{eqn:LNSvort} for $\omega \to 0$ are streamwise vortices (which can be obtained directly from resolvent analysis, for example), and these structures are independent of the mean flow chosen for the analysis. The effect of $U$ is seen directly in the equation of the streamwise velocity, via the $v \frac{\partial U}{\partial y}$ term, which is related directly with the lift-up effect: due to the presence of shear, these vortices will lead to the growth of streamwise velocity, which will assume the shape of streaks. But one should note that there are two forcing terms in the right hand side of equation \ref{eqn:LNSstreaks}, which force directly the streaks; we would like to evaluate the influence of each one in the response. For that, we can rewrite the equation as a function of the expected value of $u$ as

\begin{eqnarray}
    \mathrm{L}_{01} \mathcal{E} (u u^H)  \mathrm{L}_{01}^H = \mathcal{E}\left[ \left(f_x - v \frac{\partial U}{\partial y} \right)\left( f_x - v \frac{\partial U}{\partial y} \right)^H\right] \ \ \\
    \begin{split}
    \mathrm{S}_{xx}=\mathrm{R}_{01} \left[ \mathrm{P}_{xx} + \left(\frac{\partial U}{\partial y}\right) \mathrm{S}_{yy} \left(\frac{\partial U}{\partial y}\right)^H \right. \\ \left. - \mathcal{E} (f_x v^H)\left(\frac{\partial U}{\partial y}\right)^H - \left(\frac{\partial U}{\partial y}\right) \mathcal{E} (v f_x^H) \right] \mathrm{R}_{01}^H\end{split}, \label{eqn:ModelPqqxx}
\end{eqnarray}

\noindent where $\mathrm{R}_{01}$ is the resolvent operator associated with $\mathrm{L}_{01}$. The equation above shows that the statistics of the streamwise velocity can be educed from the statistics of the wall-normal velocity (which in turn can be obtained from the statistics of $f_y$), from the statistics of the streamwise component of the forcing $f_x$ and from the cross-term statistics. Using the values obtained from the DNS, we can evaluate the influence of each term on the right hand side of equation \ref{eqn:ModelPqqxx}: $\mathrm{P}_{xx}$ is related to the statistics of the streamwise component of the forcing; $(\frac{\partial U}{\partial y}) \mathrm{S}_{yy} (\frac{\partial U}{\partial y})'$ is related to statistics obtained using only the lift-up mechanism; the other components are the covariances between streamwise forcing and wall-normal velocity. 

\begin{figure} 
\centering
\subfigure[Simulation]{\includegraphics[clip=true, trim= 0 0 0 0, width=0.4\textwidth]{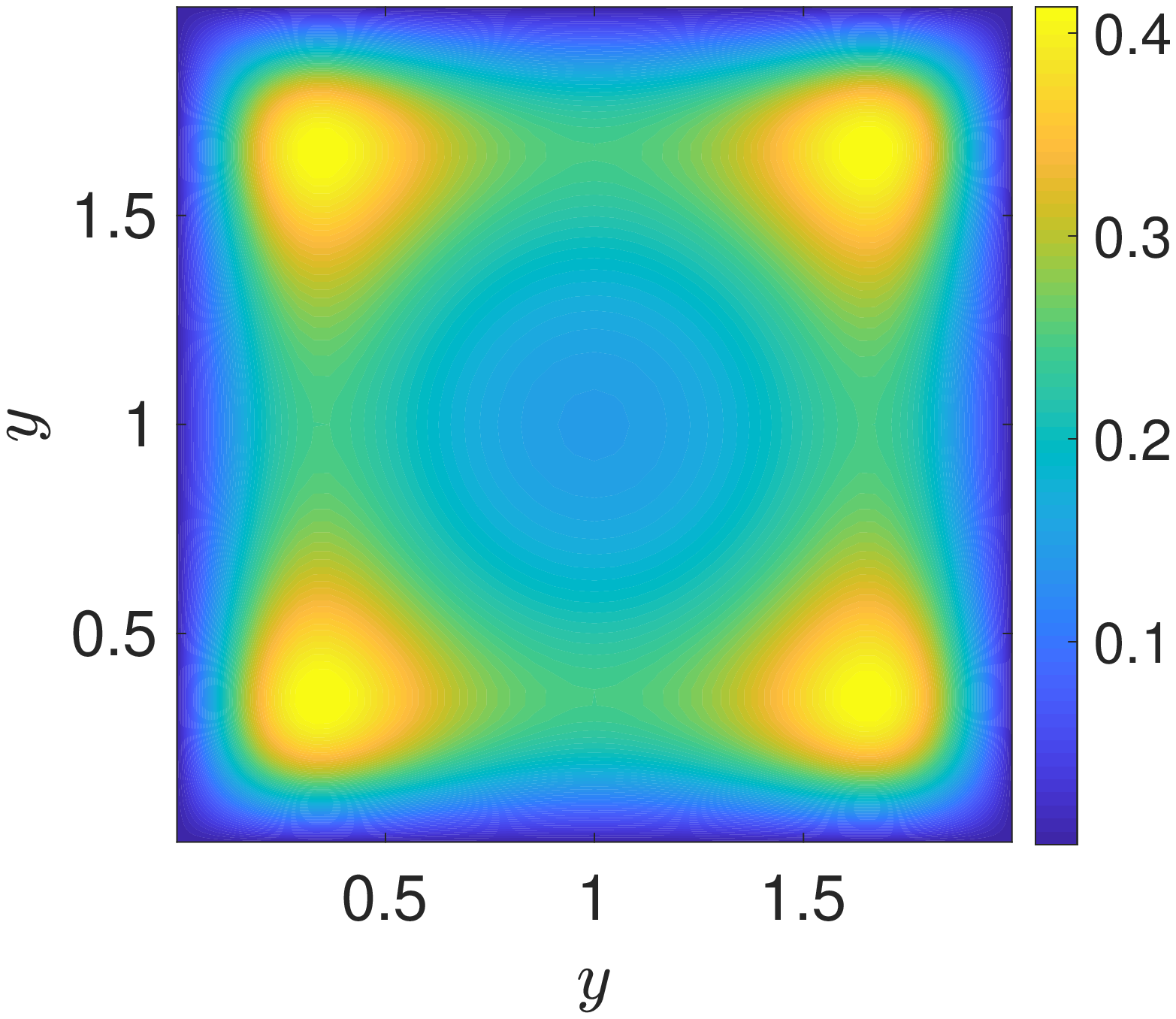}}\subfigure[All components of $\mathrm{P}_{DNS}$]{\includegraphics[clip=true, trim= 0 0 0 0, width=0.4\textwidth]{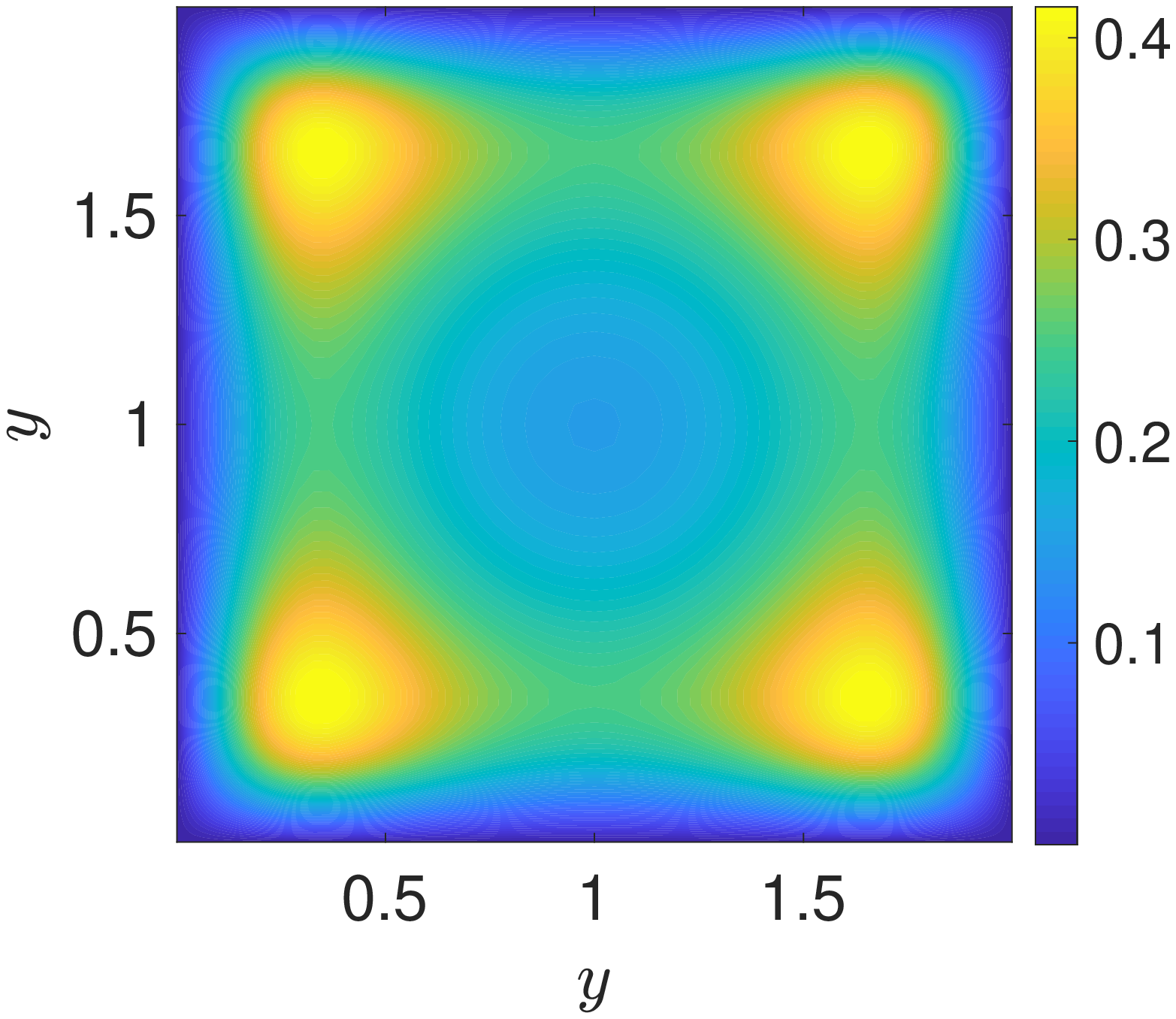}}
\subfigure[Lift-up component]{\includegraphics[clip=true, trim= 0 0 0 0, width=0.4\textwidth]{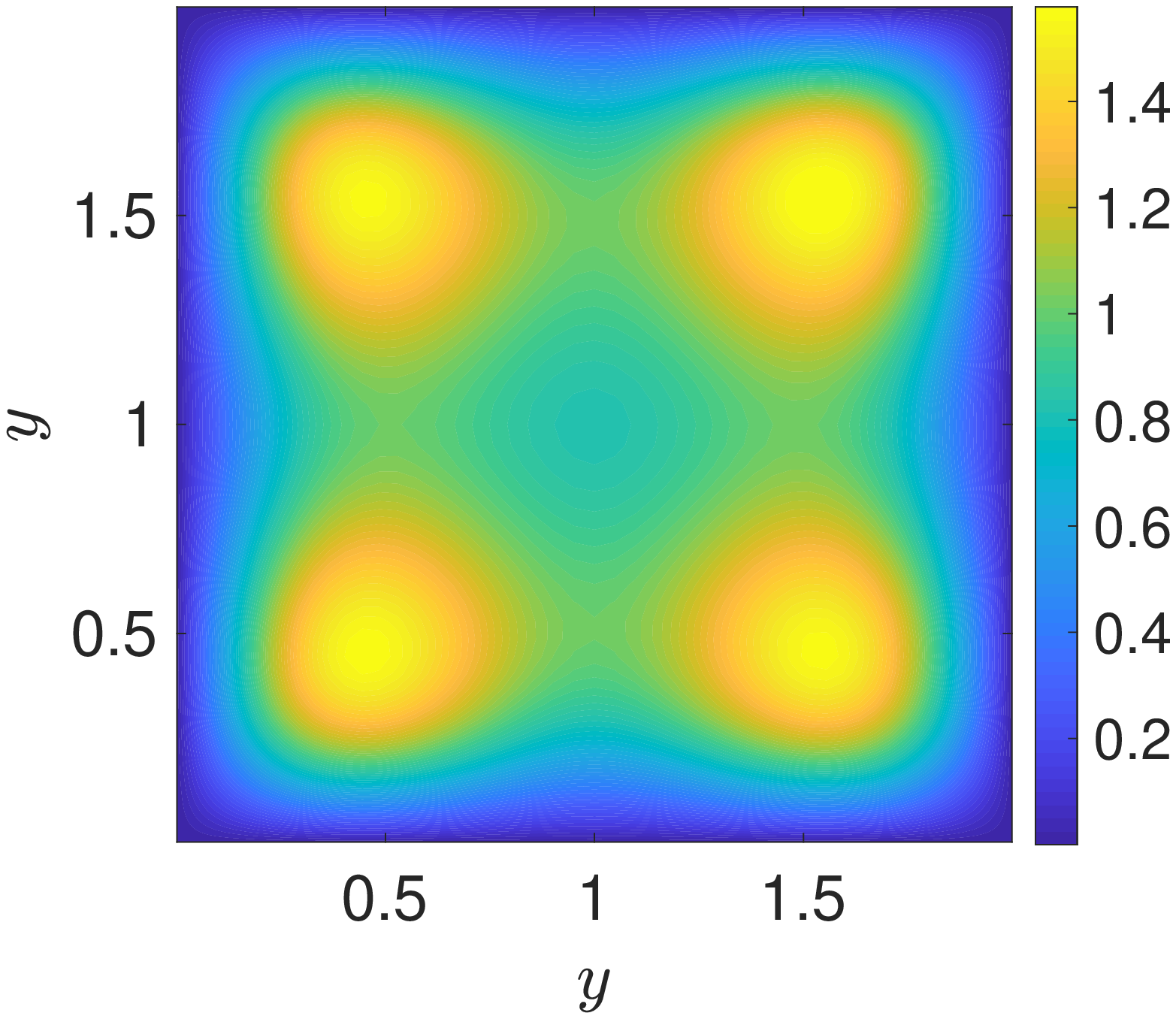}}\subfigure[Streamwise forcing component]{\includegraphics[clip=true, trim= 0 0 0 0, width=0.4\textwidth]{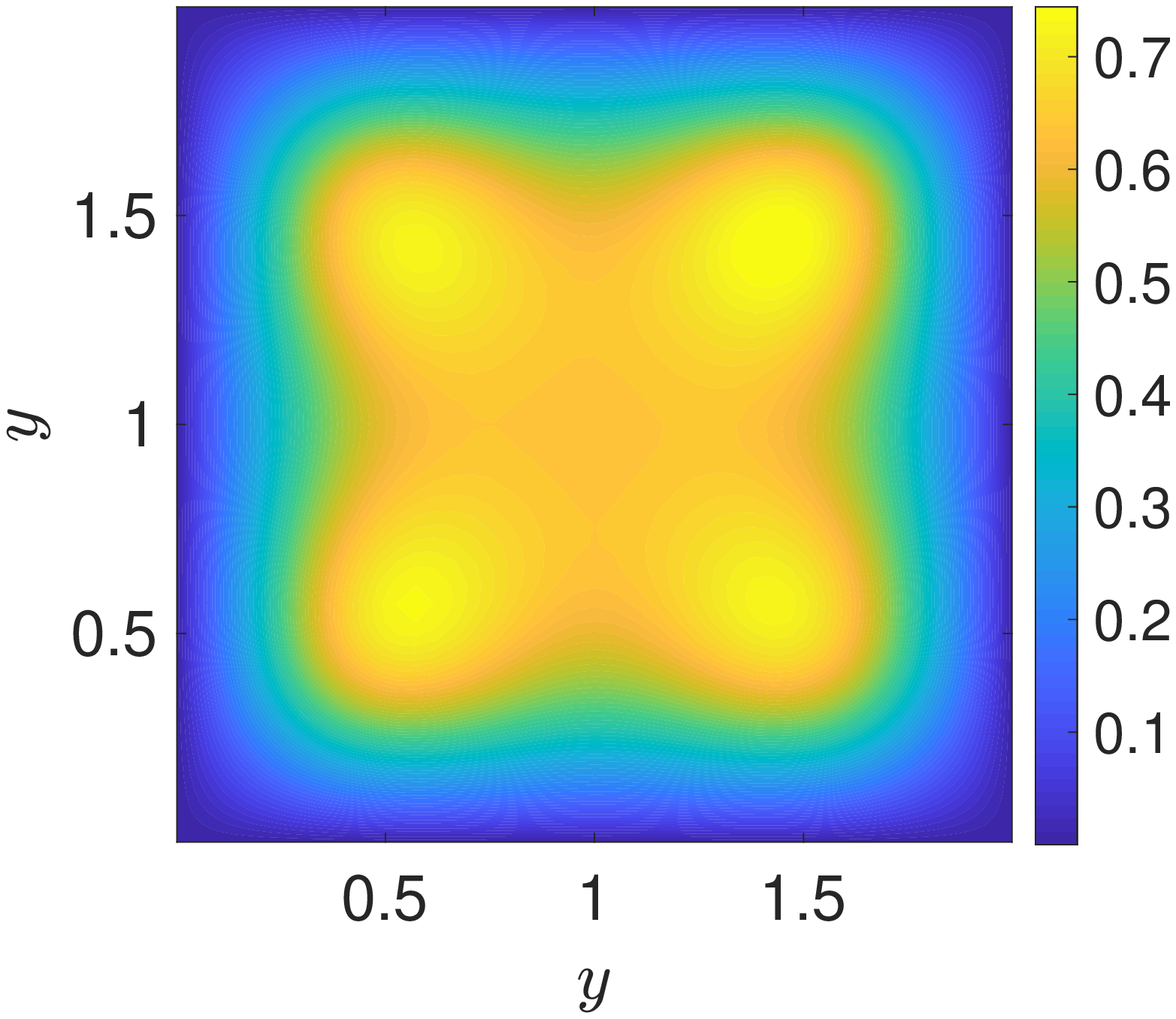}}
\subfigure[Cross-terms]{\includegraphics[clip=true, trim= 0 0 0 0, width=0.4\textwidth]{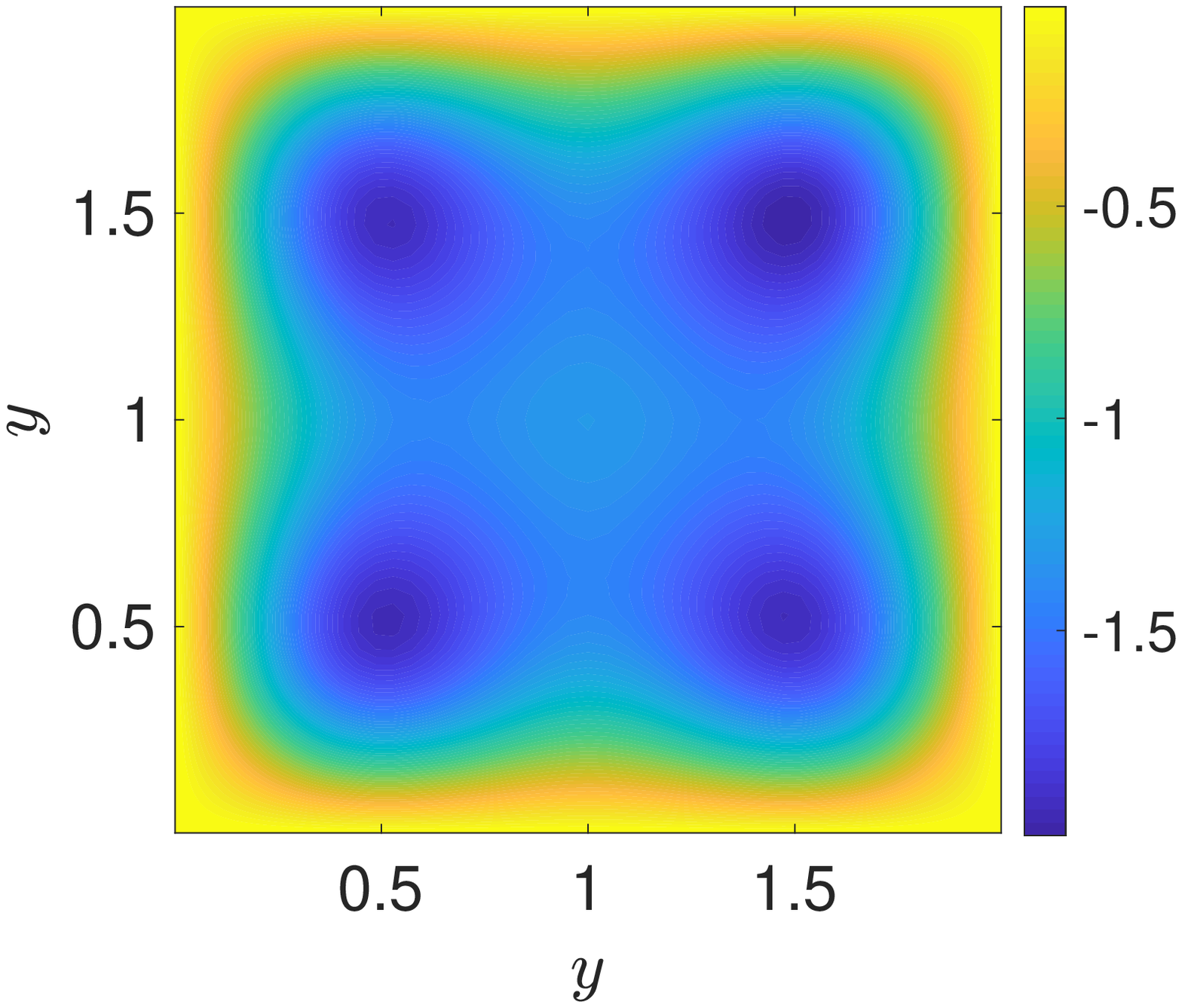}}\subfigure[Sum of the contributions]{\includegraphics[clip=true, trim= 0 0 0 0,width=0.4\textwidth]{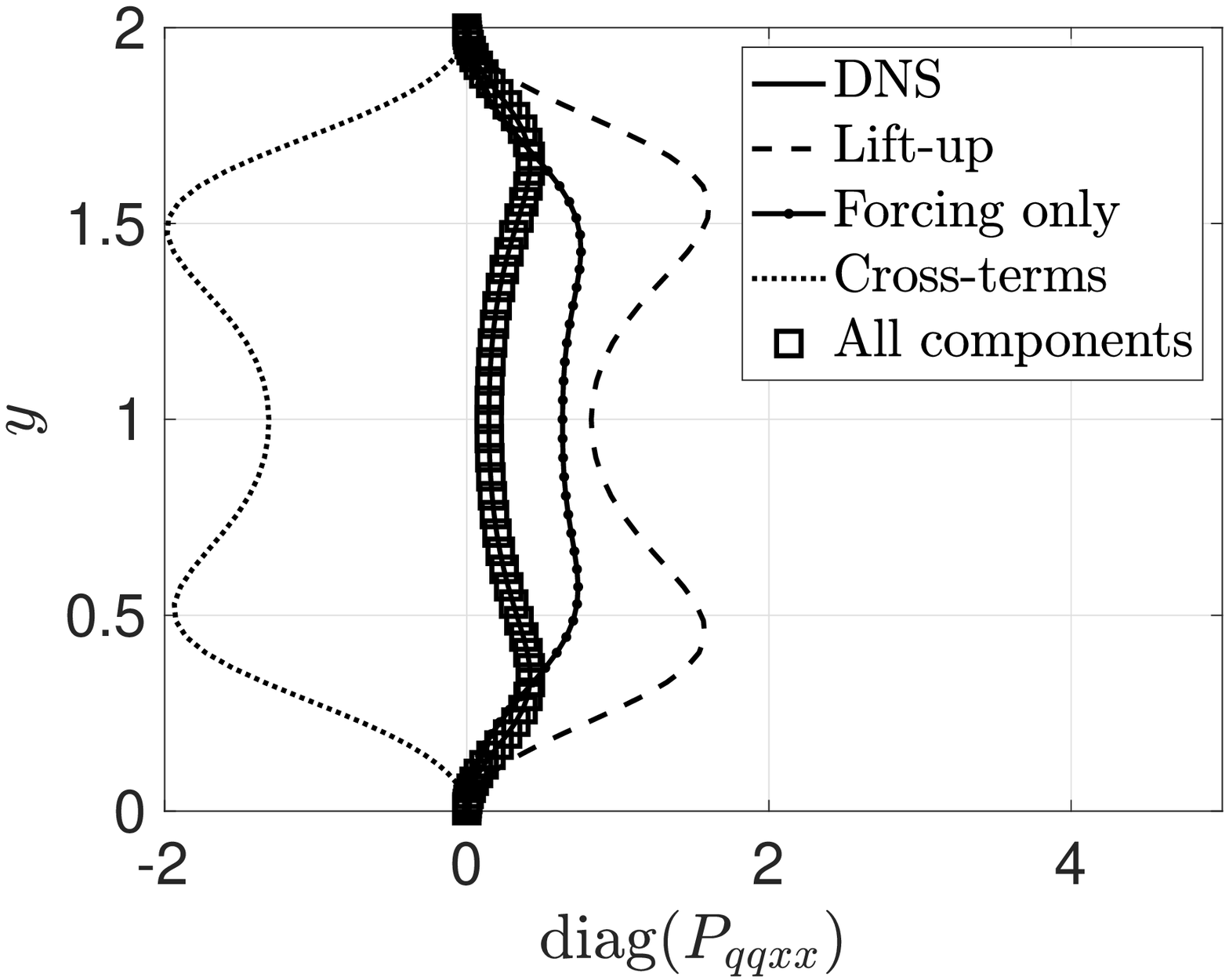}}
\caption{Real part of streamwise component of $\mathrm{S}$ from simulation (a) and the prediction using different forcing term (b-e). Contribution of each term in the reconstruction (f).}
\label{fig:Pqqom0Recons}
\end{figure}

Figures \ref{fig:Pqqom0Recons}(a-e) show the reconstruction of $\mathrm{S}_{xx}$ using each term of equation \ref{eqn:ModelPqqxx} (just the real part is shown). As expected, the reconstruction using all terms reproduces the results from the DNS; on the other hand, if we take only the term related to the lift-up mechanism, the shapes of $\mathrm{S}_{xx}$ differ, especially considering the position of the peaks. The same happens when we use only the term related to the statistics of the forcing in the streamwise direction, or when we use only the cross-terms (which have a negative contribution of the sum). Still, the sum of all these quantities generates a combination of constructive-destructive influence on $\mathrm{S}_{xx}$, leading to the correct shape and amplitude, when all terms are considered. This can be better understood by looking at power spectral density (which is the main diagonal of $\mathrm{S}$) using each term, compared to DNS results. This is shown in figure \ref{fig:Pqqom0Recons}(f), where we can see that, even though the amplitudes of each component are high, the final result considering all terms is rather small, and the contribution of the cross-term seem to be responsible for this overall reduction.  The negative effect of the cross-term thus represents a destructive interference between the lift-up mechanism and the direct excitation of streaks by the streamwise forcing component. This deterministic effect cannot be appropriately modelled when the forcing is considered as white noise, as seen in \S~\ref{sec:statisticswihtorwithout}.

This analysis highlights that consideration of isolated mechanisms may lead to quantitative errors in the prediction of flow responses. Similar results were obtained by \cite{freund2003noise,bodony2008current} and \cite{cabana2008identifying} in studies of sound generation by a sheared flow, using Lighthill's acoustic analogy. The cited works showed that when source terms, analogous to the forcing CSD $\mathrm{P}$ considered here, are decomposed into subterms, an analysis of the isolated contribution of each one may be problematic, as destructive interference between components may lead to a summed radiation which is lower than the individual contributions.

\subsubsection{Simplifications of forcing}
From the preceding analysis we can understand that all components (lift-up related and streamwise forcing) are important to obtain good predictions of $\mathrm{S}$. Nevertheless, some simplification can still be performed on the forcing terms by rewriting it as the sum of the non-linear terms of the Navier-Stokes equations. Overall, these terms can be written as

\begin{equation}
    f_i=-u_j\frac{\partial u_i}{\partial x_j}
\end{equation}

\noindent or, in vector form

\begin{equation}
    \mathbf{f}=
    \begin{pmatrix}
    f_x \\[0.9em]
    f_y \\[0.9em]
    f_z
    \end{pmatrix}
    =
    -u
    \begin{pmatrix}
    \frac{\partial u}{\partial x} \\[0.9em]
    \frac{\partial v}{\partial x} \\[0.9em]
    \frac{\partial w}{\partial x}
    \end{pmatrix}
    -
    v
    \begin{pmatrix}
    \frac{\partial u}{\partial y} \\[0.9em]
    \frac{\partial v}{\partial y} \\[0.9em]
    \frac{\partial w}{\partial y}
    \end{pmatrix}
    -
    w
    \begin{pmatrix}
    \frac{\partial u}{\partial z} \\[0.9em]
    \frac{\partial v}{\partial z} \\[0.9em]
    \frac{\partial w}{\partial z}
    \end{pmatrix}
    =\mathbf{f}_u+\mathbf{f}_v+\mathbf{f}_w.
    \label{eqn:decompforc}
\end{equation}

Individual terms are evaluated in physical space and transformed to frequency-wavenumber space afterwards. Writing the forcing this way allows us to decompose $\mathrm{P}$ into 9 components, related to each of the 3 forcing terms in equation \ref{eqn:decompforc} and extract the relevant parts of this term. Due to the low number of forcing components to evaluate, we chose to remove some of them by trial and error, in order to evaluate the influence of those in the statistics of the response. A first analysis shows that $\mathbf{f}_u$ does not play a significant role in this case, as predictions of $\mathrm{S}$ disregarding this term did not lead to any considerable mismatch. The terms $v\frac{\partial v}{\partial y}$ and $v\frac{\partial w}{\partial y}$ are also less relevant for this case. This reduces the forcing to 

\begin{equation}
    \mathbf{f}_{red}=
    -v
    \begin{pmatrix}
    \frac{\partial u}{\partial y} \\[0.9em]
    0 \\[0.9em]
    0
    \end{pmatrix}
    -
    w
    \begin{pmatrix}
    \frac{\partial u}{\partial z} \\[0.9em]
    \frac{\partial v}{\partial z} \\[0.9em]
    \frac{\partial w}{\partial z}
    \end{pmatrix}
    =\mathbf{f}_{v_x}+\mathbf{f}_w.
    \label{eqn:decompforc_simp1}
\end{equation}

This is the maximum simplification that the covariance of the forcing can suffer in order to recover the covariance of the response without introduction of significant error.

\begin{figure} 
\centering
\subfigure[All components of $f$]{\includegraphics[clip=true, trim= 0 0 0 0, width=0.4\textwidth]{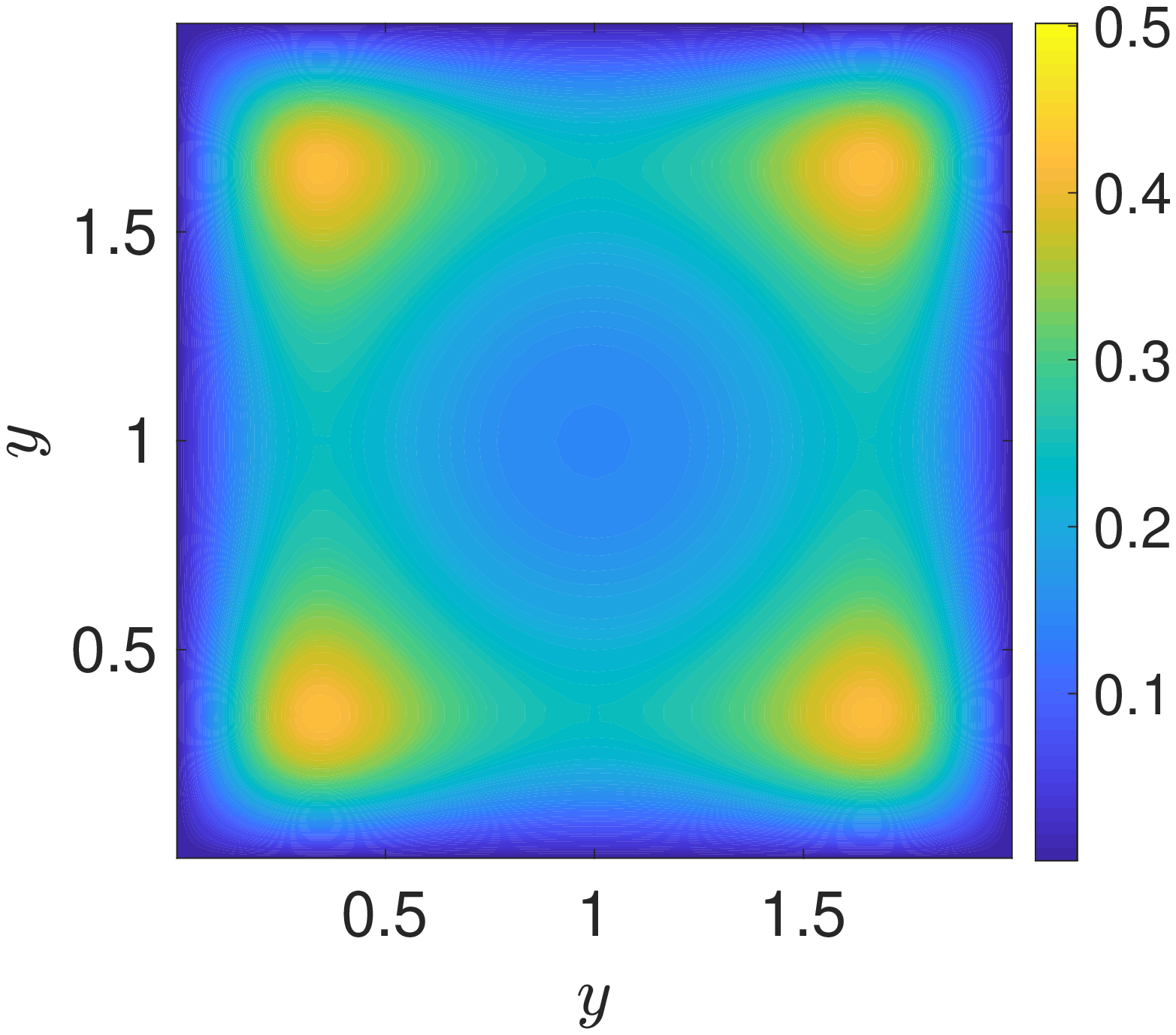}}\subfigure[$\mathrm{P}_{red}$]{\includegraphics[clip=true, trim= 0 0 0 0, width=0.4\textwidth]{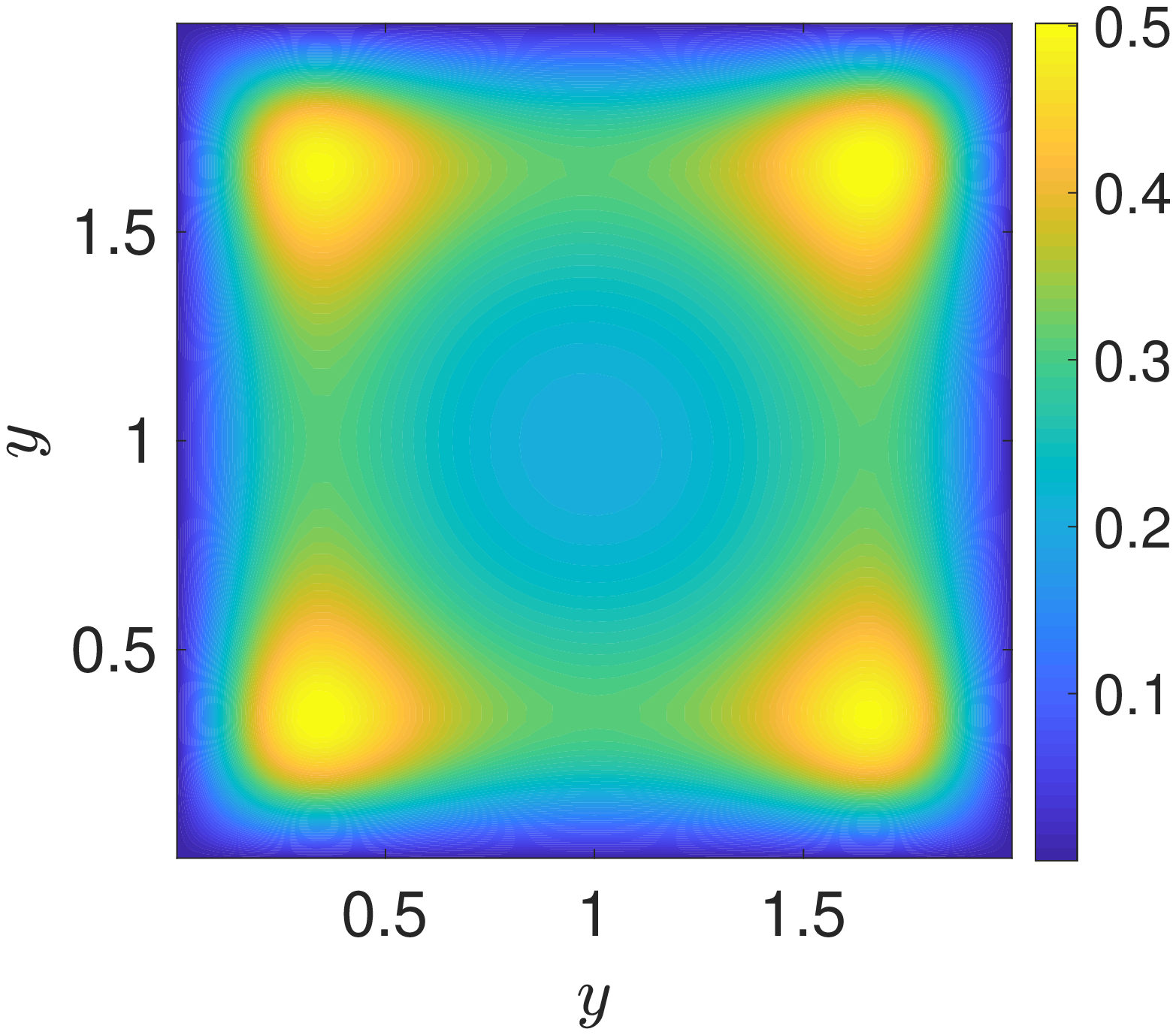}}
\subfigure[White-noise $\mathrm{P}$]{\includegraphics[clip=true, trim= 0 0 0 0, width=0.4\textwidth]{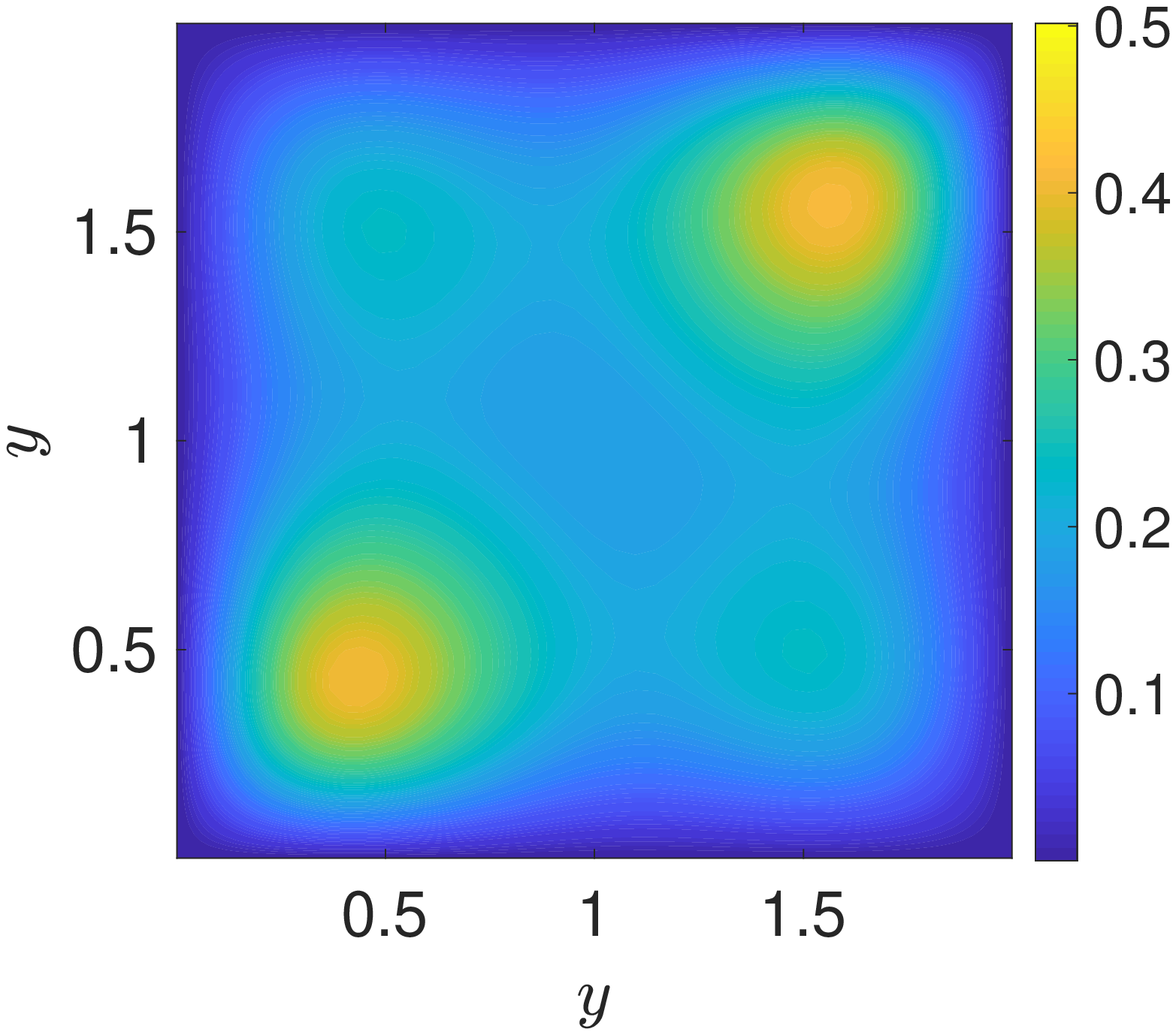}}\subfigure[Main diagonal of $\mathrm{S}$]{\includegraphics[clip=true, trim= 0 0 0 0, width=0.4\textwidth]{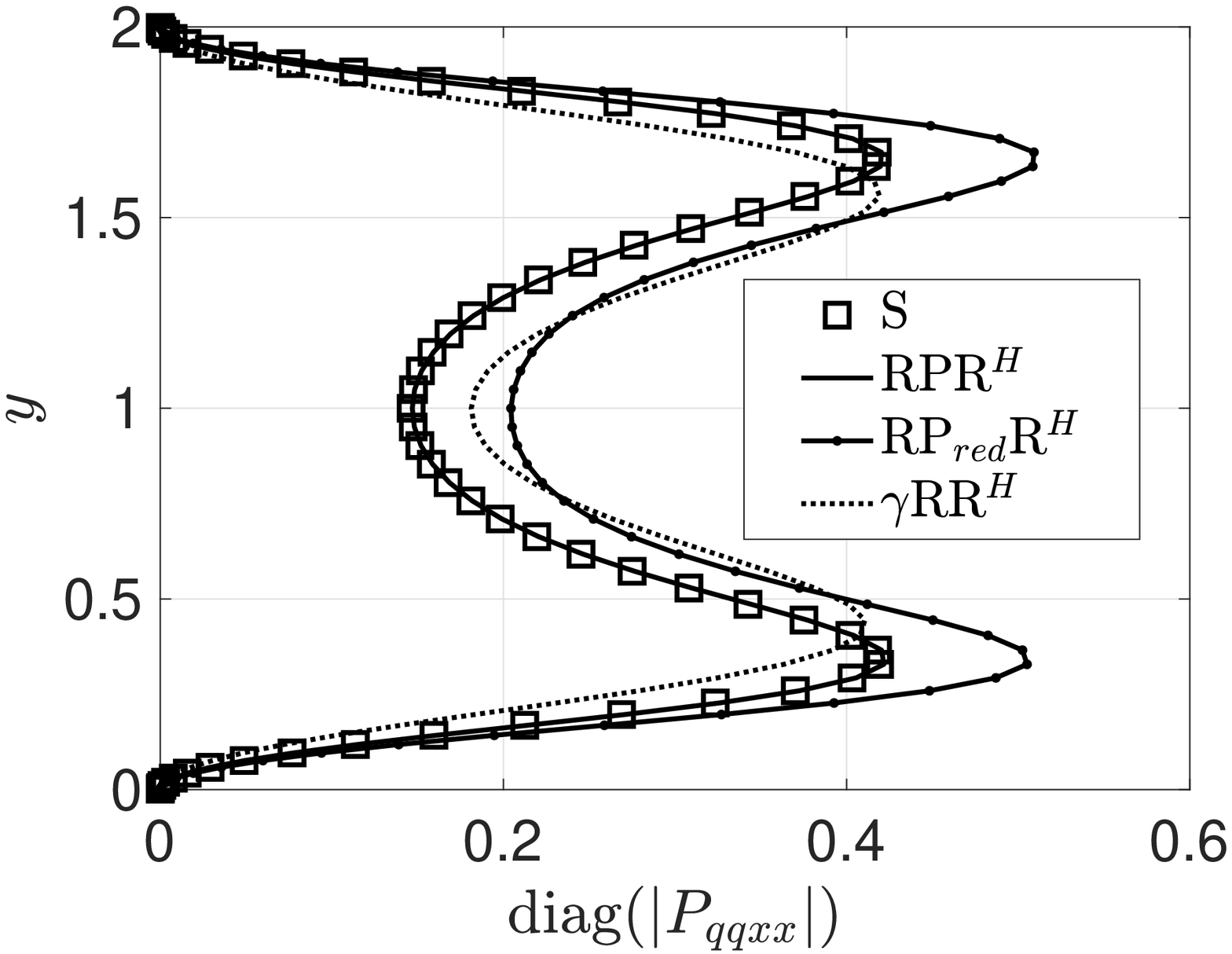}}
\caption{Streamwise component of $\mathrm{S}$ from the reconstruction using the whole forcing (a), the reduced order forcing (b), statistical white forcing (c) and the absolute value of the diagonal of each one compared with the one from the simulation (d).}
\label{fig:Pqqom0ReconsRed}
\end{figure}

The cross-spectral density $\mathrm{S}$ recovered using the total forcing, the reduced forcing and the white-noise forcing can be seen in figure \ref{fig:Pqqom0ReconsRed}. From figures \ref{fig:Pqqom0ReconsRed}(a-c), it is clear that the reduction of the forcing to the expression \ref{eqn:decompforc_simp1} leads to the correct amplitude distribution of the streamwise velocity CSD, with better agreement than consideration of white-noise forcing; in particular, the coherence between the two peaks in amplitude, which can be seen by the nearly identical values for $(y,y)$ and $(y,-y)$, is recovered from $\mathrm{P}_{red}$. Still, by retaining only the terms in eq. \ref{eqn:decompforc_simp1}, a mismatch starts to appear in the amplitudes of the reconstruction, as shown in figure \ref{fig:Pqqom0ReconsRed}(d).

\subsubsection{Low rank of forcing}

\begin{figure} 
\centering
\subfigure[SPOD eigenvalues]{\includegraphics[clip=true, trim= 0 0 0 0, width=0.5\textwidth]{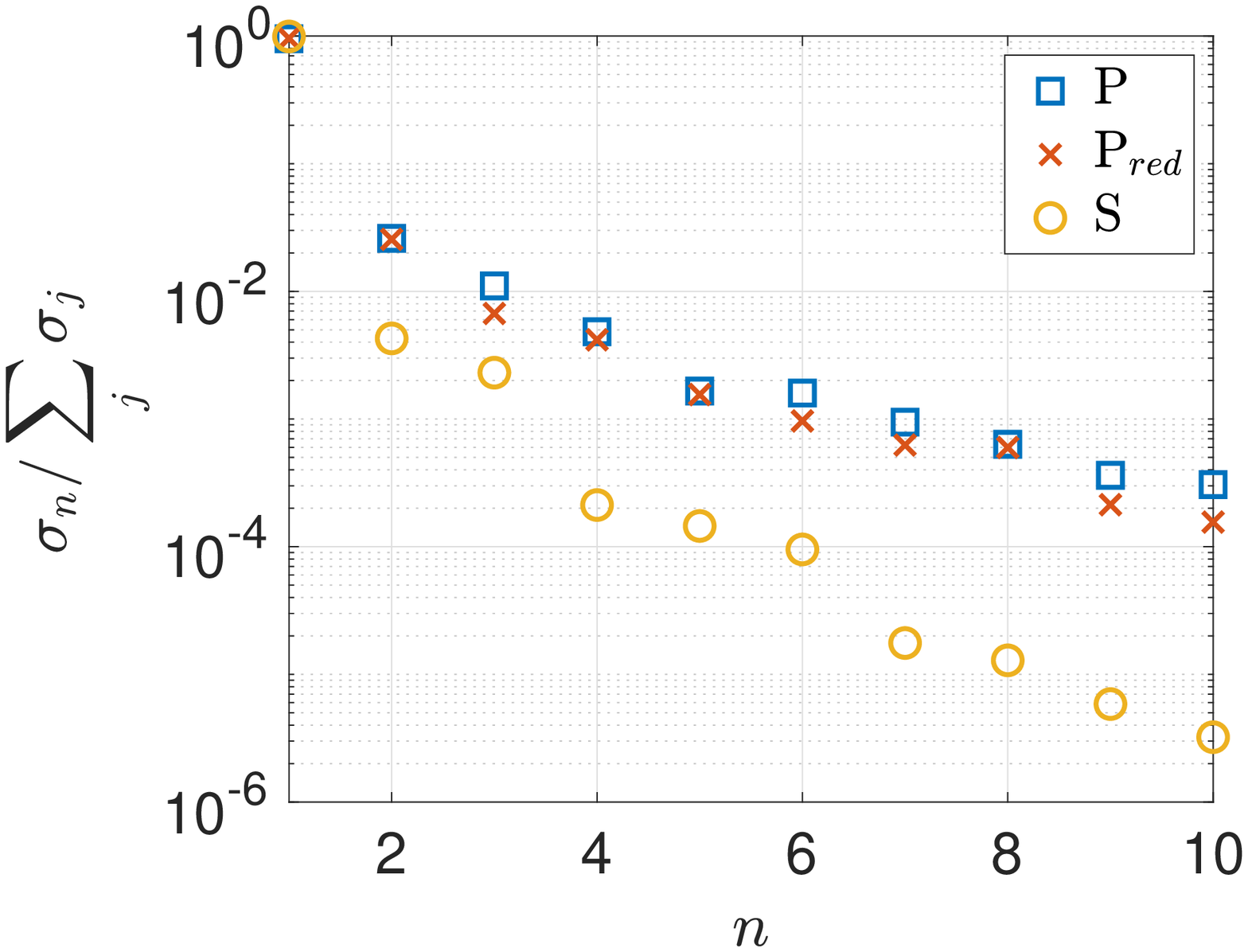}}\subfigure[Leading SPOD mode and reconstructions]{\includegraphics[clip=true, trim= 0 0 0 0, width=0.5\textwidth]{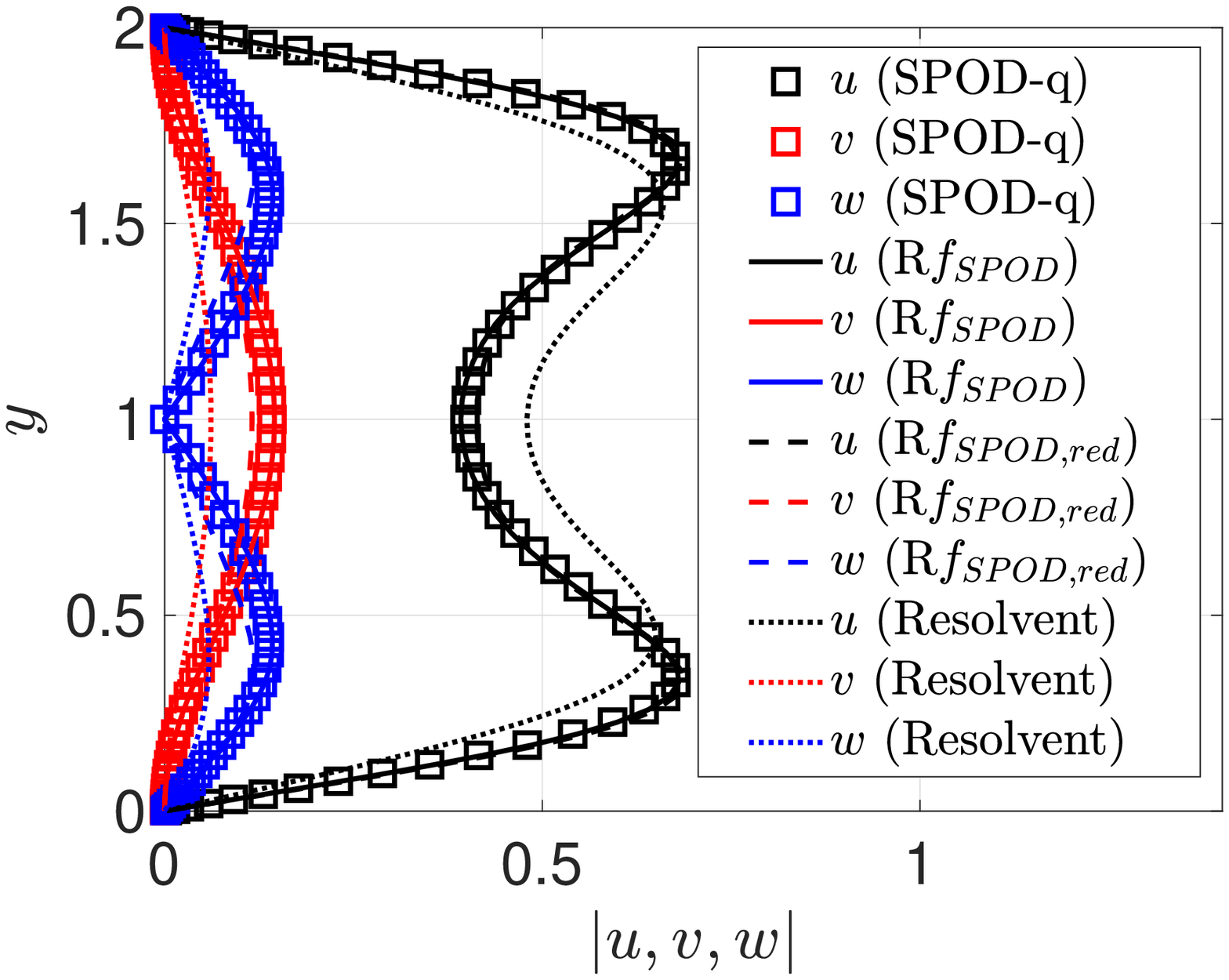}}
\caption{SPOD eigenvalues for $\mathrm{S}$ and the different $\mathrm{P}$ used here (a) and the shapes of the leading SPOD mode of the response (SPOD-q) compared with the reconstruction using the leading SPOD mode of the forcing and the first response mode from resolvent (b).}
\label{fig:Fredmaxgainsmodes}
\end{figure}

We now consider the most energetic structures in the flow for this combination of frequency-wavenumber, and, in particular, evaluate if forcings and responses have low rank, which may simplify the modelling. Figure \ref{fig:Fredmaxgainsmodes}(a) shows the eigenvalues of the Spectral Proper Orthogonal Decomposition of the full covariance of the forcing, the reduced one (both considering the correction in eq. \ref{eqn:windLNS_freq}), and the covariance of the response. SPOD here amounts simply to an eigendecomposition of the mentioned cross-spectral densities, which are Hermitian by construction. In figure \ref{fig:Fredmaxgainsmodes}(a) we can see that there is a clear separation between the first and the following modes for these matrices; therefore, the first SPOD mode would be sufficient to represent the forcing and response. Also, the energies of the covariance of the reduced forcing $\mathrm{P}_{red}$ are close to the ones of the full $\mathrm{P}$, which highlights the similarities between the two matrices. The comparison between the leading SPOD mode of the response (denoted as follows as SPOD-q) and the reconstruction using $q=\mathrm{R}f_{SPOD}$ (where $f_{SPOD}$ is the first SPOD mode of the corrected forcing, both full and reduced) is shown in figure \ref{fig:Fredmaxgainsmodes}(b). These plots show clearly that the leading SPOD mode of both $\mathrm{P}$ and $\mathrm{P}_{red}$ lead to close agreement with the first SPOD mode of $\mathrm{S}$, which confirms that the terms in equation \ref{eqn:decompforc_simp1} are, indeed, the dominant ones in this problem. Figure \ref{fig:Fredmaxgainsmodes}(b) also show that the first resolvent mode, computed under the hypothesis of white-noise forcing, does not match the SPOD mode from the simulation, showing that the statistics of the non-linear terms are important to match exactly the shapes of the most energetic structures of the flow. The main shapes are nonetheless retrieved in the leading response mode of resolvent analysis, with $v$ and $w$ forming a streamwise vortex, as seen for instance in the amplitude distribution of $w$, with two lobes in phase opposition (not shown), and an amplified streak in $u$. There is a mismatch in the relative amplitudes of the streamwise vortex and the streak, which is corrected when the forcing statistics are considered.

\begin{figure} 
\centering
\subfigure[Absolute value of first SPOD mode of the forcings]{\includegraphics[clip=true, trim= 0 0 0 0, width=0.5\textwidth]{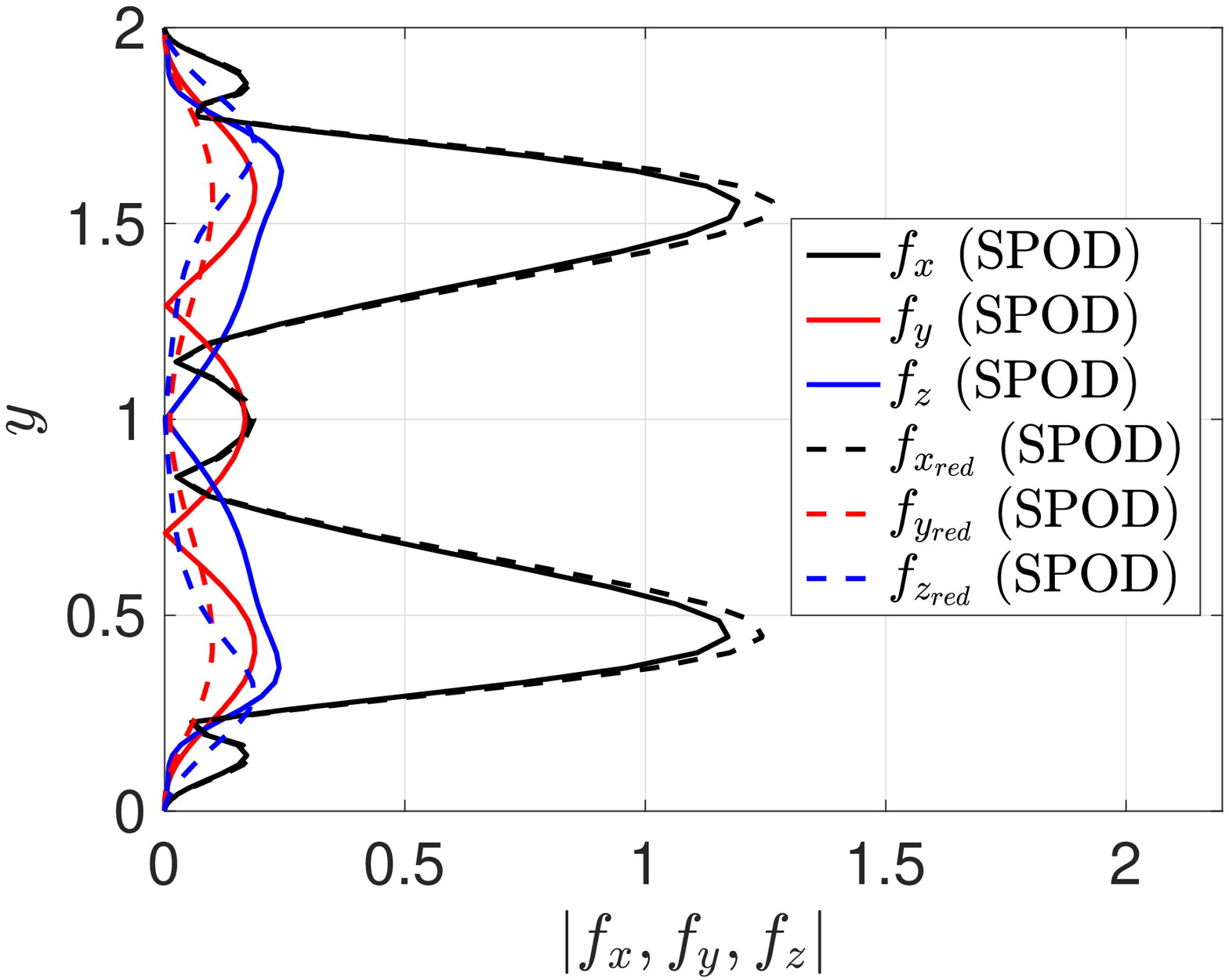}}\subfigure[Reduced forcing in the physical space]{\includegraphics[clip=true, trim= 0 0 0 0, width=0.5\textwidth]{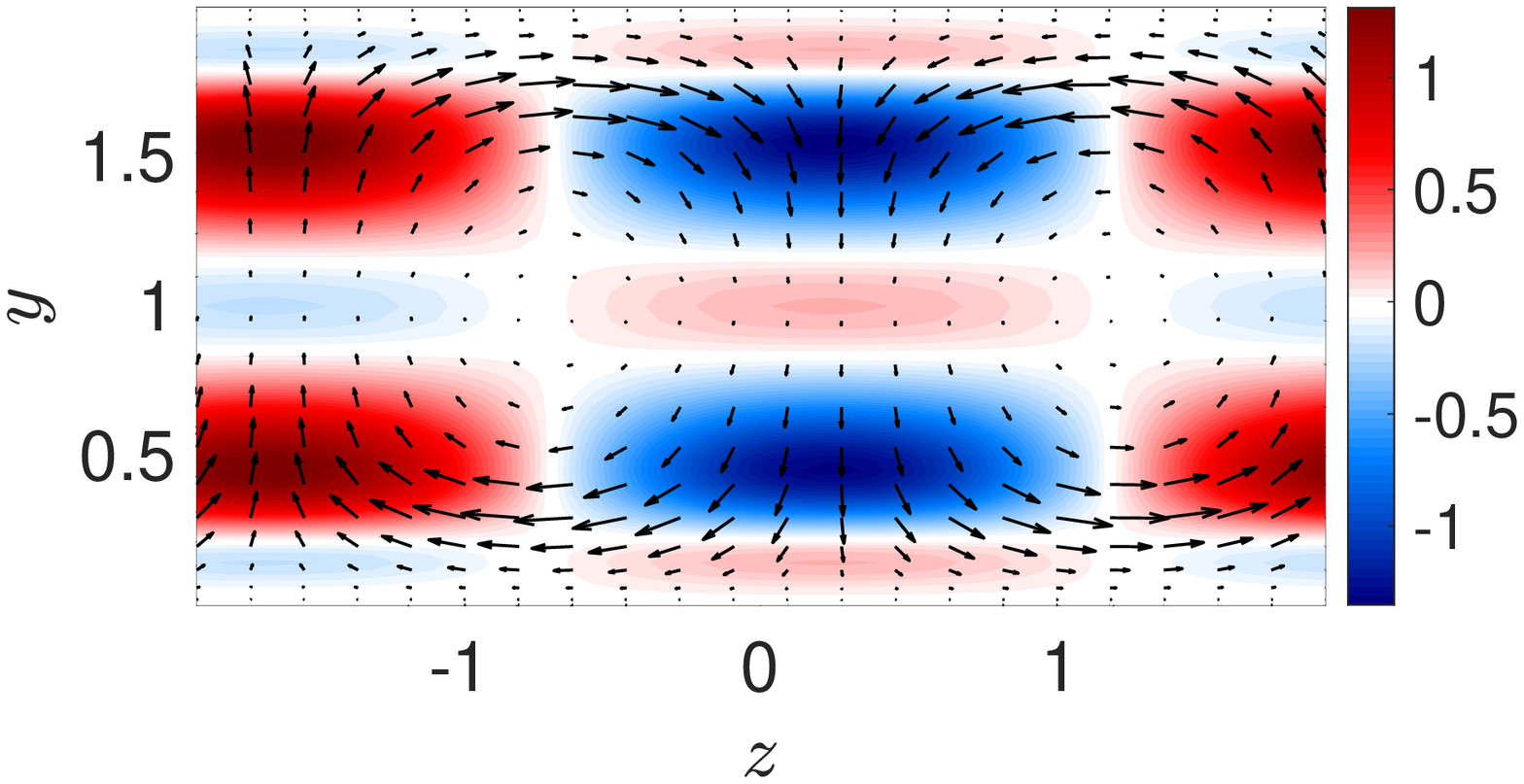}}
\caption{Absolute value of first SPOD mode of the full and reduced forcings (a) and the shape of the reduced forcing in the physical space. Colours: streamwise forcing; arrows: wall-normal and spanwise forcings.}
\label{fig:Fredmax}
\end{figure}

We can also evaluate the most energetic structure of the non-linear terms; considering that we are interested in the physical shapes of these, we calculated the SPOD of $\mathrm{P}$ without the correction term. The first SPOD mode of the forcings for the considered frequency is shown in figure \ref{fig:Fredmax}. Even though the leading mode of the full forcing has an intricate structure (especially due to the presence of an extra oscillation in the centre of the domain in the wall-normal velocity), the first SPOD mode of the reduced forcing, from eq. \ref{eqn:decompforc_simp1} is clear, at least considering the spanwise and wall-normal components, with the shape of a streamwise vortex (also shown in figure \ref{fig:Fredmax}(b), where the mode was reconstructed using the wavenumbers for this case). This connects directly with the conclusions drawn previously: the streamwise vortices of the response are excited by streamwise rotational forcing from the non-linear terms; these vortices, in turn, feed the lift-up effect such that streaks appear in the velocity field. Considering that we are dealing with the non-linear term of the Navier-Stokes equations, this result may be seen as surprising. This term gather the contribution of all triadic interactions that affect the considered frequency and wavenumber; therefore, this term will be fed by a myriad of combinations of frequencies and wavenumbers. Still, the dominant part of these are the ones that generate a streamwise vortex, related to the $w\partial v / \partial z$ and $w\partial w / \partial z$ non-linear terms. This shows that the spanwise velocity fluctuations of other wavenumber combinations greatly affects the forcing term for the mode $(0,1)$, meaning that any combination dominated by this component is likely relevant for this forcing term. However, the streamwise vortical forcing is not sufficient in order to match the response; a distribution of streamwise forcing, related to $v\partial u / \partial y$ and $w\partial u / \partial z$, is essential to recover the correct response.

\begin{figure} 
\centering
\includegraphics[clip=true, trim= 0 0 0 0, width=0.8\textwidth]{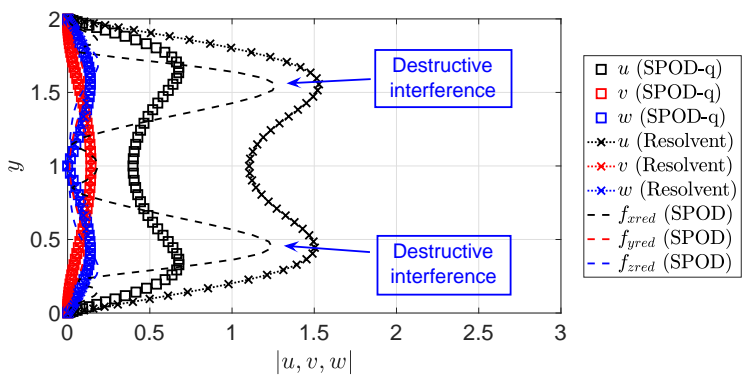}
\caption{Sketch of the action of the non-linear terms on the optimal response of mode $(0,1)$.}
\label{fig:ConstDestsketch}
\end{figure}

One can also think of the most energetic structure of the forcing in light of the lift-up effect. As shown in figure \ref{fig:ConstDestsketch}, by rescaling the streamwise vortices obtained via resolvent analysis to match the amplitudes of these structures in the first SPOD mode, shapes of $v$ and $w$ components are well reproduced, but the streak associated to the resolvent has a much larger amplitude than the one from SPOD. Here is where the streamwise forcing acts: looking at its structure in figure \ref{fig:Fredmax}(b), there are large portions of negative streamwise forcing at positions where the vertical forcing would induce a positive streak. Therefore, the effect of the forcing in this case is mostly to cancel the streak generated by the linear mechanism, leading to the structure found in the SPOD. This is shown schematically in figure  \ref{fig:ConstDestsketch}.

\subsection{Case $(n_\alpha,n_\beta)=(1,0)$}
\label{sec:mode10}
\subsubsection{Contribution of each component of the non-linear terms}

{We now turn our eyes to the case $(1,0)$, also taken at the limit $\omega \to 0$ ($\omega=0.0123$, as in the previous section). Analysis of the energy related to each velocity component for this case points to a dominance of $w$ fluctuations for the present mode. Since the dynamics of $w$ is uncoupled from the rest, it is sufficient to analyse the non-linear terms related to $f_z$ to recover the correct statistics of the velocity, which will also be dominated by the spanwise component.}

\begin{figure} 
\centering
\subfigure[Simulation]{\includegraphics[clip=true, trim= 0 0 0 0, width=0.5\textwidth]{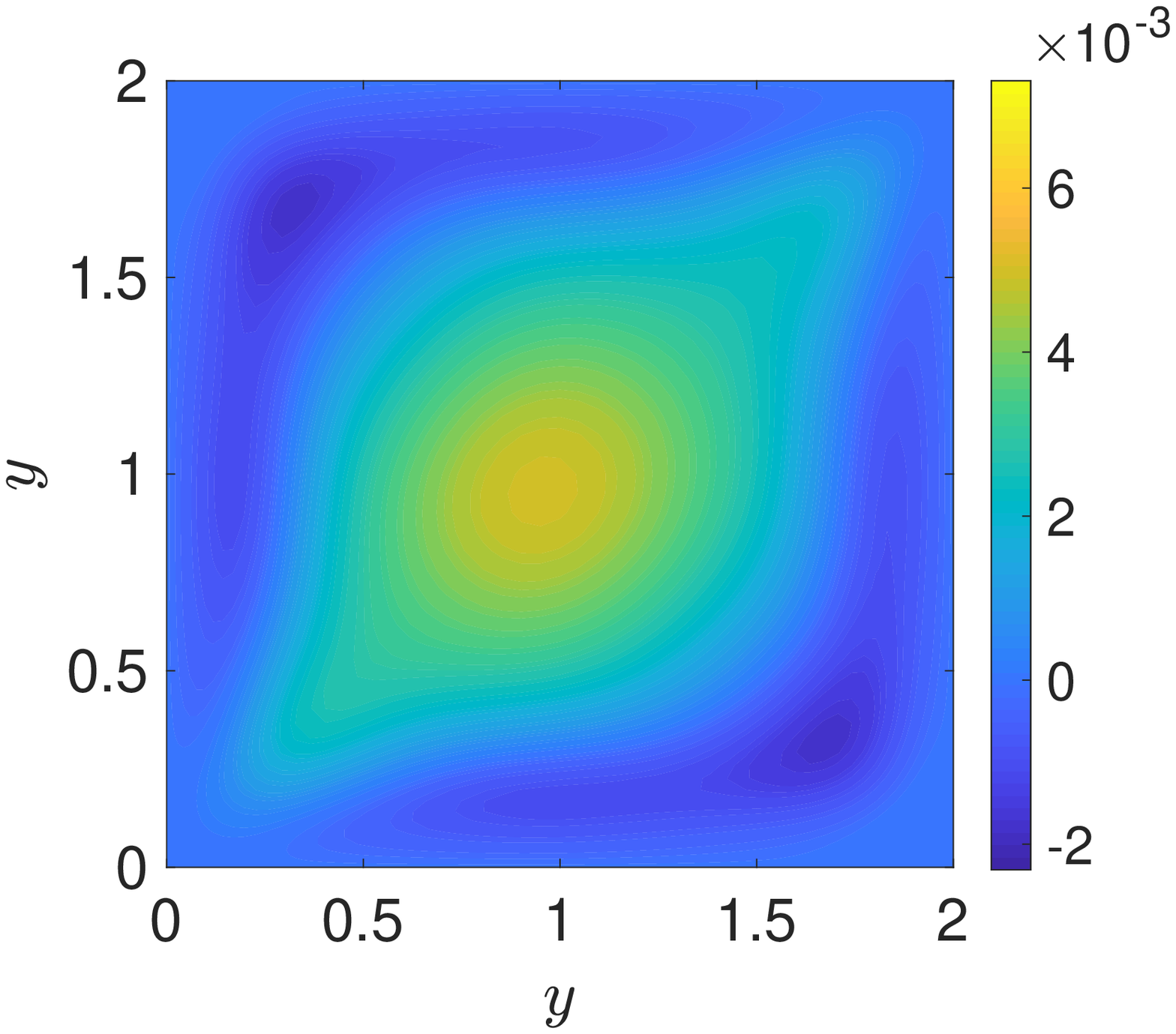}}\subfigure[All components of $f$]{\includegraphics[clip=true, trim= 0 0 0 0, width=0.5\textwidth]{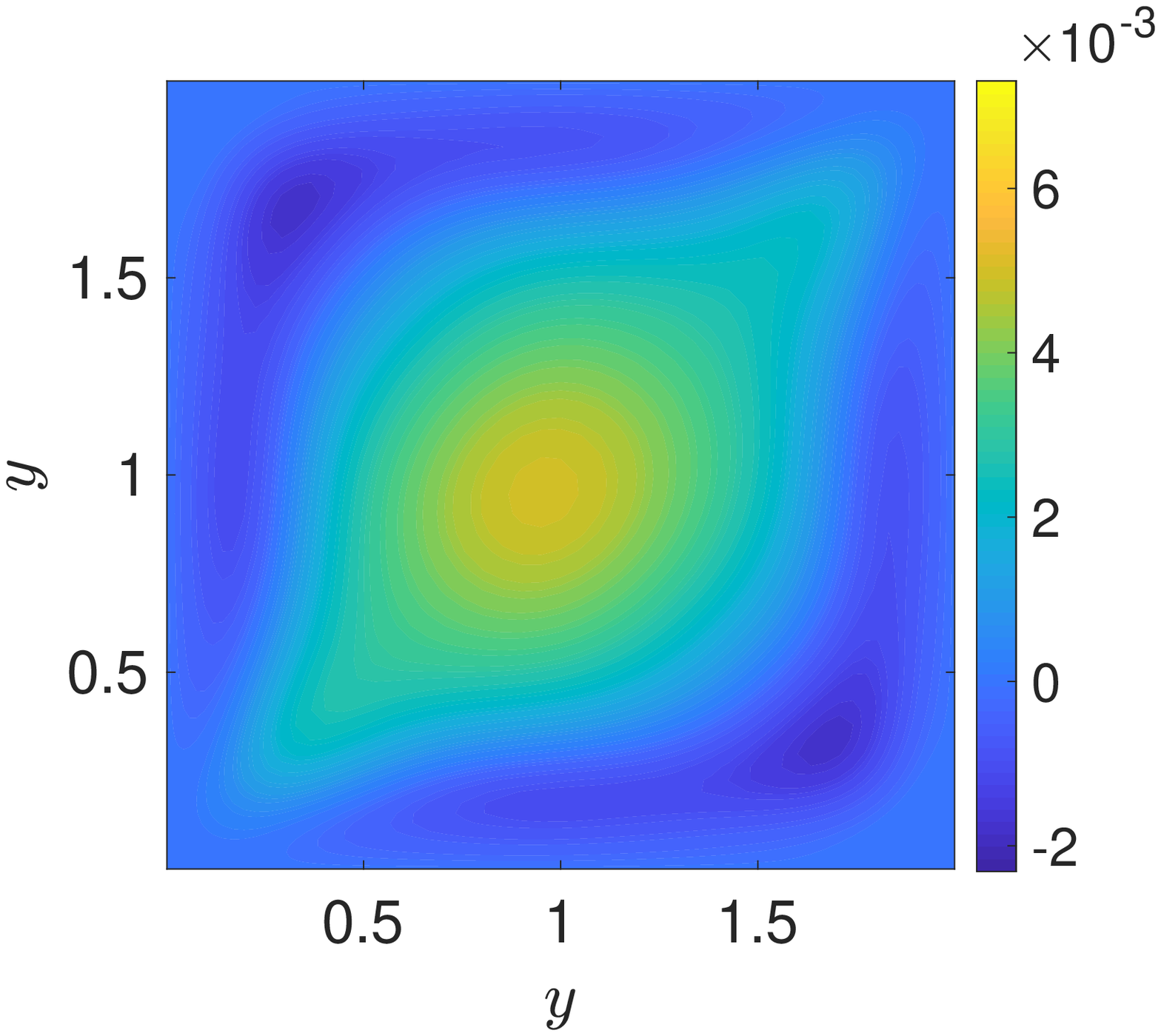}}
\subfigure[Using $f_u+f_v$]{\includegraphics[clip=true, trim= 0 0 0 0, width=0.5\textwidth]{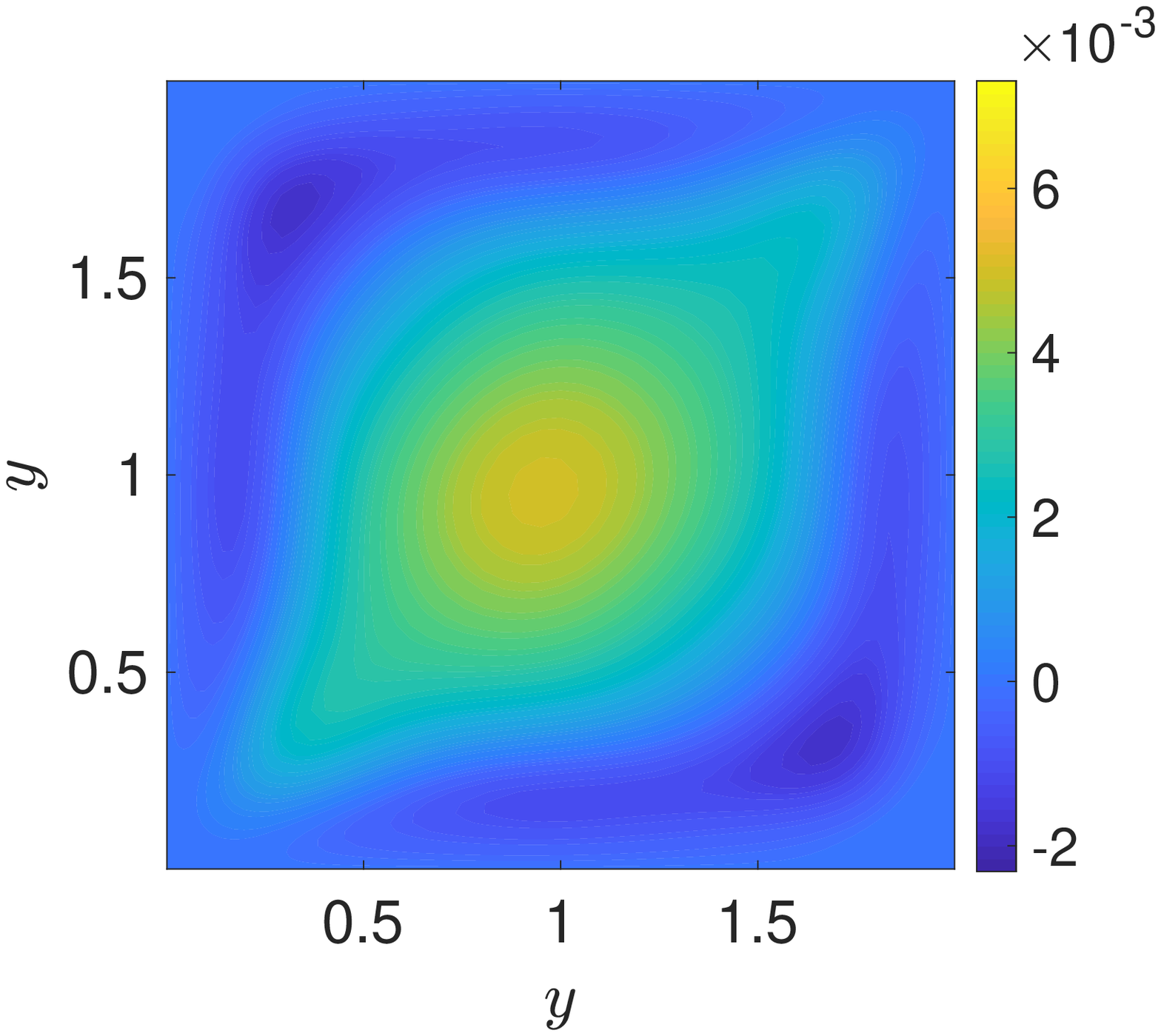}}\subfigure[Using $f_u$]{\includegraphics[clip=true, trim= 0 0 0 0, width=0.5\textwidth]{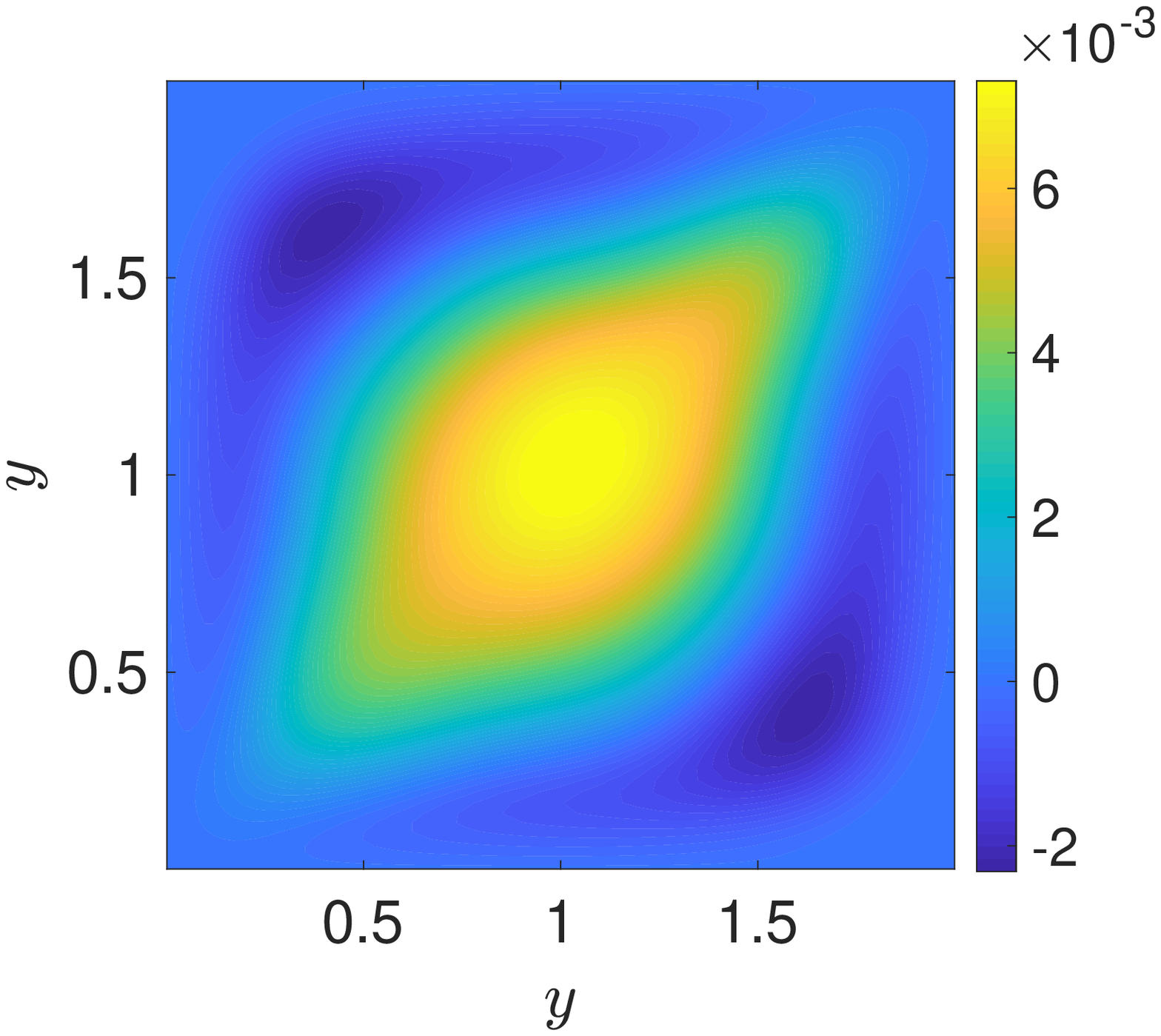}}
\subfigure[White-noise $\mathrm{P}$]{\includegraphics[clip=true, trim= 0 0 0 0, width=0.5\textwidth]{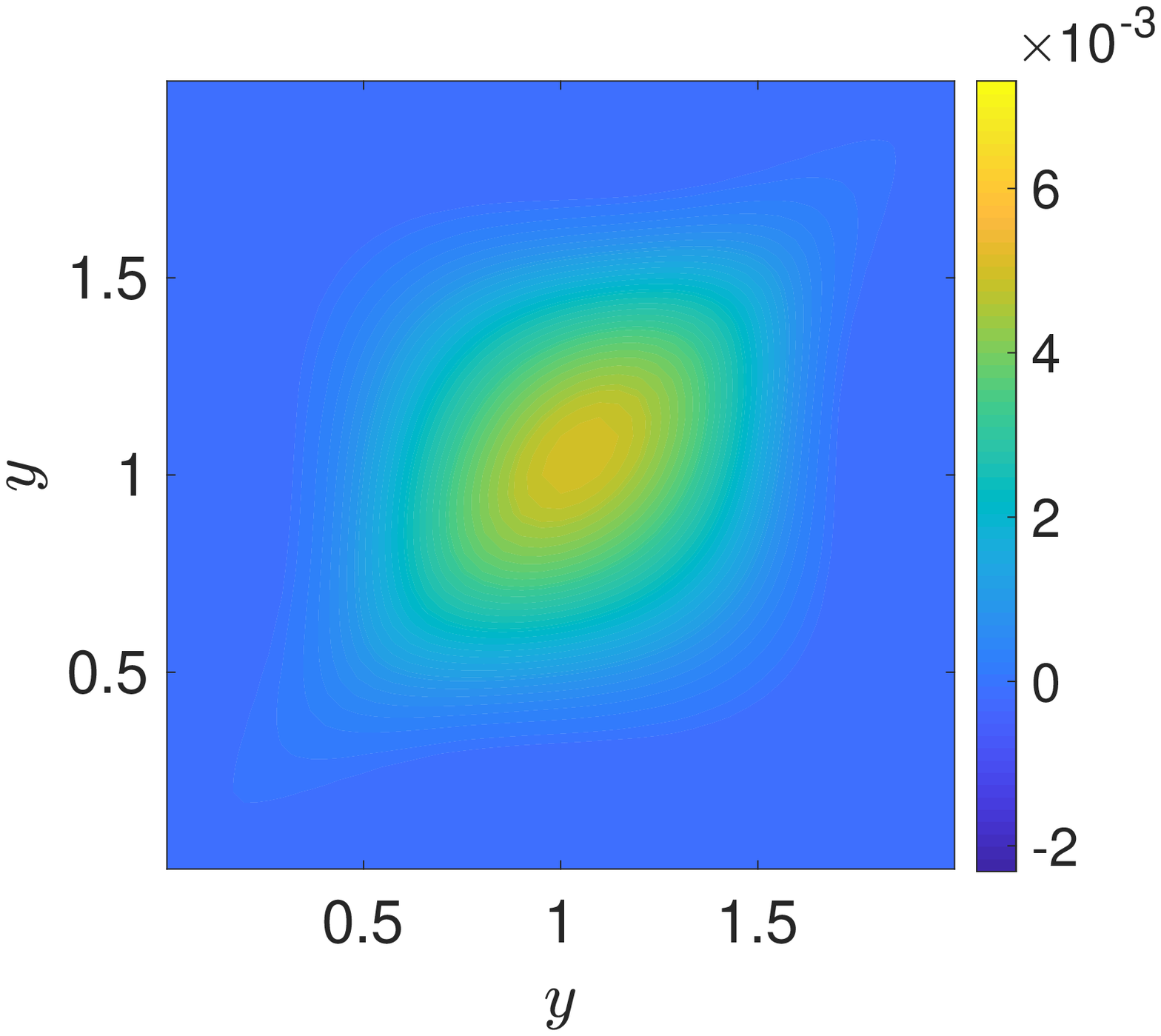}}\subfigure[Main diagonal of $\mathrm{S}$]{\includegraphics[clip=true, trim= 0 0 0 0, width=0.5\textwidth]{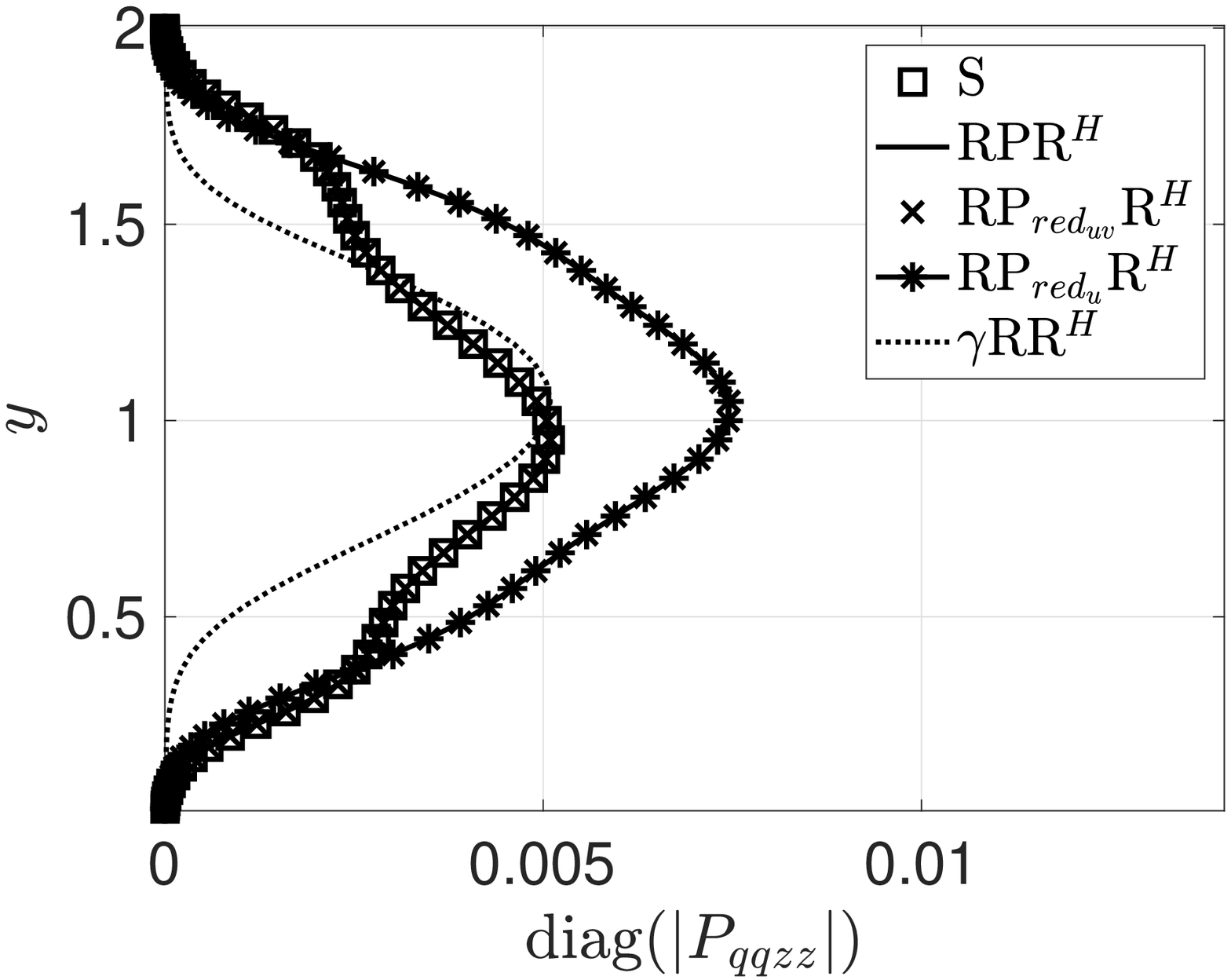}}
\caption{Streamwise component of $\mathrm{S}$ from the velocity field (a), from the reconstruction using the whole forcing (b), the reduced order forcing with $f_u+f_v$ (c), the reduced order forcing with $f_u$ (d), statistical white forcing (e) and the absolute value of the diagonal of each one compared with the one from the simulation (f).}
\label{fig:PqqPffred10}
\end{figure}

The $z$ component non-linear terms can be written as

\begin{equation}
    f_z=\underbrace{-u\frac{\partial w}{\partial x}}_{f_u}\underbrace{-v\frac{\partial w}{\partial y}}_{f_v}\underbrace{-w\frac{\partial w}{\partial z}}_{f_w},
    \label{eqn:Forcing10z}
\end{equation}

\noindent which is already simpler than the previous case, since the full forcing term only has the contribution of only three terms. We can perform the same analysis as in the previous case and try to remove some of the terms in order to check the impact of each one in the statistics of the response. The reductions that led to similar shapes for the statistics of the response are the combinations $f_u+f_v$ (nearly perfect match) and $f_u$ (slight differences). The reconstructions of $\mathrm{S}$ using each of these combination of terms is shown in figure \ref{fig:PqqPffred10}.

As expected, the reconstruction using all components of the forcing term recovers very accurately $\mathrm{S}$ from the simulation, as well as by using just the terms $f_u$ and $f_v$. A further reduction (using just $f_u$) also gives overall correct shapes, but an amplitude mismatch appears for the computed $\mathrm{S}$, as shown in figure \ref{fig:PqqPffred10}(f). No other simplification led to accurate recovery of $\mathrm{S}$. These plots also show the need of using the statistics of the forcing for the prediction of the response: figures \ref{fig:PqqPffred10}(e,f) show that considering white-noise statistics for the forcing leads to wrong shapes for the statistics of the response, even though the peak is roughly captured. 

\subsubsection{Low rank of forcing and response}

\begin{figure} 
\centering
\subfigure[Gains]{\includegraphics[clip=true, trim= 0 0 0 0, width=0.5\textwidth]{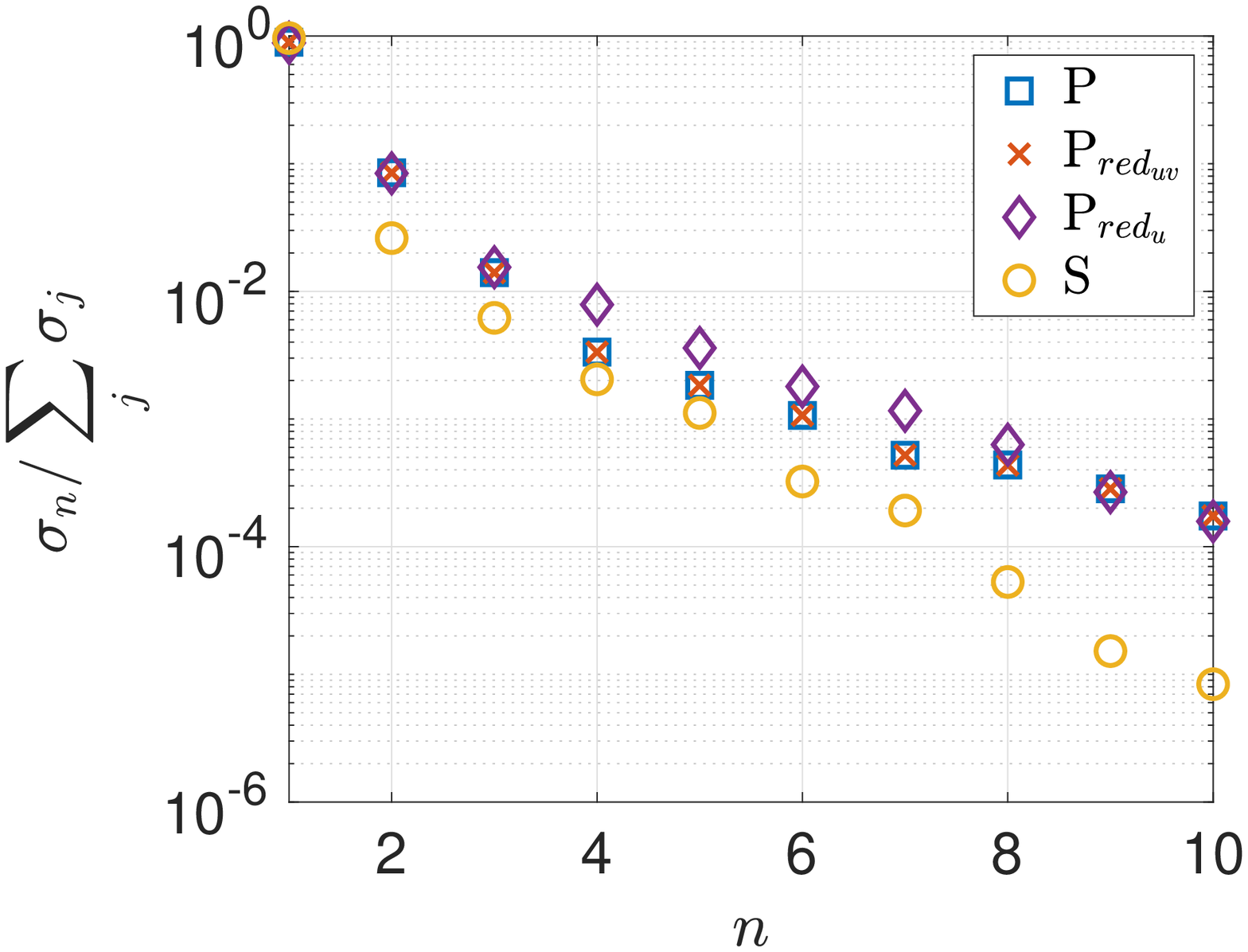}}\subfigure[SPOD modes and reconstructions]{\includegraphics[clip=true, trim= 0 0 0 0, width=0.5\textwidth]{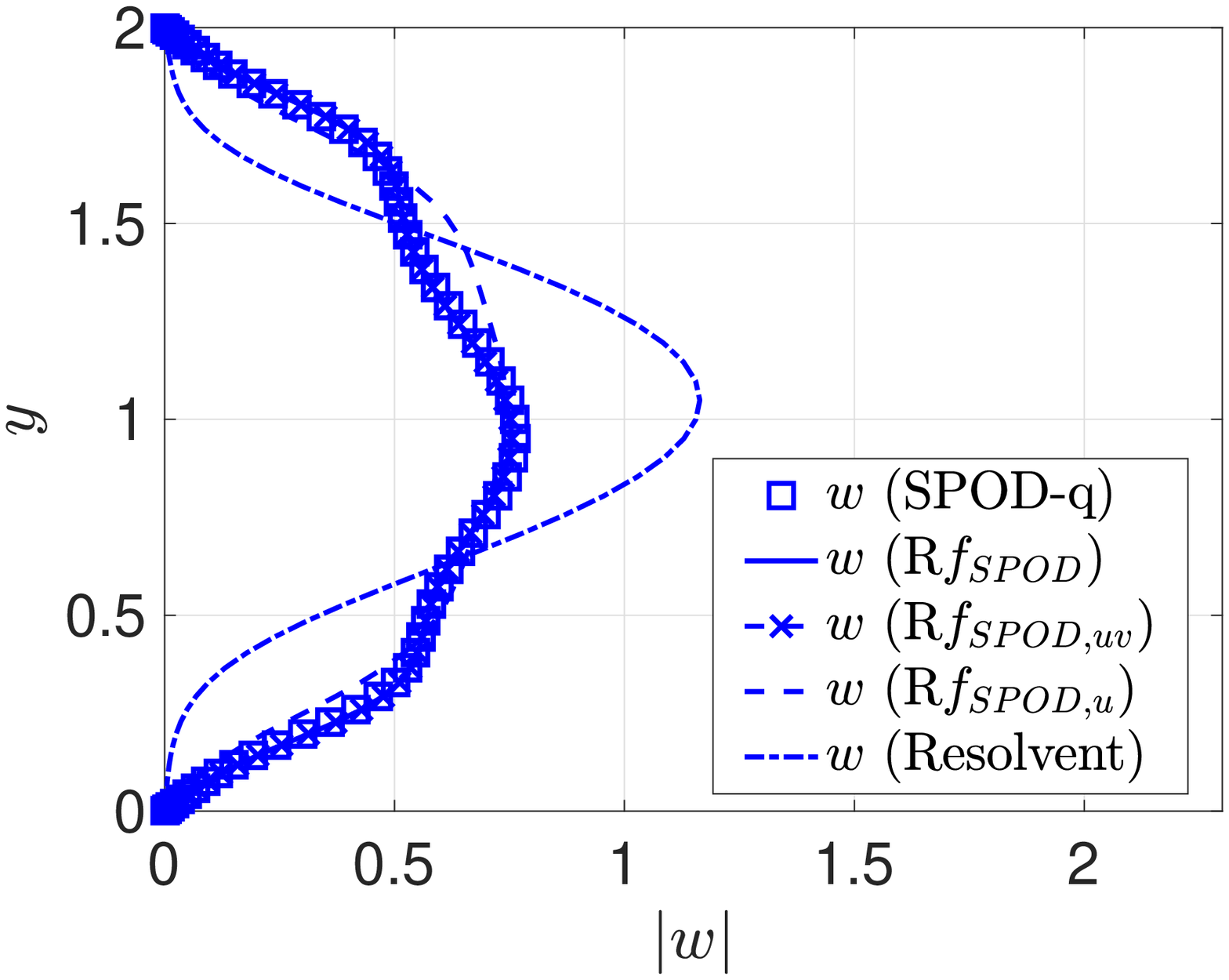}}
\caption{Energies from SPOD for $\mathrm{S}$ and the different $\mathrm{P}$ used here (a) and the shapes of the SPOD mode of the response compared with the reconstruction using the first SPOD mode of the forcing (b).}
\label{fig:Fredmaxgainsmodes10}
\end{figure}

Taking SPOD modes of $\mathrm{S}$ and of the corrected $\mathrm{P}$ (both full and reduced), we obtain the gains shown in figure \ref{fig:Fredmaxgainsmodes10}(a). It is clear that, for the present frequency, these matrices are also low-rank: the first SPOD mode is at least one order of magnitude higher than the other ones, pointing out that using only the first SPOD mode is sufficient to represent the forcing and the response. For a reconstruction using the first SPOD mode of the forcings, we obtain modes very close to the first SPOD mode of the response for all cases; even the more drastic reduction, considering only $f_u$, leads to a close agreement with the response statistics for most positions, with a slight mismatch above the centerline. As expected, given the differences between the prediction using the white-noise $\mathrm{P}$ and the covariance of the response from the simulation, the first resolvent mode does not capture the correct shape of the most energetic structure in the flow, especially close to the wall. 

\begin{figure} 
\centering
\subfigure[Absolute value of first SPOD mode of the forcings]{\includegraphics[clip=true, trim= 0 0 0 0, width=0.5\textwidth]{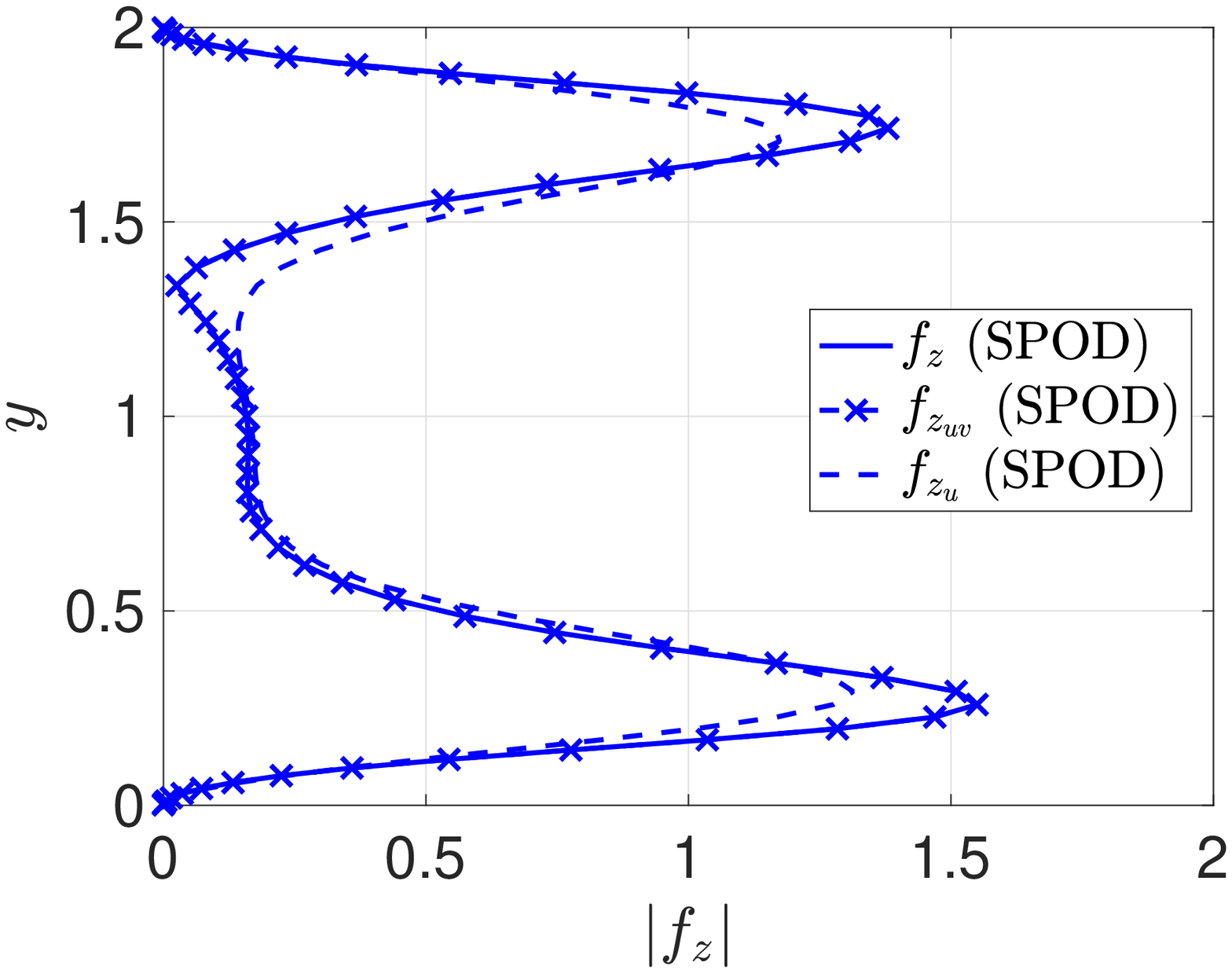}}\subfigure[Reduced forcing $f_u$ in the physical space]{\includegraphics[clip=true, trim= 0 0 0 0, width=0.5\textwidth]{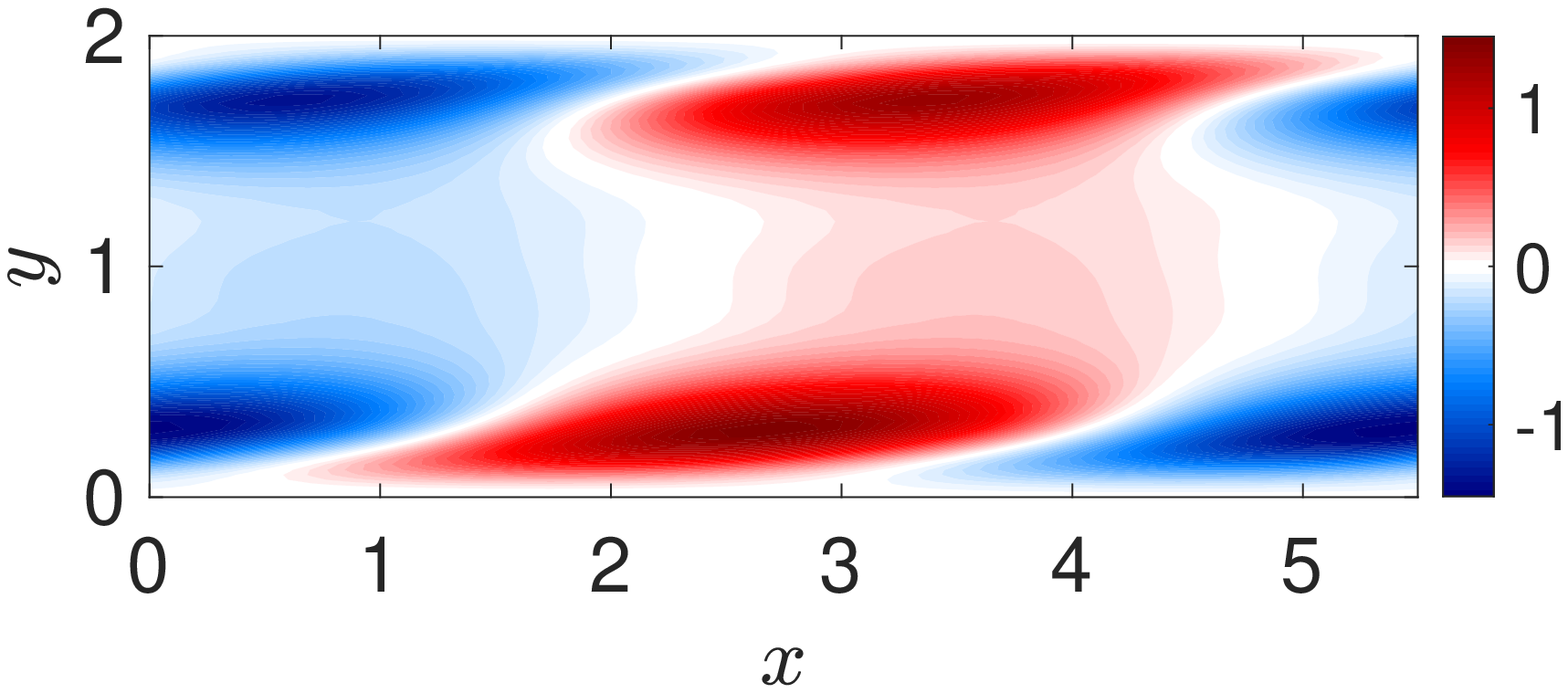}}
\caption{Absolute value of first SPOD mode of the full and reduced forcings (a) and the shape of the reduced forcing in the physical space. Colours: spanwise forcing.}
\label{fig:Fredmax10}
\end{figure}

Finally, we can also look at the shapes of the leading SPOD mode of the forcing for this case. Figure \ref{fig:Fredmax10} shows that the reduction of the forcing to only one of the terms does not lead to any substantial changes in the optimal forcing; the $f_v$ component adjusts the shape of forcing structure in some specific regions, without a major role in the bulk shape. In figure \ref{fig:Fredmax10}(b) we can see the reduced optimal forcing in the physical space. The peaks, for this case, are concentrated in regions close to the wall, which explains why the difference between resolvent and SPOD modes is more evident in these regions; since $f_z$ is higher in that region, the first SPOD mode also has higher amplitudes closer to the wall. 

From these results, we can see that the case $(1,0)$ is mostly dependent on the term $f_z$, meaning that the triadic interactions that lead to spanwise forcings with higher magnitudes near the wall will be more relevant to the forcing. Also, since it is mainly dependent on $u\partial w / \partial x$, it is expected that the interactions composed by terms with wavenumbers pairs with high streamwise velocity will affect the non-linear term more substantially.

\subsection{Relation with the self-sustaining process}
\label{sec:selfsust}
This work was based on the minimal flow unit for Couette flow, the minimal computational box in which turbulence can be sustained. This case was developed by \cite{hamilton1995regeneration}, who also proposed a self-sustaining process for the turbulence in this simple shear flow. Considering that the structures present in this flow are ubiquitous to all shear flows, the proposed mechanism has also been extended to several other cases (see \cite{jimenez1999autonomous,schoppa2002coherent}, for example). This can be summarised as: (1) streaks are produced as a result of the lift-up effect \citep{ellingsenpalm1975}; (2) the growth of the streaks leads to an instability in the flow, which triggers the breakdown of these structures; (3) streamwise vortices are regenerated via a non-linear mechanism, thus leading to a regeneration of the streaks. 

\begin{figure} 
\centering
\includegraphics[clip=true, trim= 0 0 0 0, width=\textwidth]{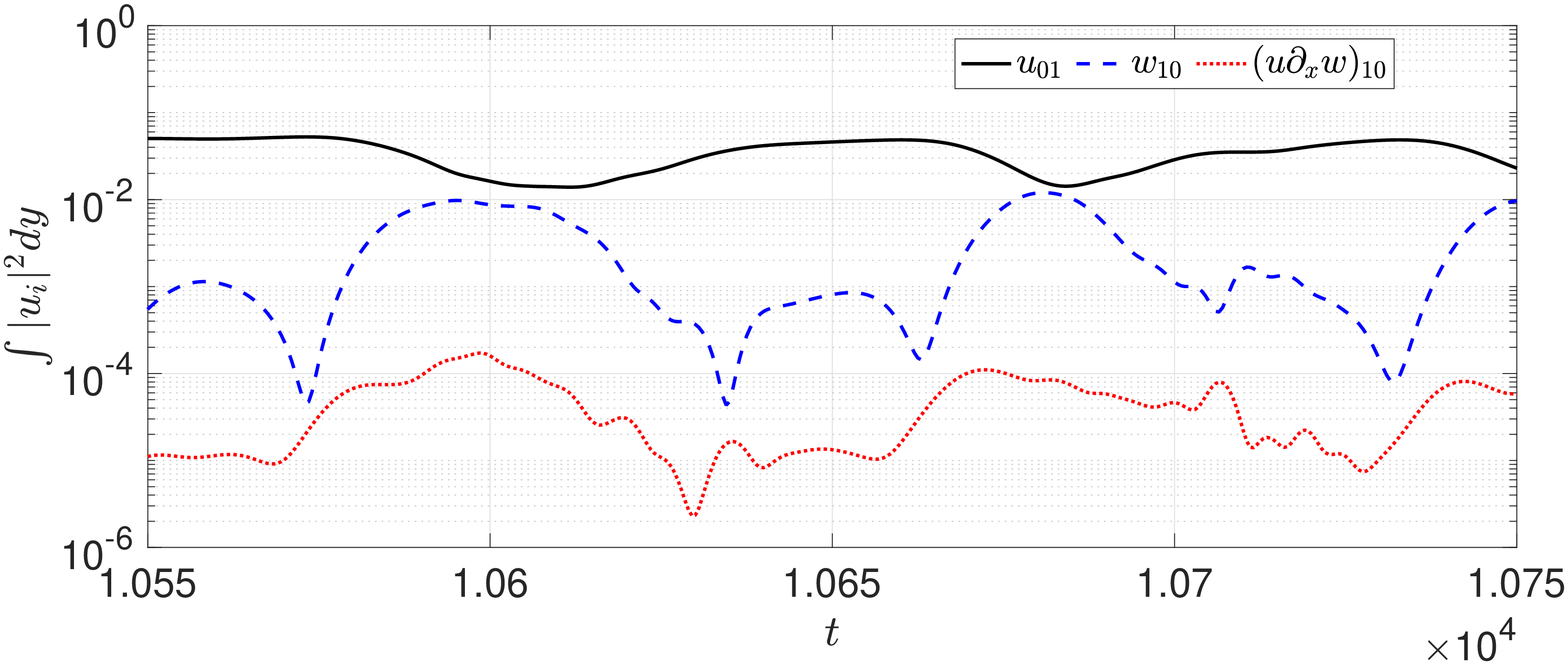}
\caption{Integrated energy as a function of time related to the streamwise velocity and the non-linear terms $\partial(uv)/\partial y$ and $v\partial u/\partial y$  (mode (0,1)), spanwise velocity and the non-linear term $u\partial w/\partial x$ (mode (1,0)).}
\label{fig:timeseries}
\end{figure}

Figure \ref{fig:timeseries} shows some of the relevant quantities for two of the cycles of the present simulation. The process of generation and breakdown of streaks can be tracked by looking at the streamwise component of the velocity for mode $(0,1)$; the energy of mode $(1,0)$ can be approximated by the contribution of the spanwise component of this mode. The dominant part of the forcing term for mode $(1,0)$ is also shown. 

From figure \ref{fig:timeseries}, it is clear that the dynamics of the spanwise velocity for mode $(1,0)$ (followed closely by its main forcing component) opposes the trend of the streamwise velocity of mode $(0,1)$, which agrees with the analysis of streak breakdown for Couette flow performed by \cite{schoppa1999formation}. The authors show that the streak instability is triggered by streamwise varying, low-amplitude spanwise velocity disturbances; therefore, it is expected that an increase in the $w$ component of the velocity will lead to the breakdown of the streaks. Interestingly, the amplitudes of this component can reach the same order of magnitude of the streamwise component during streak breakdown, highlighting the importance of the underlying dynamics followed by $w$ in this flow. The increase of the spanwise component can also be connected to the streak regeneration process. The analysis in section \ref{sec:mode01} has shown that this component is essential for the recovery of the velocity statistics of streaks; the simplified forcing of eq. \ref{eqn:decompforc_simp1} has its streamwise vortical component, $f_w$, strongly related to fluctuations in $w$. Therefore, in order to recover the energy of mode $(0,1)$, higher amplitudes of spanwise velocity should be present in the flow. This may be connected to the time series shown in figure \ref{fig:timeseries}.

\section{Conclusions}

Using a direct numerical simulation of a minimal flow unit, taken for turbulent Couette flow at $Re=400$, we carried out an analysis of the terms of the Navier-Stokes equations that are non-linear in flow fluctuations around the turbulent mean. {It is shown that, even though resolvent response modes coming from linear analysis correspond to SPOD modes if the non-linear terms are uncorrelated, the statistics of these terms actually play a substantial role in some cases, changing the shape of the resulting energetic modes}. By detailing the process of recovering the frequency-domain statistics of the velocity from the statistics of the non-linear terms (treated as a forcing term, in a resolvent formulation), we managed to understand the influence of the windowing in the equations, which gives rise to new terms that must be included in the formulation, as proposed by \cite{martini2019accurate}; in the present work, we further show the validity of this result by applying the methodology for the analysis of a turbulent flow. Most of the comparisons between SPOD and resolvent modes in the literature do not consider, to the best of our knowledge, the error due to windowing. For lower frequencies, this may lead to unexpected errors that may deteriorate the comparisons. In order to correctly evaluate the validity of the models using linear analysis, a correction term should be considered as an additional forcing (as done in the present work), or as a correction of the statistics of the velocity. From the present analysis, it is clear that this error mainly affects low frequencies/wavenumbers. When the appropriate correction is included, relative errors of about $10^{-5}$ are obtained, which ensures the accuracy of the obtained cross-spectral densities of forcing and response for this flow.

Considering that the CSDs of forcing ($\mathrm{P}$) and response ($\mathrm{S}$) are accurately related to each other by the resolvent operator, we further analysed the linearised Navier-Stokes equations in order to evaluate which parts of the forcing were relevant for the prediction of the statistics of the velocity. This was done for the two most energetic cases at low frequencies: the case $(n_\alpha,n_\beta)=(0,1)$ (streaks), and $(n_\alpha,n_\beta)=(1,0)$ (spanwise velocity modes). These analyses were performed for $\omega \to 0$, as spectra for the two wavenumbers have highest levels for the lowest frequencies, a consequence of the zero phase speed of dominant modes in Couette flow with walls moving in opposite directions. The first mode, $(0,1)$, is characterised by the appearance of streaks and streamwise vortices. We have shown that using spatial white-noise as statistics of the forcing leads to a partial agreement between the prediciton of the covariance of the response using the resolvent operator and the one obtained from the simulation; even though streamwise vortices and streaks are obtained from white-noise forcing, there is a mismatch in their relative amplitudes. The forcing is shown to act with a destructive interference between direct forcing of streaks by streamwise forcing and the lift-up mechanism, where the streamwise vortical part of the forcing leads to streamwise vortices and streaks. Simplification of non-linear terms is also possible, with four of them (among nine possible ones) leading to the bulk of the response statistics; of particular relevance is the contribution of spanwise velocity in forcing, as it appears in three of the four dominant terms.

The same process was performed for the wavenumber $(1,0)$. The equations for this case decouple for the spanwise velocity. As this component is the most energetic one for this case, only the spanwise forcing would be needed to recover most of the covariance matrix of the response. By applying the same process as the previous case, we manage to reduce the forcing to only one term, $u\partial w/\partial x$. Simplification of the forcing as white noise again led to a mismatch in the predicted flow responses; here, the amplitude distribution of the forcing, which is stronger closer to the walls, is a salient feature, leading to the observed flow response.

The results presented herein aim to clear the usual complexity related to dealing with the non-linear terms. Even though these may be considered as an external forcing of the system for a simplified analysis, leading for instance to interesting conclusions about optimal responses of the system, we must keep in mind that these ``forcing'' terms come from the dynamics of the flow. Turbulence thus leads to a particular structure, or colour, to these forcing terms, and such structure has been shown to be relevant if one wishes to obtain accurately the flow responses via resolvent analysis. It is notable that for the two considered wavenumbers, at low frequencies, the forcing CSD is of low rank, dominated by its first eigenfunction. Hence, despite the apparent complexity of the bulk of non-linear terms from this flow, some clear order can be found. Such organisation contributes to the aforementioned constructive or destructive interferences in leading to flow responses. {In light of the present results, it is thus relevant to understand and model how coherent structures in turbulent flows give rise to such organised non-linear terms.}

Acknowledgements: Petr\^onio Nogueira was funded by a CNPq scholarship. Andr\'e Cavalieri acknowledges financial support by CNPq (grant number 310523/2017-6).
Declaration of Interests. The authors report no conflict of interest.

\appendix
\section{Correction due to windowing}
\label{sec:corrwind}
The equations shown in the section \S \ref{sec:PqqfromPff} are exact when no windowing function is applied to the turbulent signals; cross-spectral densities (CSD) can be obtained as Fourier transforms of the correlation function. Still, considering most applications, windowing and segment averaging are applied to time series in order to apply Welch's method for a faster determination of cross-spectral densities. As windowing is unavoidable when dealing with large datasets, one can choose the window in order to minimise spectral leakage and/or aliasing. The inclusion of a window in the signal processing, in turn, leads to appearance of new terms in the equations. Following the formulation of \cite{martini2019accurate}, the linearised, time-invariant, Navier-stokes equations,

\begin{equation}
    \mathrm{H} \frac{\partial  \mathbf{\tilde{q}}(t)}{\partial t} = \mathrm{L} \mathbf{\tilde{q}}(t) + \mathbf{\tilde{f}}(t),
\end{equation}

\noindent when multiplied by a window function $h(t)$, can be re-written as

\begin{equation}
    \mathrm{H} \frac{\partial (h(t) \mathbf{\tilde{q}}(t))}{\partial t} = \mathrm{L}(h(t) \mathbf{\tilde{q}}(t)) + (h(t) \mathbf{\tilde{f}}(t)) + \mathrm{H} \left(\frac{\partial h}{\partial t} \mathbf{\tilde{q}}\right)(t),
    \label{eqn:windLNS}
\end{equation}

\noindent where $\mathbf{\tilde{q}}=[\tilde{u} \ \tilde{v} \ \tilde{w} \ \tilde{p}]^T$, $\mathbf{\tilde{f}}=[\tilde{f}_x \ \tilde{f}_y \ \tilde{f}_z \ 0]^T$ are the response and forcing in time domain, and $\mathrm{H}$ is defined to disregard the derivative of the pressure in the equations. The Fourier transform of the windowed signal in each segment is given by

\begin{equation}
    \mathbf{\hat{q}}(\omega)=\frac{1}{T}\int_{t_0}^{t_0+T} h(t) \mathbf{\tilde{q}}(t) \ee^{\ii\omega t}dt,
    \label{eqn:fouwind}
\end{equation}

\noindent and equivalently for $\mathbf{f}$, where $t_0$ denotes the initial time for each segment and $T$ the duration of time segments. Defining 

\begin{equation}
    \overline{{\mathbf{q}}}(\omega)=\frac{1}{T}\int_0^T \frac{d h(t)}{dt} \mathbf{\tilde{q}}(t) \ee^{\ii\omega t}dt,
    \label{eqn:fouwind2}
\end{equation}

\noindent equation \ref{eqn:windLNS} can be written as

\begin{equation}
    (-\ii \omega \mathrm{H} - \mathrm{L}) \mathbf{\hat{q}} = \left(\hat{\mathbf{f}} + \mathrm{H}\overline{\mathbf{q}} = \right) = \mathbf{\hat{f}}_{corr},
    \label{eqn:windLNS_freq}
\end{equation}

\noindent where the right-hand side becomes the effective force of the windowed signal, which is the sum of the external driving force and the external extra term due to windowing. Equation \ref{eqn:windLNS_freq} indicates that, even if we manage to obtain converged statistics for the forcing and the response, the fact that we multiply the signal by a window leads to a mismatch between the statistics of the response computed from the velocity fields, and the recovered statistics from the non-linear terms (by using $\mathrm{R}\mathrm{P}\mathrm{R}^H$). This error is well defined and is a function of the window used in the Welch method, of the operator used in the analysis and of the chosen frequency. Fundamentally the error comes from the difference between $\mathbf{q}$, the true Fourier transform of the signal, and $\hat{\mathbf{q}}$, its estimate  obtained with using the window $h(t)$, the difference also being present for $\mathbf{f}$. Such error is reduced when long segments are taken, which decreases the magnitude of the time derivative of the windowing function in eq. \ref{eqn:fouwind2}; however, the use of sufficiently long segments for negligible correction terms is potentially prohibitive.

\section{Simple coloured forcing statistics}\label{appA}
\begin{figure} 
\centering
\subfigure[Cross-spectral density matrix of the response]{\includegraphics[clip=true, trim= 0 0 0 0, width=0.5\textwidth]{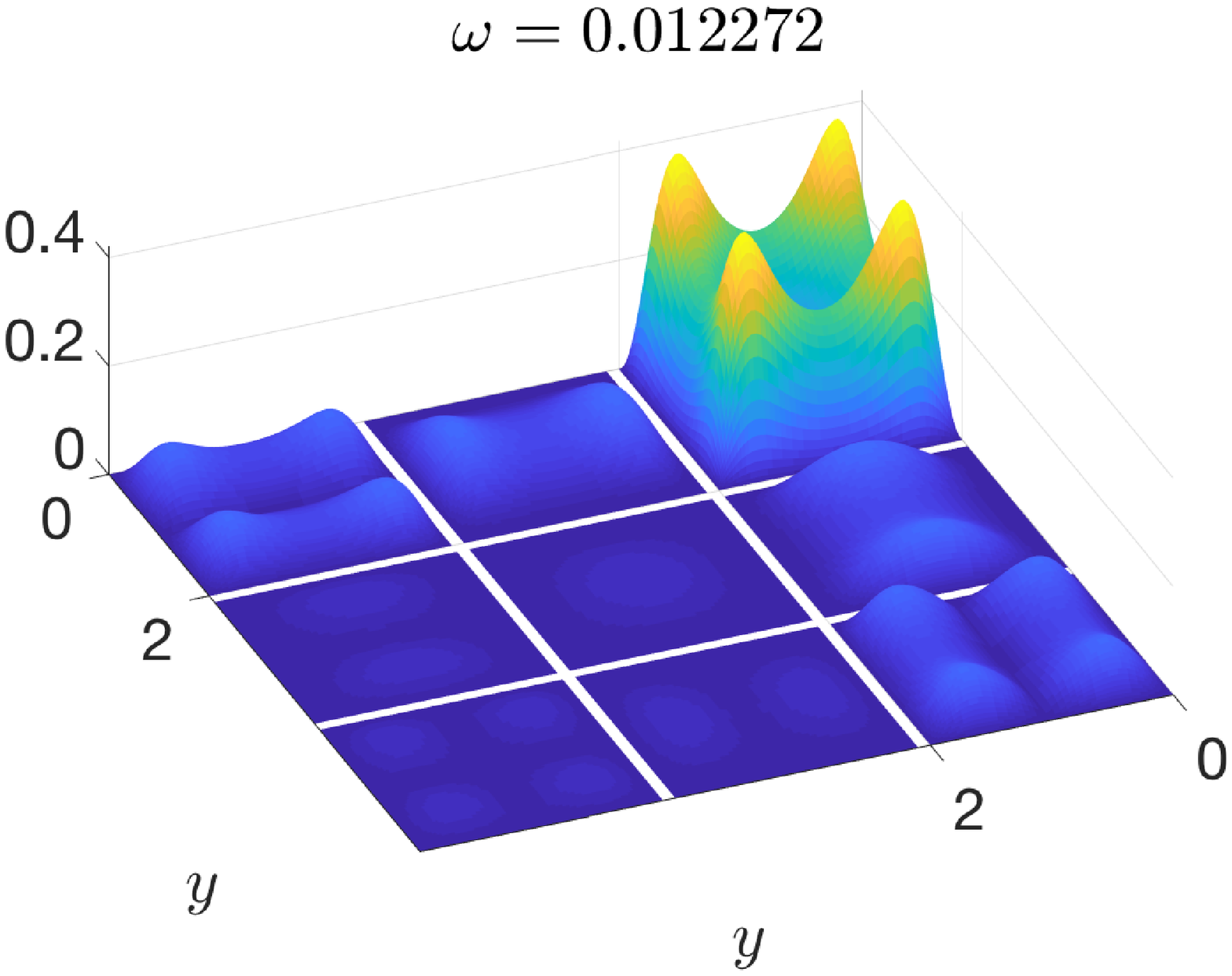}}\subfigure[Cross-spectral density matrix of the forcing]{\includegraphics[clip=true, trim= 0 0 0 0, width=0.5\textwidth]{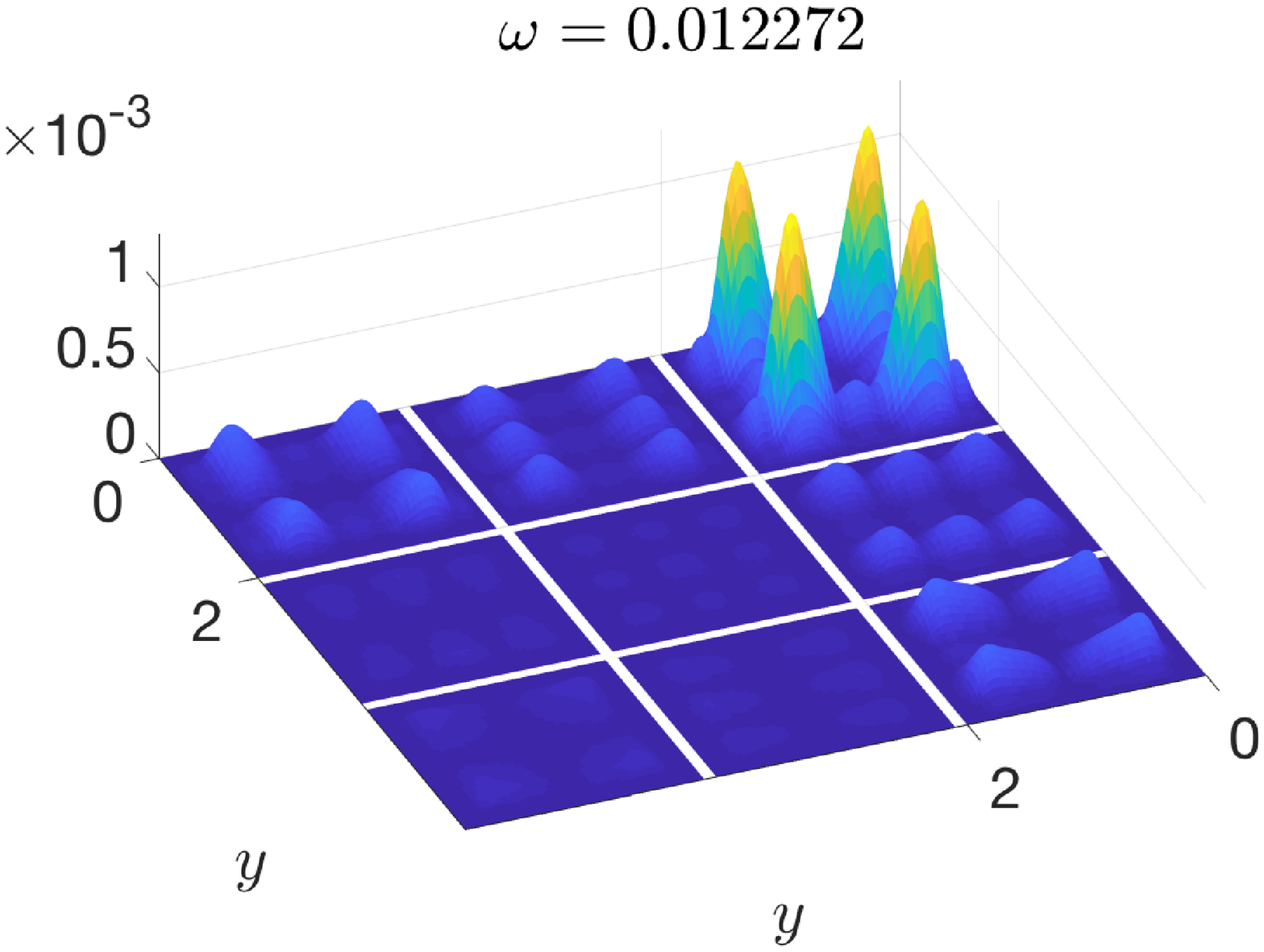}}
\subfigure[Coherence of the response]{\includegraphics[clip=true, trim= 0 0 0 0, width=0.5\textwidth]{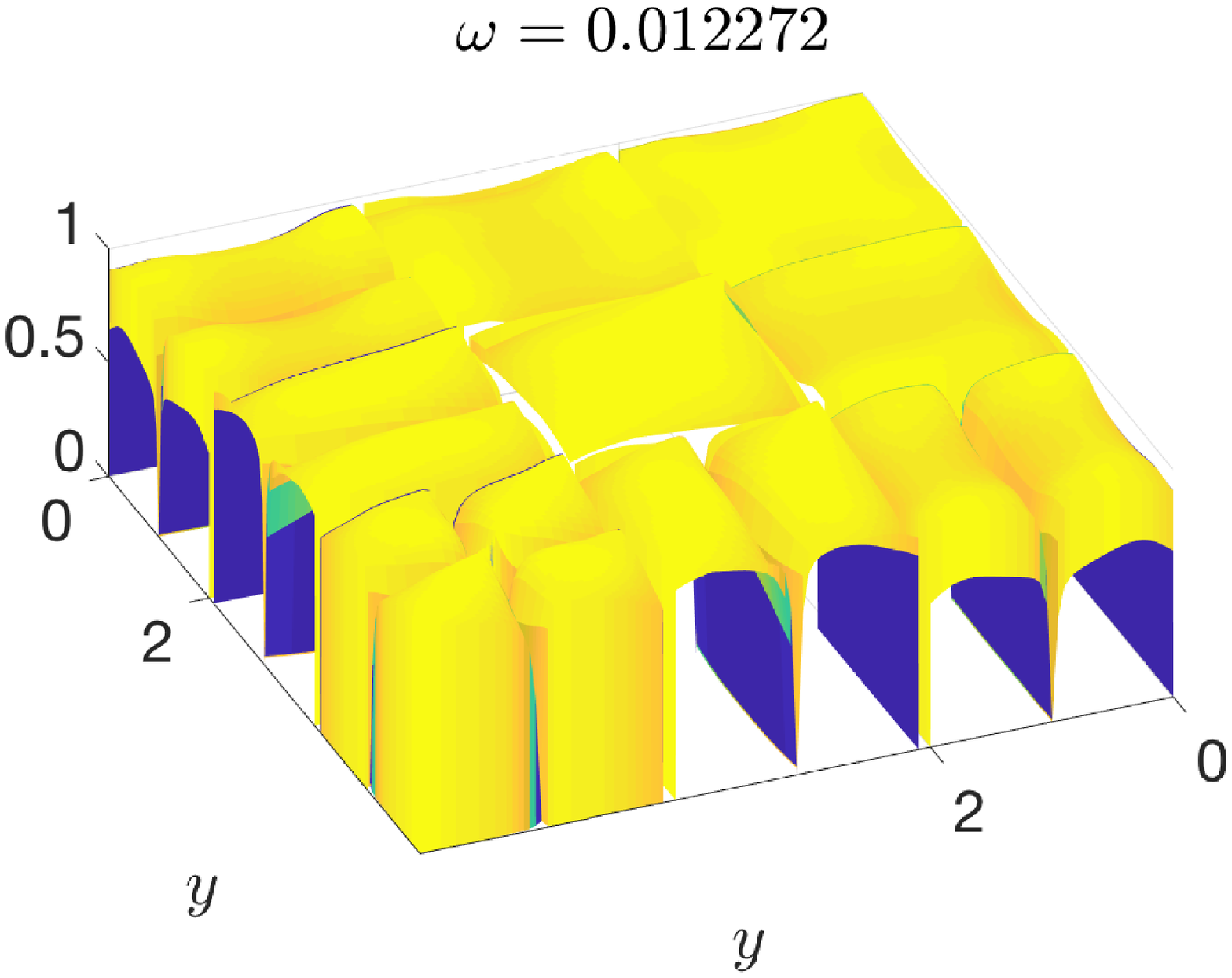}}\subfigure[Coherence of the forcing]{\includegraphics[clip=true, trim= 0 0 0 0, width=0.5\textwidth]{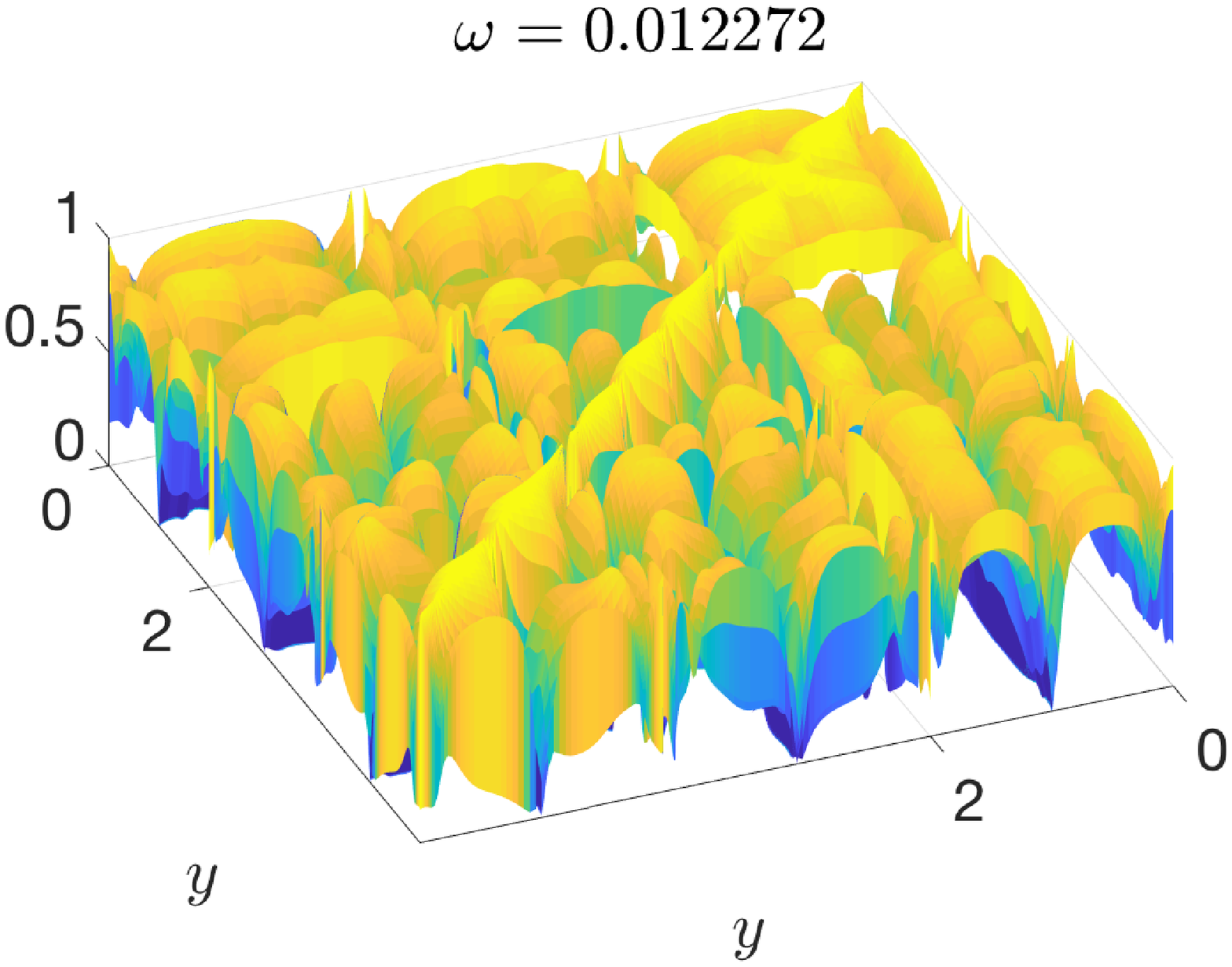}}
\caption{Cross-spectral density matrix of the response (a) and of the forcing (b) for mode $(0,1)$ and $\omega=0.0123$. The coherence of each quantity is shown in (c,d). Each square corresponds to the absolute value of one of the matrix components of the given quantity, starting from the streamwise component (farther from the viewer) to the spanwise component (closer to the viewer).}
\label{fig:PffPqqAll001}
\end{figure}

The paper focused on the analysis of the forcing statistics in the minimal Couette flow. From the full forcing statistics, reduced order forcings based on the identification of the relevant parts of the non-linear terms were proposed in order to simplify the analysis. One of the main outcomes of this survey was that coherent structures in the non-linear terms could be clearly distinguished, revealing some of the dynamics behind the streak generation. The results presented herein can also be used to guide modelling of the forcing statistics, since a modelled $\mathrm{P}$ should keep the main characteristics of the one identified in the simulation. To exemplify such a process, we devote this appendix to the analysis of some of the relevant characteristics of $\mathrm{P}$, its effect in $\mathrm{S}$ and how does a simple model for the forcing behave in some frequencies. We start by looking at the frequency studied in the paper ($\omega=0.0123$) and wavenumbers $(n_\alpha,n_\beta)=(0,1)$, corresponding to streaks and streamwise vortices studied in section \ref{sec:mode01}. The covariance matrices and their respective coherences are shown in figure \ref{fig:PffPqqAll001}.

As previously observed, the response for this combination of wavenumbers is mostly concentrated at the streamwise direction (leading to a clear signature of streaks), but the other components also have non-negligible amplitudes due to the presence of streamwise vortices in the flow.  The forcing also has the same overall behaviour, with higher amplitudes in the streamwise directions and lower in the wall-normal and spanwise ones. Amplitudes of the forcing in the main diagonal of $\mathrm{P}$ are comparable to the ones outside this region, pointing to the presence of coherent structures that extent throughout domain (as in the response). This is confirmed by the coherence computed for each quantity, shown in figure \ref{fig:PffPqqAll001}(c,d). From these plots, it is clear that both forcing and response are completely coherent throughout the domain (except for the regions where this quantity decays to zero, closer to the walls). For this frequency, it is expected that a model for the forcing that does not consider this high coherence between the components will not adequately represent $\mathrm{P}$, leading thus to errors in predictions of $\mathrm{S}$.

\begin{figure} 
\centering
\subfigure[Cross-spectral density matrix of the response]{\includegraphics[clip=true, trim= 0 0 0 0, width=0.5\textwidth]{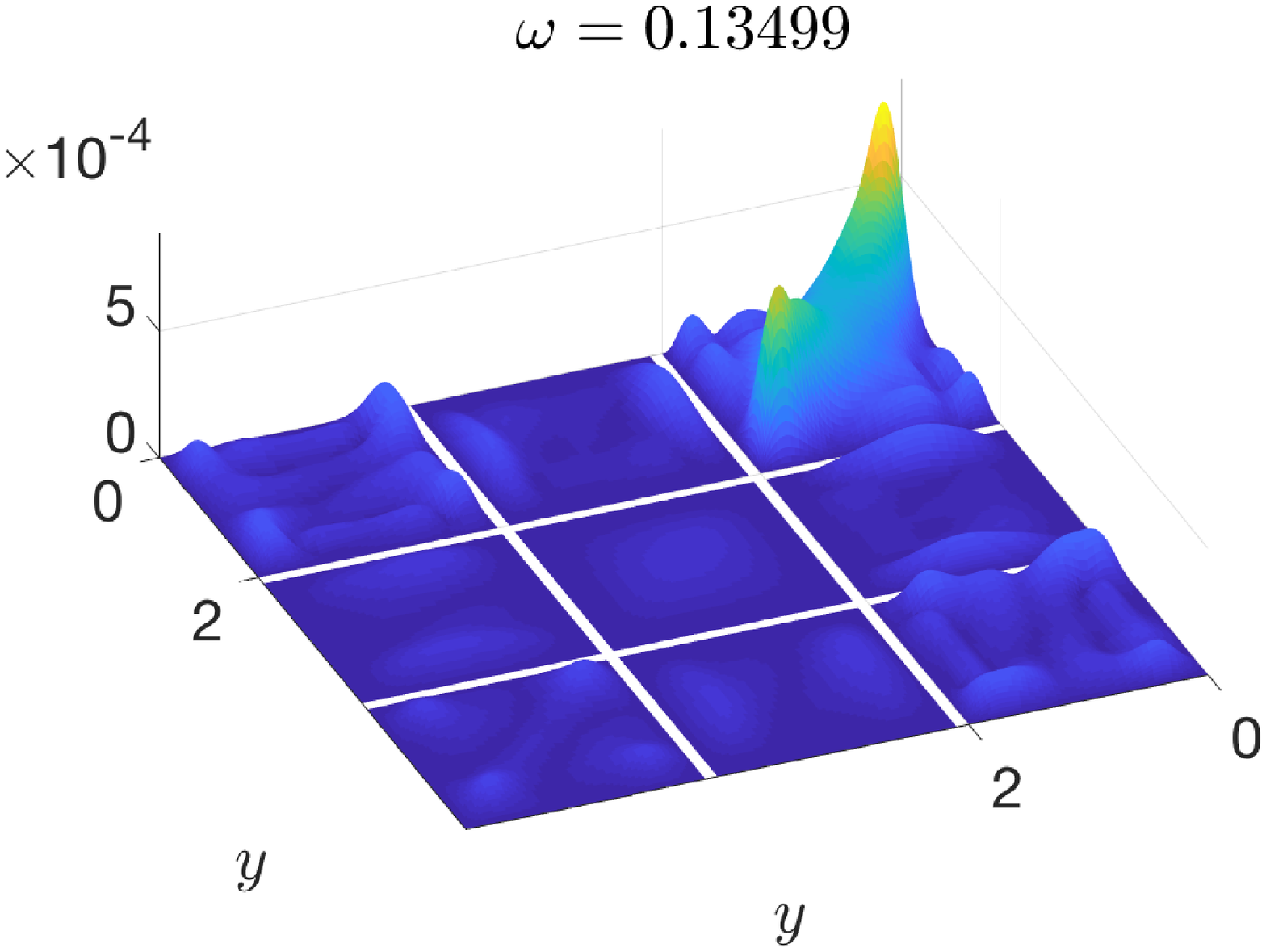}}\subfigure[Cross-spectral density matrix of the forcing]{\includegraphics[clip=true, trim= 0 0 0 0, width=0.5\textwidth]{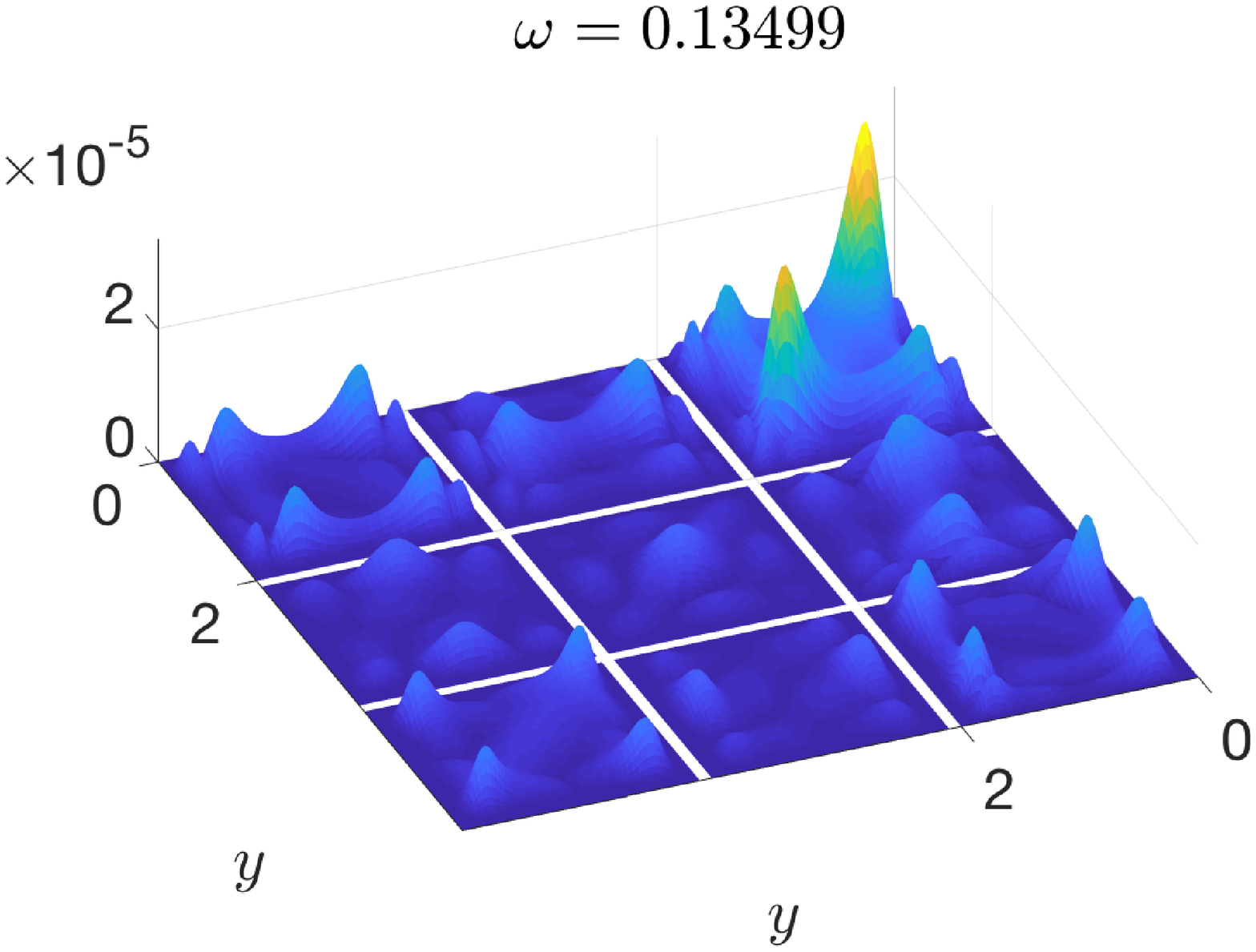}}
\subfigure[Coherence of the response]{\includegraphics[clip=true, trim= 0 0 0 0, width=0.5\textwidth]{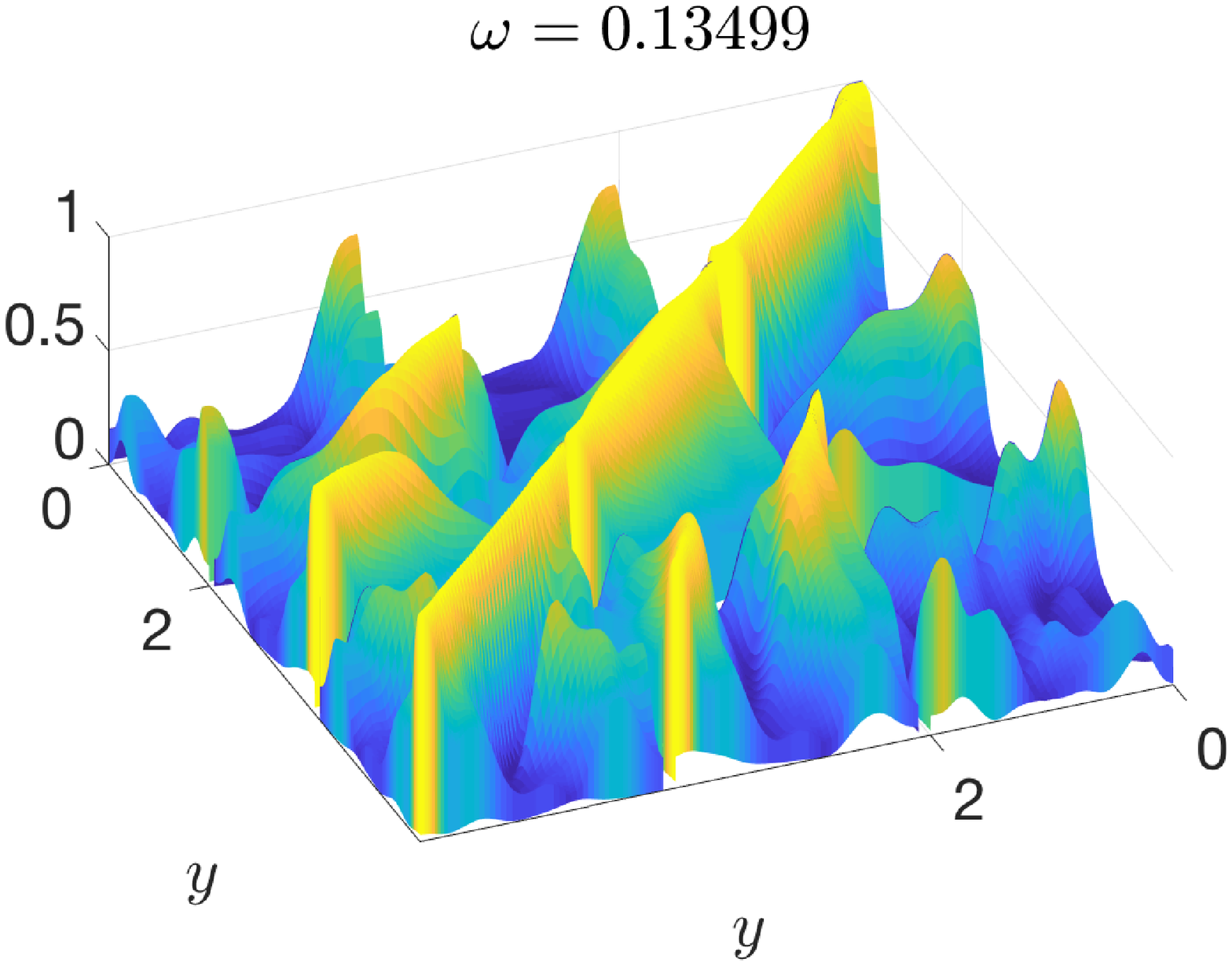}}\subfigure[Coherence of the forcing]{\includegraphics[clip=true, trim= 0 0 0 0, width=0.5\textwidth]{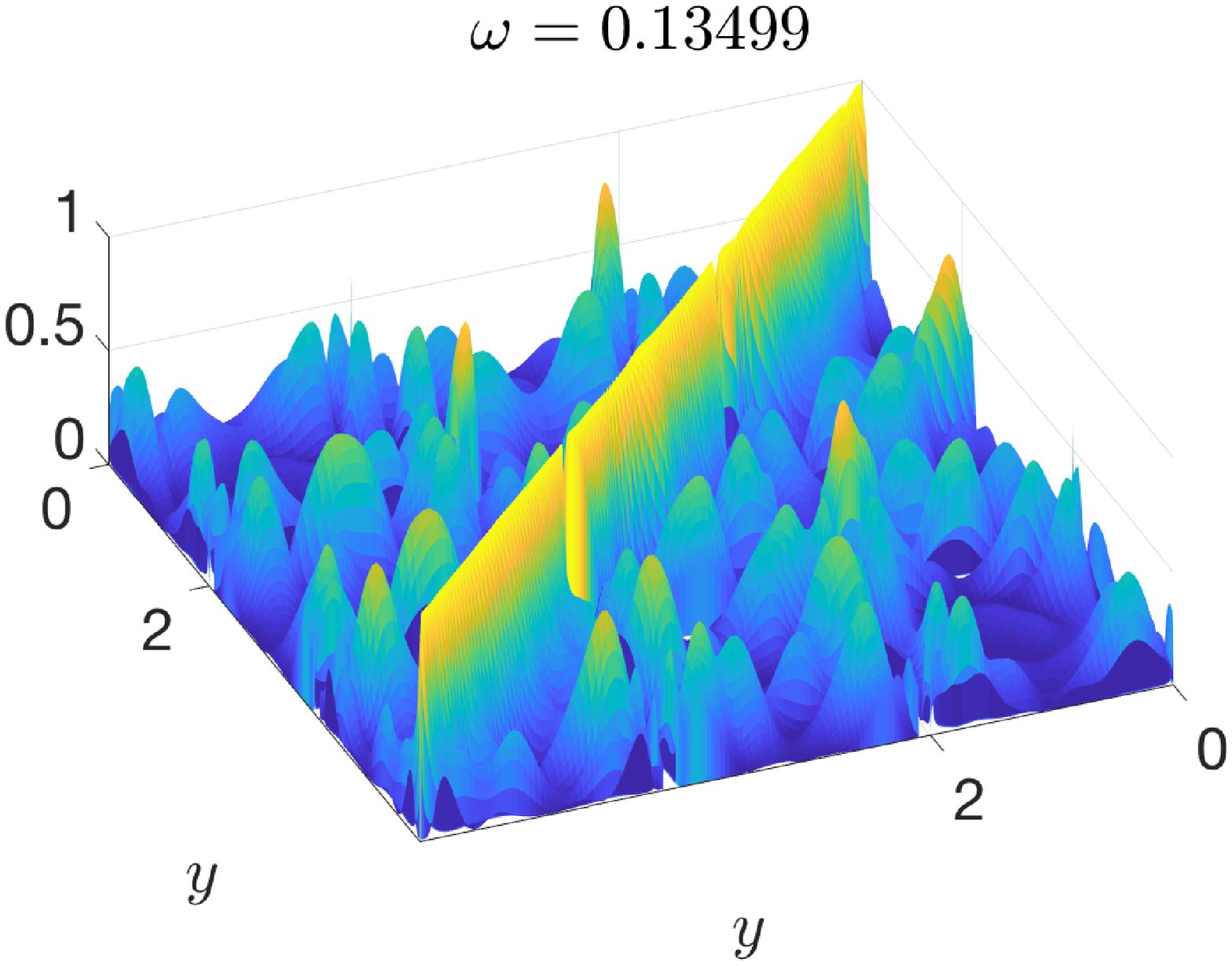}}
\caption{Cross-spectral density matrix of the response (a) and of the forcing (b) for mode $(0,1)$ and $\omega=0.135$. The coherence of each quantity is shown in (c,d). Each square corresponds to the absolute value of one of the matrix components of the given quantity, starting from the streamwise component (farther from the viewer) to the spanwise component (closer to the viewer).}
\label{fig:PffPqqAll01}
\end{figure}

A similar analysis can be performed for a slightly higher frequency still associated to an energetic region of the spectrum ($\omega=0.135$). Figure \ref{fig:PffPqqAll01} shows the same plots for this frequency. As in the case $\omega \to 0$, there is also a dominance of the streamwise component in both $\mathrm{P}$ and $\mathrm{S}$, but the off-diagonal terms are much lower than the ones at the main diagonal of the matrix. Therefore, structures for this case could be assumed as less coherent (or to have a smaller coherence length) than the previous ones. This is seen by the coherence plots in figure \ref{fig:PffPqqAll01}(c,d); as usual, the main diagonal has values equal to one, but coherence of both forcing and response decays fast as we move away from that line. Another feature that differentiate this case from the previous one is that the coherence of the cross-components of forcing/velocity for this case is also low. The observed difference between frequencies highlights that the forcing is coloured in time, with different statistics for each frequency. Models based on the white-in-time hypothesis \cite{chevalier_hoepffner_bewley_henningson_2006} would thus be unable to describe the present behaviour.

\begin{figure} 
\centering
\subfigure[Cross-spectral density matrix of the response (coloured $\mathrm{P}$)]{\includegraphics[clip=true, trim= 0 0 0 0, width=0.5\textwidth]{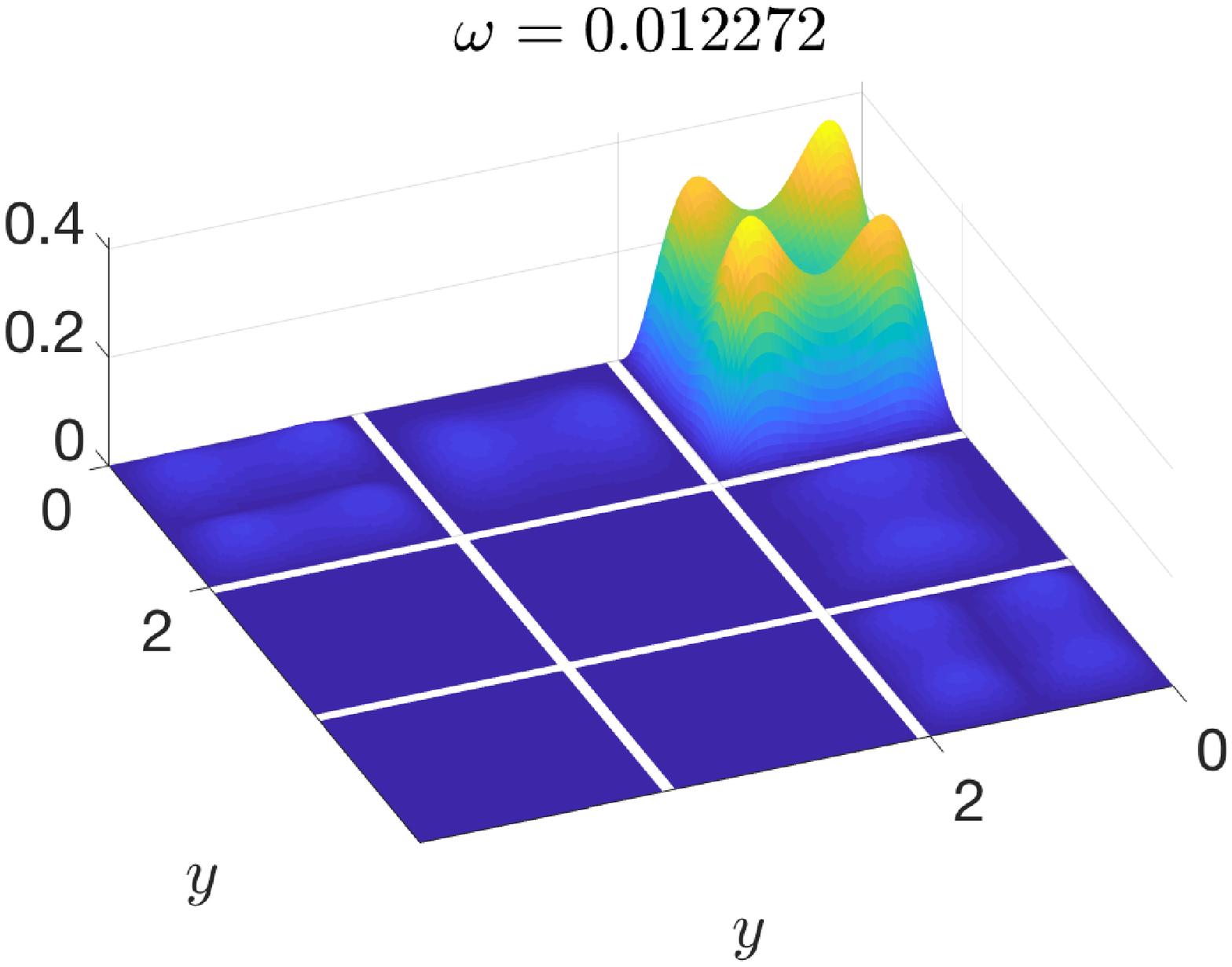}}\subfigure[Cross-spectral density matrix of the response (white $\mathrm{P}$)]{\includegraphics[clip=true, trim= 0 0 0 0, width=0.5\textwidth]{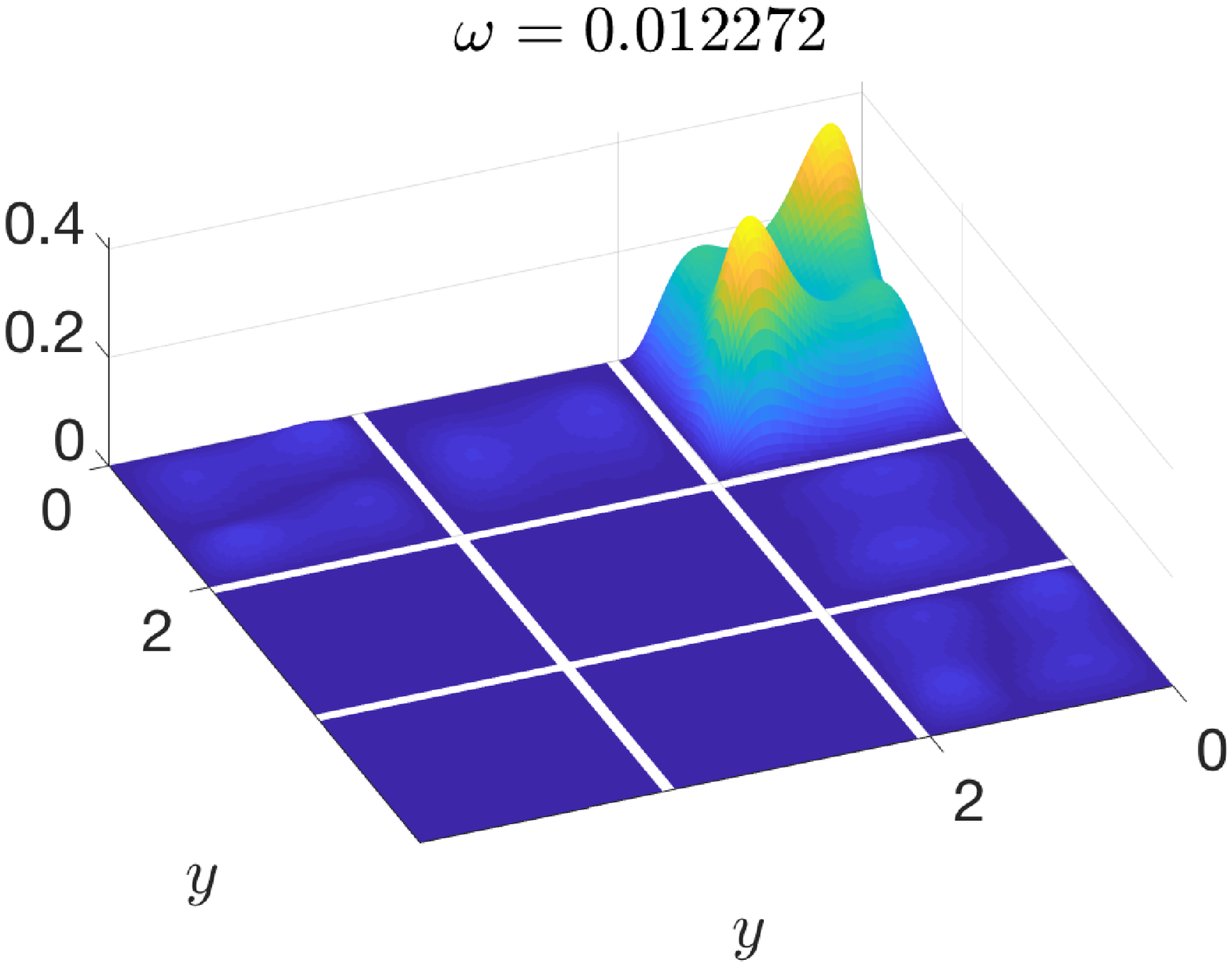}}
\subfigure[Cross-spectral density matrix of the response (coloured $\mathrm{P}$)]{\includegraphics[clip=true, trim= 0 0 0 0, width=0.5\textwidth]{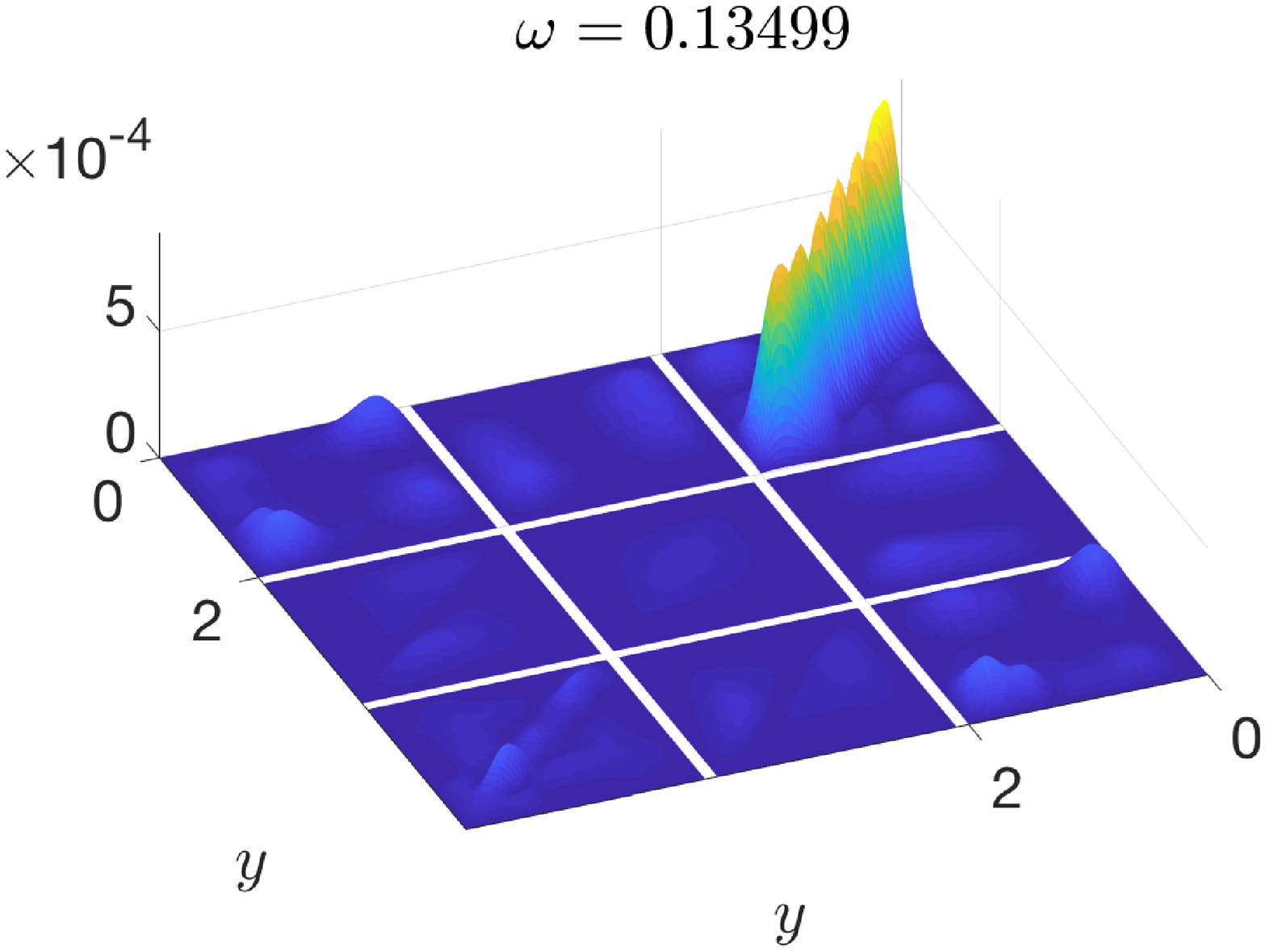}}\subfigure[Cross-spectral density matrix of the response (white $\mathrm{P}$)]{\includegraphics[clip=true, trim= 0 0 0 0, width=0.5\textwidth]{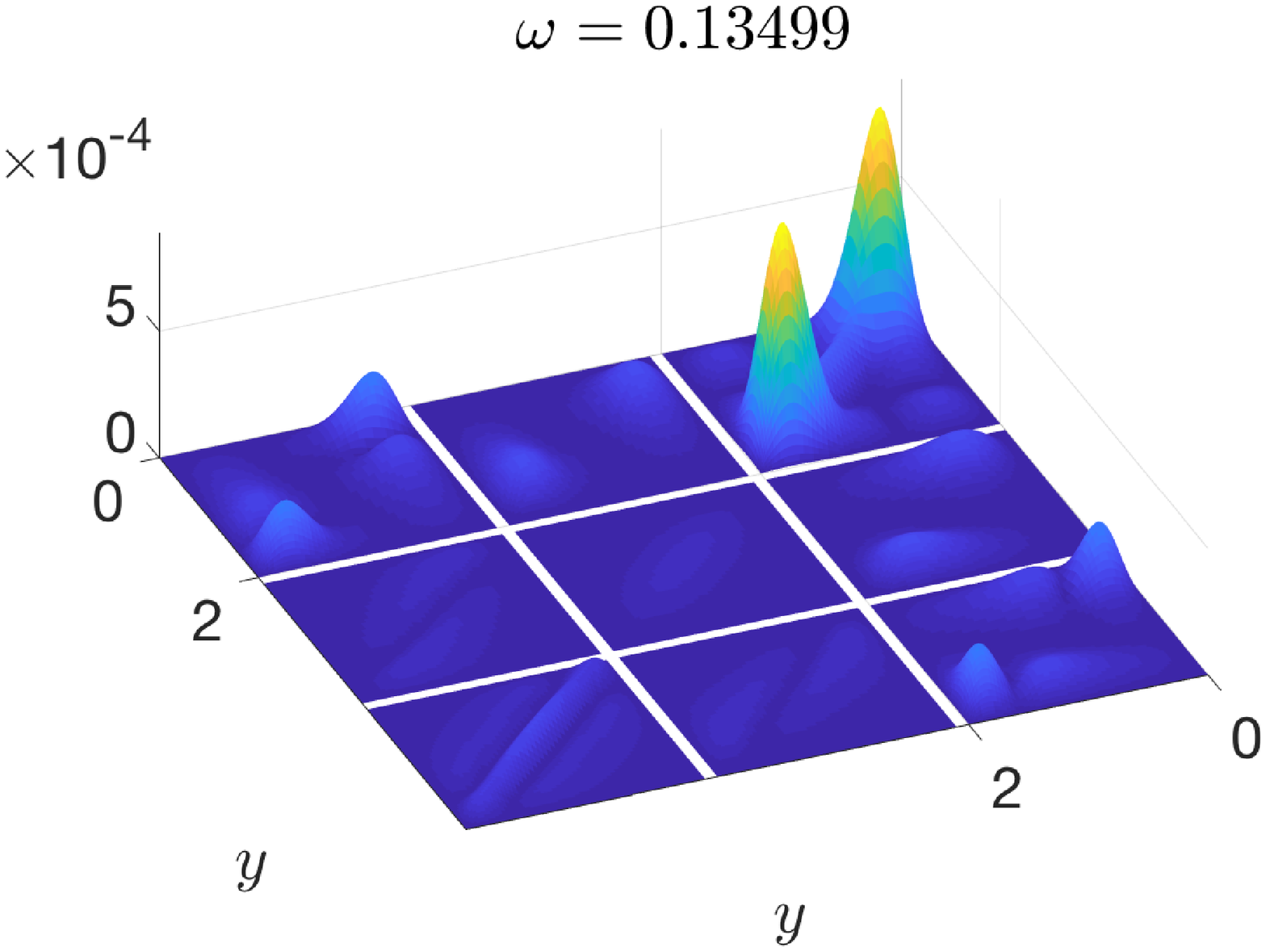}}
\caption{Predicted cross-spectral density matrix of the response using a coloured $\mathrm{P}$ (a,c) and the white-noise $\mathrm{P}$ (b,d) for the two frequencies of the study. Each square corresponds to the absolute value of one of the matrix components of the given quantity, starting from the streamwise component (farther from the viewer) to the spanwise component (closer to the viewer).}
\label{fig:PffPqqAllColour}
\end{figure}

Consider a simple coloured statistics of the forcing given by the diagonal matrix $\mathrm{P}_{{colour}}=diag(g_x(y) \ g_y(y) \ g_z(y))$, where $g_{x,y,z}(y)$ are functions that satisfy the boundary condition of the problem. These functions can be approximated, for example, by a sum of sine functions in the shape $g_{x,y,z}(y)=|\sum_{n=1}^{N_s}a_{n,x,y,z}\sin{n \pi y / 2}|$, which will lead to real positive values for the main diagonal of $\mathrm{P}$, and zero values at the boundaries. This corresponds to spatially incoherent forcing with amplitude varying in $y$. An optimisation using the \emph{fminsearch} function in MATLAB using eight sine functions for each $g_{x,y,z}(y)$ (focusing on minimising the error between the predicted $\mathrm{S}$ and the one from DNS) led to forcing statistics whose $\mathrm{S}$ predictions are shown in figure \ref{fig:PffPqqAllColour}. We call these slightly coloured statistics, since only the main diagonal is modelled, and no coherence between cross-components is considered. In figures \ref{fig:PffPqqAllColour}(a,b) we compare the results of the prediction using this model and the white-noite forcing statistics. The model seems to improve the results for the streamwise component, improving the amplitudes of the off-diagonal peaks. Still, adding a colour to the forcing in that fashion led to virtually no difference for the other components. Considering that the main problem of the prediction for this frequency is the relative amplitude of the streaks compared to the vortices, as discussed in section \ref{sec:reducedforcing} this simple model does not seem to be sufficiently accurate to recover accurately the said behaviour. 

For higher frequencies ($\omega=0.135$), the action of the present model is more concentrated in the main diagonal of $\mathrm{S}$, having low influence on the lateral lobes (which are very low for this case). By using the coloured $\mathrm{P}$, we managed to obtain higher amplitudes between the two main peaks of the streamwise component, which qualitatively follows the behaviour found in the data. Considering that the coherence associated to the covariance of the forcing decays sharply from the diagonal for this frequency, the model was expected to work better; still, the present results lack quantitative correspondence. These results point out that simple models for the statistics of the non-linear terms can be useful in order to recover some of the characteristics of the response. However, a more complete description of the statistics of the forcing is required to reproduce results from simulation and experiments when high fidelity levels are needed.

\bibliographystyle{jfm}
\bibliography{NonlinearTerms.bbl}

\end{document}